\documentclass[twocolumns]{aa}  
\usepackage{caption}
\usepackage{subcaption}
\usepackage{lineno}

\usepackage{multirow}
\usepackage{amsmath}
\usepackage{physics}
\usepackage{natbib}
\usepackage[hidelinks]{hyperref}
\hypersetup{colorlinks,linkcolor={black},citecolor={blue},urlcolor={black}} 
\bibpunct{(}{)}{;}{a}{}{,} 
\usepackage{graphicx}
\usepackage{txfonts}
\renewcommand{\arraystretch}{1.5}

\begin{document}

   \title{Perspectives for multi-messenger astronomy\\ with the next generation of gravitational-wave detectors and high-energy satellites}

   \author{S. Ronchini
          \inst{1,2}
          \and
          M. Branchesi\inst{1,2}
          \and
          G. Oganesyan\inst{1,2}
          \and
          B. Banerjee\inst{1,2}
          \and
          U. Dupletsa\inst{1,2}
          \and
          G. Ghirlanda\inst{3,4}
          \and
          J. Harms\inst{1,2}
          \and
          M. Mapelli\inst{5,6}
          \and
          F. Santoliquido\inst{5,6}
          }

   \institute{Gran Sasso Science Institute (GSSI), I-67100 L'Aquila, Italy
         \and
            INFN, Laboratori Nazionali del Gran Sasso, I-67100 Assergi, Italy
        \and
            INAF—Osservatorio Astronomico di Brera, Merate, Italy
        \and
            Università degli Studi di Milano-Bicocca, Dip. di Fisica “G. Occhialini”, Milano, Italy
        \and
            Dipartimento di Fisica e Astronomia ’G. Galilei’, Università degli studi di Padova, vicolo dell’Osservatorio 3, I-35122, Padova, Italia
        \and
            INFN, Sezione di Padova, via Marzolo 8, I-35131, Padova, Italia
             }

\abstract{
The Einstein Telescope (ET) is going to bring a revolution for the future of multi-messenger astrophysics. 
In order to detect the counterparts of binary neutron star (BNS) mergers at high redshift, the high-energy observations will play a crucial role.
Here, we explore the perspectives of ET, as single observatory and in a network of gravitational-wave (GW) detectors, operating in synergy with future $\gamma$-ray and X-ray satellites. We predict the high-energy emission of BNS mergers and its detectability in a theoretical framework which is 
able to reproduce the properties of the current sample of observed short GRBs (SGRB). We estimate the joint GW and high-energy detection rate for both the prompt and afterglow emissions, testing several combinations of instruments and observational strategies. We find that the vast majority of SGRBs detected in $\gamma$-rays will have a detectable GW counterpart; the joint detection efficiency approaches $100\%$ considering 
a network of third generation GW observatories. The probability of identifying the electromagnetic counterpart of BNS mergers is significantly enhanced if the sky localization provided by GW instruments is observed by wide field X-ray monitors. We emphasize that the role of the future X-ray observatories will be very crucial for the detection of the fainter
emission outside the jet core, which will allow us to probe the yet unexplored population of low-luminosity SGRBs in the nearby Universe, as well as to unveil the nature of the jet structure and the connections with the progenitor properties. 
}
\keywords{(stars:) gamma-ray bursts - general --
                gravitational waves --
                astroparticle physics
               }
\titlerunning{Multi-messenger astrophysics in the era of 3G GW detectors}
\authorrunning{Ronchini et al.}
\maketitle

\section{Introduction} \label{sec:intro}
The first detection of gravitational-waves from the merger of two neutron stars, GW 170817, by the Advanced LIGO \citep{Aasi2015} and Virgo \citep{Acernese2015} interferometers and the observed electromagnetic (EM) counterparts represented a milestone for multi-messenger astronomy, in particular for probing the nature of EM transients originated from BNS mergers (\citealt{GW170817GW,GW170817MM,GW170817GRBGW}). The extensive study of the EM emission allowed us to obtain an exhaustive picture of the outflow properties of both the relativistic structured jet (\citealt{Tro2017}, \citealt{Hall2017}, \citealt{Moo2018}, \citealt{Ghi2019}) and the isotropic sub-relativistic kilonova outflow (\citealt{Pian2017}, \citealt{Sma2017}, \citealt{Kas2017}). Considering the distance range expected to be reached by LIGO, Virgo and KAGRA in the next runs \citep{Abbott2020LRR} for the detection of BNS mergers and the astrophysical rate of this type of system $R_{BNS}=[10-1700]$ Gpc$^{-3}$ yr$^{-1}$ (\citealt{lvk_pop}) derived in the first three runs of observations by the Advanced GW detectors, the detection rate results to be from a few tens to a few hundreds per year in the next observational runs.Therefore, taking into account that only a fraction of these produces a jet and only a fraction of them points along the line of sight, the probability of joint GW and SGRB detection with current instruments is quite limited (\citealt{Metzger2012}, \citealt{Chen2013}, \citealt{Clark2015}, \citealt{Patricelli2016}, \citealt{Ghirlanda2016}, \citealt{Beniamini2019}, \citealt{Sal2018}, \citealt{Sal2018a}, \citealt{Sal2020}, \citealt{Perna2022},\citealt{Patricelli2022},\citealt{Colombo2022}). The advent of third generation GW observatories, such as the Einstein Telescope (ET) (\citealt{ET}) and Cosmic Explorer (CE) (\citealt{CE}), will be a substantial leap forward compared to the current instruments for the cosmological census of GW sources. 
ET is an underground observatory, designed as three low and three high frequency interferometers nested in a triangular shape forming three 10 km arms. CE, instead, is an L-shape surface interferometer of 40 km length. ET and CE are expected to detect $\sim 10^5$ BNS per year, reaching redshifts well above the star formation peak (\citealt{Maggiore:2019uih}, \citealt{CE2021}); for optimally located and oriented systems, ET will be able to detect BNS up to $z\sim5$, while CE up to $z\sim10$. For comparison, the current GW detectors are expected to reach $z\sim 0.2$, even considering new upgrades planned for the fifth run of observations in 2025 (\citealt{Abbott2020LRR}). Furthermore, accessing low frequencies makes ET capable to detect BNSs well before the merger and follow their inspiral not only for minutes but also for several hours in the case of closeby events (\citealt{Chan2018}, \citealt{Grimm2020}, \citealt{Tito2021}, \citealt{Li2021}). This enables us to use the Earth rotation to determine the sky localization even when ET is operating as a single detector \citep{Chan2018,Grimm2020,Tito2021,Li2021}. These authors found that a fraction of a few percents of BNS detected by ET will have a sky localization uncertainity $\Delta \Omega<100$ deg$^2$. This fraction increases to a few tens of percents when ET is operating in network with CE. Considering a network of three third generation detectors located in Europe, USA, and Australia a few tens of percents of detected BNS will be localized within 10 deg$^2$ and almost the totality within 100 deg$^2$ (\citealt{Li2021},\citealt{Ssohrab22}, \citealt{Mills2018}). A detailed study of  performance metrics of several configurations of network including CE, ET, and the second generation detectors and their upgrade, are given in \citealt{Ssohrab22}. 
Another proposed GW observatory to be mentioned is the Neutron star Extreme Matter Observatory \citep[NEMO,][]{nemo}, a 4-km detector whose sensitivity at high frequencies will be relevant for the investigation of neutron star physics and for multi-messenger purposes \citep{Sarin2022mm}.\\
Wide Field Of View (FOV) $\gamma$-ray and X-ray observatories will become more and more crucial to work in synergy with gravitational-wave detectors which will make it possible to reach even larger distances. They  represent a unique way to detect high-z counterparts. Being intrinsically fainter than the high energy emission, the kilonova emission will be hardly detected at distances larger than a redshift of 0.2-0.3 even with next generation optical wide FOV observatories, such as the Vera Rubin Observatory (see e.g. \citealt{Maggiore:2019uih}). Larger distances  can be reached with extremely large telescopes, such as ELT, but only arcmin-arcsec localized GW sources can be effectively pointed by these telescopes. The next generation of GW detectors observing with sensitive wide FOV high-energy satellites will enable {\it i)} to build large samples of joint detections, {\it ii)} to have joint detections at high redshift, and {\it iii)} to observe, in the local Universe, counterparts from larger viewing angles with respect to the jet axis which  will enable precise parameter estimation of the source progenitor properties. This will allow us to deeply investigate the connection between BNS mergers and short GRBs, to unveil the jet structure \citep{Hayes2020,Beniamini2019,Biscoveanu2020}, to estimate how many BNS mergers produce jets \citep{Om2022}, how the production of the jet is connected to the progenitor properties, and what is the relation between merger remnants and the properties of the high-energy emission. Accessing a large sample of joint detections at high-redshift will enable us to evaluate cosmological parameters, including the Hubble constant and the dark energy equation of state \citep{Zhao2018,Josi2021,Yu2021,Chen2021}, and test modified gravity at cosmological distances by comparing the GW luminosity distance and the one derived from the electromagnetic side \citep{Belgacem2019,Mancarella2021}. Identifying the BNS host galaxies and having a redshift will be of primary importance for these science cases. $\gamma$-ray and X-ray telescopes can potentially give precise (arcmin-arcsec) sky localization to drive the ground-based follow-up.\\ 

To maximize the scientific return of the next generation GW detectors in the multi-messenger context it is of paramount importance to evaluate the expected joint detections, to identify instrument scientific requirements and optimal observation strategy.
In this work, we focus on the synergy of third generation (3G) GW detectors with high-energy ($\gamma$-ray and X-ray) telescopes.
The novelty of our study consists in the development of a theoretical framework that starting from an astrophysically-motivated population of BNSs and the modelling of the associated high-energy EM signal is able to reproduce the statistical properties of current SGRB catalogs. This approach, based on calibrating the BNS population and the jet properties using observed SGRBs, allows us to be less dependent on the uncertainties related to the BNS population synthesis models about the local merger rate and its redshift distribution. 
On the GW side, we evaluate the detection efficiency of 3G GW detectors, as well as their capabilities in localizing the source. On the EM side, we provide a comprehensive overview of the detectability of prompt and afterglow emission associated with the SGRB.\\

The population synthesis model for the cosmic BNS merger rate density, the GW detector configurations and the modelling of the GW signal, the modelling and calibration of the prompt and afterglow emission are described in Sec.~\ref{modeldata}. In Sec.~\ref{results} we show the predictions for joint detection of GW along with prompt and afterglow emission exploring several observational strategies. We consider ET as single observatory and in different network of 3G GW detectors observing together with future high-energy facilities, which can operate in survey and pointing modes. We conclude highlighting the perspectives for joint GW and high-energy detections, discussing the optimal instruments and observational strategies to completely exploit the synergy between 3G GW detectors and $\gamma-$ and X-ray satellites, and the impact on the multi-messenger science case. 

\section{Modelling and data analysis} 
\label{modeldata}
\subsection{The BNS population model}
\label{population}

The population of BNS mergers is produced as detailed in \cite{Santoliquido2021}. The catalog of isolated compact binaries is generated with MOBSE \citep{Mapelli2017,Giacobbo2018,GiacobboMapelli2018}, which is a population synthesis-code taking into account the metallicity dependence of mass-loss rate of massive stars, the uncertainty related to the common envelope evolution, natal kicks and mass transfer efficiency. The BNS merger rate density as a function of redshift is obtained with the {\sc cosmo$\mathcal{R}$ate} code \citep{Santoliquido2020}. Here, {\sc cosmo$\mathcal{R}$ate} adopts the star formation rate density and average metallicity evolution of the Universe from \cite{MadauFragos2017}. We assume a metallicity spread $\sigma_{\rm Z}=0.3$. Redshift evolution and overall normalisation of BNS merger density depend on several assumptions regarding the common envelope prescription (parametrised by the common envelope ejection efficiency $\alpha_{\rm CE}$), the natal kick model, the supernova explosion, the mass transfer via accretion, and the shape of the initial mass function. Here, we describe electron-capture supernovae as in \cite{2019MNRAS.482.2234G} and assume the delayed supernova model \citep{Fryer2012} to decide whether a core-collapse supernova produces a black hole or a neutron star. When a neutron star forms from either a core-collapse or an electron-capture supernova, we randomly draw its mass according to an uniform distribution between 1 and 2.5 M$_\odot$.
The assumption of a uniform and broad mass distribution is consistent with the last results from GW observations \citep{GWTC3pop}. In the Appendix we discuss how using a mass distribution sharply peaked around 1.33 M$_\odot$ as observed in Galactic double neutron-star systems \citep{Ozel2016} can impact our results.
We generate the natal kicks according to $v_{\rm kick}\propto{}m_{\rm ej}/m_{\rm rem}\,{}v_{\rm H05}$, where $m_{\rm ej}$ is the mass of the ejecta, $m_{\rm rem}$ is the mass of the compact remnant (neutron star or black hole) and $v_{\rm H05}$ is a random number extracted from a Maxwellian distribution with one-dimensional root-mean square $\sigma=265$ km s$^{-1}$ \citep{Hobbs2005}. This formalism ensures low kicks for most electron-capture and ultra-stripped supernovae \citep{2020ApJ...891..141G}. We assume a value of $\alpha_{\rm CE}=3$ for common envelope, we calculate the concentration parameter $\lambda$ as in \cite{2014A&A...563A..83C} and the mass transfer formalism from \cite{Hurley2002}. We draw the mass of the progenitor primary star from a Kroupa mass function \citep{2001MNRAS.322..231K}. For the initial mass ratios, orbital periods and eccentricities we use the distributions inferred by \cite{2012Sci...337..444S}.
With such combination of assumptions the entire population consists of $\sim9\times10^5$ BNS mergers per year (observed at Earth by an ideal instrument), and the local BNS merger rate ($R_{BNS}=365\, \rm Gpc^{-3} yr^{-1}$) is consistent with the most updated constraints provided by GW observations during the first, second and third run of observations by the Advanced LIGO and Virgo detectors ($R_{BNS}\in [10-1700]\rm Gpc^{-3} yr^{-1}$, \citealt{GWTC3, GWTC3pop, GWTC2, GWTC2pop}). Moreover, the local rate assumed here is perfectly consistent with the collection of local BNS merger rates given by \cite{floorilya2021} and including both observed (from GW, kilonovae, SGRBs, and Galactic pulsar binaries' observations) and theoretically predicted rates. The BNS systems are isotropically distributed in the sky and randomly oriented.

\subsection{GW detector configuration, GW signal modeling and parameter estimation}

In this work, we consider three GW detector configurations: 1) ET as a single detector located in Sardinia (Italy), 2) ET in a network with CE (40 km arms) located in the LIGO-Livingston site, 3) ET in a network with 2CE of 40 km arms, with the second CE located in Australia. We assume the full sensitivity configuration for ET (referred as ET-D configuration, \citealt{Hild2011etd}). For CE we use the sensitivity of the 40 km arms detector given by \cite{Evans2021}. For CE as well as for each of the three combinations of high and low frequency interferometers of ET we assume a duty cycle of 0.85.  Starting from the BNS population described in Section \ref{population}, we inject GW signals constructed using a post-Newtonian formalism, in particular assuming a TaylorF2 waveform \citep{Buo2009}. Since the spin is expected to be small in binaries of neutron stars that merge within a Hubble time \citep{Burgay2003}, the effects due to spins are neglected. The signal to noise ratio (SNR) of each inspiral is then computed through a matched-filter. A network SNR threshold equal to 8 is used to select GW detections. The parameter estimation is obtained using the {\it GWFish} code \citep{Harms2022} and computing the elements of the inverse of the Fisher matrix (\citealt{Grimm2020}). The code is publicly available at this repository\footnote{\url{https://github.com/janosch314/GWFish}}.  In {\it GWFish} the waveforms, modeled in frequency domain, are implemented from scratch and the code includes an interface with LALSimulation\footnote{\url{https://lscsoft.docs.ligo.org/lalsuite/lalsimulation/}}.
Detector networks consider Earth's rotation, which has an impact on the localization capabilities. The most important computational aspects (waveform derivatives and Fisher matrix inversion) are carried out using an hybrid approach for derivatives (analytic + numeric) and a combination of matrix normalization and singular value decomposition for matrix inversion (which takes care of the quasi singularity of the Fisher matrices). Such an approach approximates the true likelihood with a Gaussian profile, which is valid in the limit of a high information content in the data.
In this regime the likelihood is highly peaked and the role of priors is negligible. Hereafter, the uncertainty on the sky localization $\Delta \Omega$ is given as 90$\%$ credible region, while the uncertainty on other GW parameters, such as $\theta_v$ and luminosity distance, is given at $1\sigma$ confidence level.  

\subsection{Prompt emission modeling}
\label{sec_prompt}
For the estimation of the high-energy EM emission from the BNS merger we assume that only a fraction $f_j$ of the mergers produces a jet which is able to penetrate the post-merger ejecta and to successfully break-out. In principle, $f_j$ could depend on the properties of the binary system, but here we do not assume any dependency. The EM counterpart is evaluated in the assumption of a universal jet structure (\citealt{Lip2001}, \citealt{Om2015}, \citealt{Om2019}, \citealt{Om2020}), which is the same of GRB 170817A. In particular, we adopt:
\begin{equation} \label{stru_1}
    \epsilon(\theta)=\frac{dE}{d\Omega}\propto \frac{1}{1+(\theta/\theta_c)^{s_{\epsilon}}}
\end{equation}
and
\begin{equation}\label{stru_2}
    \Gamma(\theta)=1+ \frac{\Gamma_0-1}{1+(\theta/\theta_c)^{s_{\Gamma}}}
\end{equation}
where $\epsilon(\theta)$ is the angular structure of the local emissivity and $\Gamma(\theta)$ the bulk Lorentz factor profile ($\theta$ is the polar angle measured from the jet symmetry axis). The choice of $\theta_c$ and $\Gamma_0$ influences the number of prompt emission detections, while the off-core specific shape of the jet has more impact on the brightness of the afterglow component and the prompt tail. From the modeling of radio afterglow of GRB 170817A \citep{Ghirlanda2019}, the best fit parameters of the jet structure are:
\begin{equation}
    s_{\epsilon}=5.5^{+1.3}_{-1.4}, \,
    s_{\Gamma}=3.5^{+2.1}_{-1.7}, \,
    \theta_c=(3.4 \pm 1.0) \text{ deg}
\end{equation}
For our modeling, we fix for simplicity $s_{\epsilon}=s_{\Gamma}=s$ and we adopt:
\begin{equation}\label{stru_3}
    s=4, \,
    \theta_c=3.4 \text{ deg}
\end{equation}
Moreover, we fix $\Gamma_0=500$, which is in agreement with usual values inferred for GRBs (e.g. \citealt{GG2018}).
These parameters are the same for the modeling of both prompt and afterglow emission. A broader structure with $ s=2$ and same $\theta_c$ is also tested \citep{Rossi2002}. Hereafter we refer to:
\begin{enumerate}
    \item \emph{Stru1} for the case $s=4$ and $\theta_c=3.4$ deg
    \item \emph{Stru2} for the case $s=2$ and $\theta_c=3.4$ deg
\end{enumerate}
The dependency of our results on the choice of the structure parameters is discussed later in the paper. The aperture angle assumed here is consistent with the values found in SGRBs with well constrained jet break (typical values of $\theta_c$ are in the interval $\sim 1^{\circ}-8^{\circ}$, see \citealt{Fong2014}, \citealt{Jin2018} and references within). Similar values are found in hydrodynamical simulations of jet propagation in NS mergers, e.g. \citealt{Laz2018} predicts a jet aperture angle of $\sim 5^{\circ}$ (see also \citealt{Geng2019}, \citealt{Wu2018} ). Also \cite{Lamb2022} find a rather small jet core ($\theta_c\sim 2.2^{\circ}$) and a jet profile compatible with eq.~\ref{stru_1} and \ref{stru_2}, with best fit values $s_{\epsilon}\sim3.1$ and $s_{\Gamma}\sim 1.8$.  \\

We define $E_t$ the beaming corrected energy budget of the jet, which is connected to the isotropic energy measured by an observer aligned to the jet (on-axis) in the following way:
\begin{equation}
    E_{\rm iso}=\frac{4 \pi}{\Delta \omega} E_{t}
\end{equation}
where $\Delta \omega\sim 4\pi (1-\cos(\theta_c))$. $E_t$ is related to $\epsilon(\theta)$ through the following relation:
$$
E_t=\int_{\ \Omega}\epsilon(\theta)d \Omega.
$$
$E_t$ does not represent the energy content of the jet, but rather the energy which is converted into radiation through dissipation mechanisms (such as internal shocks or magnetic reconnection). 
We extract the value of $E_{t}$ from the following probability distribution:
\begin{equation} \label{distr_E}
    P\left(E_{t} / E^{*}\right) \propto\left(E_{t} / E^{*}\right)^{-\lambda_E} e^{-E^{*} / E_{t}}
\end{equation}
which is a power law with a low-energy exponential cut-off. The prompt emission spectrum is assumed to be angle-independent in the jet comoving frame and follows a smoothly broken power law, with low- and high- energy photon indexes of $2/3$ and 2.5, respectively. These are typical values measured for SGRBs \citep{Lara2011}. In the frame of an on-axis observer, we assume that the spectral peak energy has a log-normal distribution, i.e.:
\begin{equation}
    P(\log(E_p)) \propto \exp\left(-\frac{(\log(E_p)-\log(\mu_E))^2}{\sigma_E^2}\right)
\end{equation}
where $\mu_E$ and $\sigma_E$ are the mean value and the standard deviation of the distribution, respectively.
The assumed distribution function corresponds to that adopted by \cite{Ghirlanda2016}.
Such distribution in the observer frame is equivalent to assume a monochromatic distribution of $E_p'$ in the jet comoving frame and a log-normal distribution of $\Gamma_0$ (indeed $E_p=\Gamma_0 E_p'$). For the computation of the burst duration, we compute the time $t_{90}$ such that the energy emitted from 0 to $t_{90}$ is equal to the 90$\%$ of the total energy released by the GRB. The light curve is approximated with a linear rising phase from $t=0$ to the peak time $t=t_p$, followed by a decreasing phase which evolves according to high latitude emission (HLE). Therefore we assume that a single pulse dominates the SGRB emission. For $t_p$ we consider a log-normal distribution as well, as measured in the rest frame of the burst:
\begin{equation}
    P(\log(t_p)) \propto \exp\left(-\frac{(\log(t_p)-\log(\mu_{\tau}))^2}{\sigma_{\tau}^2}\right)
\end{equation}
The angle-dependent prompt emission light curve is computed adopting the approach of \cite{Ascenzi2020}. Specifically, we work in the approximation of an infinitesimal duration emission at a fixed radius $R_0$, namely the burst duration in the rest frame is negligible compared to the time needed by the jet to propagate up to $R_0$. The jet moves with a bulk Lorentz factor $\Gamma_0$. As shown by \cite{Oga2020}, the presence of a jet structure like the one adopted here produces an X-ray light curve with a steep decay followed by a shallower phase, known as plateau. The timescale $\tau_p=R_0/2c\Gamma_0^2$, which corresponds to the typical prompt emission timescale, determines the duration of the plateau and larger values of $\tau_p$ produce a longer plateau. In order to find a fiducial combination of $R_0$ and $\Gamma_0$, we considered all the SGRBs from the \emph{Swift}-XRT catalog with a plateau phase. We verified that the end time of the plateau typically spans in the range $10^2-10^3$ s. Having fixed  $\Gamma_0=500$, we therefore adopted $R_0=5\times 10^{14}$ cm in order to obtain a typical plateau duration around few hundred seconds. With this approach we are able to predict the prompt emission output, as well as the contribution at later time given by the jet wings due to HLE effects, which can be relevant for the detectability of the SGRB afterglow, especially in the X-rays and for off-axis observers.

\subsection{Calibration of the prompt emission model}
\label{calib}
As illustrated in the previous section, our prompt emission model depends on seven free parameters which determine the observational properties of our SGRB sample. Here, we show how our prompt emission model is calibrated such that the BNS population is able to reproduce the number of SGRBs detected by \emph{Fermi}-GBM and their statistical properties reported in the \emph{Fermi}-GBM catalog\footnote{https://heasarc.gsfc.nasa.gov/W3Browse/fermi/fermigbrst.html}.
In particular, the choice of our fiducial parameters is done comparing our simulated observables with the distributions of observed peak energy, duration, peak flux and fluence reported by \cite{Ghirlanda2016}. This calibration sample is obtained by selecting SGRBs in the \emph{Fermi} catalog with peak flux $F_p> 5 \text{ ph }\text{cm}^{-2}\text{ s}^{-1}$ to ensure the completeness of the sample. For the parameter estimation we use Goodman $\&$ Weare’s Affine Invariant Markov chain Monte Carlo (MCMC) Ensemble sampler \footnote{We adopted 16 walkers and 30000 steps}. The likelihood function is $\log{\mathcal{L}}=\log(P_C(N_{\rm pre}|N_{\rm det}))+\sum_{i=1}^4 \log(P_{\rm KS,i})$, where $P_C(N_{\rm pre}|N_{\rm det})$ is the Poissonian probability that \emph{Fermi}-GBM detects $N_{pre}$ SGRBs per year given that the average observed number is $N_{det}$, while $P_{KS,i}$ is, for each observable $i$, the Kolmogorov-Smirnov probability that the predicted and observed distributions come from the same population. The adopted likelihood is valid in the approximation that each observable is independent from the others.\\

The detection rate of SGRBs $R_{\rm SGRB, det}$ can be related to the BNS rate $R_{\rm BNS}$ as:
\begin{equation}
    R_{\rm SGRB, det}=\int [\frac{dN_{\rm SGRB}}{dz dt}|_{\rm det}]dz
\end{equation}
where 
\begin{equation}\label{f_j}
    \frac{dN_{\rm SGRB}}{dz dt}|_{\rm det}=\expval{\Omega_{\rm obs}}\times \phi_{\rm det}(z)\times f_{j}\frac{dN_{\rm BNS}}{dz dt}.
\end{equation}
$\expval{\Omega_{\rm obs}}$ is the time-averaged sky coverage of the instrument and $f_{j}$ is the fraction of BNS able to form a jet.
The quantity $\frac{dN_{\rm BNS}}{dz dt}$ is the number of BNS mergers observed at Earth per unit time and redshift and it is provided by the population synthesis model.  
$\phi_{\rm det}(z)$ is the variable that combines the detection efficiency of the instrument (which takes into account the k-correction) and the beaming of the SGRB emission. Namely, $\phi_{\rm det}(z)$ represents the probability that a SGRB at redshift $z$ is detectable by a given instrument at Earth, i.e.:
$$
\phi_{\rm det}(z)=\frac{N_{\rm SGRB}(F>F_{\rm lim},z)}{N_{SGRB}(z)}
$$
where $N_{\rm SGRB}(z)=\dfrac{dN_{\rm SGRB}}{dz dt}$,  $N_{\rm SGRB}(F>F_{\rm lim},z)=\dfrac{dN_{\rm SGRB}}{dz dt}(F>F_{\rm lim})$ and $F_{\rm lim}$ is the limiting flux in the band of the instrument. In turn, $N_{\rm SGRB}(F>F_{\rm lim},z)$ depends on the luminosity function of SGRBs, their average spectral properties, as well as on the jet structure.\\

The inclusion of $f_j$ as free parameter of our model makes our estimates of absolute numbers of detections independent on the overall normalization of the BNS population model which is still subject to large uncertainties \citep{Santoliquido2021}. Indeed the likelihood function is written in such a way that, whatever the BNS merger rate normalisation is, the value of $f_j$ is optimized to reproduce the current average rate of \emph{Fermi}-GBM detections. Namely, in eq.~\ref{f_j} the quantity $\frac{dN_{BNS}}{dz dt}$ is uniquely determined by the BNS population model, $\phi_{det}(z)$ depends on the prompt emission model and therefore the MCMC converges to that value of $f_j$ that reproduces $R_{\rm SGRB, det}$, which is known.\\

The MCMC sampling has been performed for two different jet structures \emph{Stru1} and \emph{Stru2}.
For both structures, we obtain the posterior distribution of the model parameters. The confidence intervals of the parameters are reported in Tab.~\ref{tab_mc}, while the corner plots of the posterior distributions are reported in Figs.~\ref{corner_s1} and \ref{corner_s2} for \emph{Stru1} and \emph{Stru2}, respectively. The modification of the off-core slope of the jet structure does not influence the best fit parameters, except for $f_j$, which tends to be smaller for $s=2$. Such result is justified by the fact that a broader structure implies a larger detectability of the prompt emission at larger viewing angles, which corresponds to a larger number of off-axis detections. Therefore, for a given BNS population and an average SGRB detection rate, a broader structure requires a smaller fraction of BNS to be able to produce SGRB. The confidence interval we obtain for $f_j$ is not particularly tight and it is compatible with other works which combine EM observations of SGRBs and BNS rates from GW observations \citep{Ghirlanda2019, Om2022, Sarin2022, Ben2019, Lamb2017}.

\begin{table}[t]
\centering
\begin{tabular}{|c|c|c|}
\hline
parameter & Stru 1 & Stru 2  \\\hline
$\lambda_E$             & $6.7^{+3.6}_{-3.3}$   & $6.4^{+3.7}_{-3.5}$   \\\hline
$\log_{10}{(E_t/10^{49} \text{erg/s})}$         & $-0.30^{+0.40}_{-0.34}$ & $-0.23^{+0.48}_{-0.69}$\\\hline
$\log_{10}{(\mu_E/\text{keV})}$         & $3.2^{+0.1}_{-0.1}$       & $3.2^{+0.1}_{-0.1}$        \\\hline
$\sigma_E$              & $0.36^{+0.10}_{-0.09}$    & $0.37^{+0.10}_{-0.10}$  \\\hline
$\log_{10}{(\mu_{\tau}/s)}$     & $0.05^{+0.24}_{-0.31}$    & $0.02^{+0.09}_{-0.10}$  \\\hline
$\sigma_{\tau}$             & $0.60^{+0.10}_{-0.11}$    & $0.59^{+0.09}_{-0.10}$  \\\hline
$\log_{10}{f_j}$        & $-0.59^{+0.37}_{-0.32}$   & $-0.95^{+0.52}_{-0.41}$  \\\hline

\end{tabular}
    \caption{Prompt emission model parameters from the MCMC sampling for the two jet structures analysed in the present work. The interval of confidence from the posterior distribution of the prompt emission parameters is given as 1$\sigma$.}
    \label{tab_mc}
\end{table}

\subsection{Forward shock modeling}

\begin{figure}
    \centering
    \includegraphics[width=1.0\columnwidth]{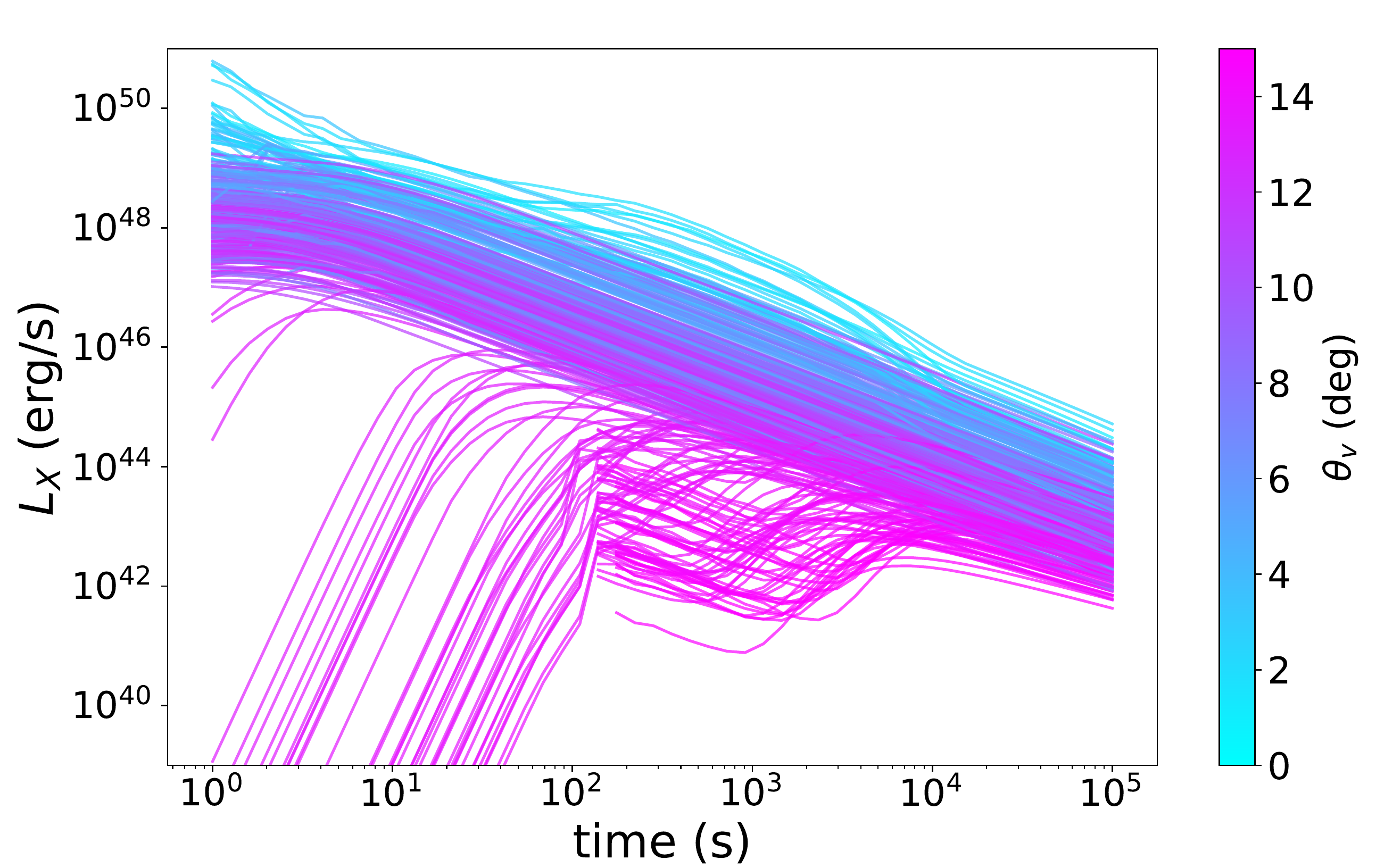}
    \caption{X-ray light curves of SGRBs with viewing angle $0^{\circ}<\theta_v<15^{\circ}$ predicted by our model. The lines are produced with a random extraction from the posterior distribution of prompt parameters. In the computation we assume \emph{Stru1} and we include both HLE and FS (configuration \emph{FS-SGRB}). The color of each line indicates the viewing angle. The X-ray luminosity is computed at 5 keV.}
    \label{lc_sample}
\end{figure}

The forward shock (FS) light curves are produced using the Python package \emph{afterglowpy} \citep{Ryan2020}. For the jet kinetic structure we assume the same profile adopted for the structure of the comoving frame emissivity reported in the prompt emission modeling (see eq.~\ref{stru_1}). This implies that we are assuming that the conversion efficiency from kinetic energy to radiated energy has no angular dependence. The detectability of X-ray emission is evaluated for both \emph{Stru1} and \emph{Stru2}. \emph{Afterglowpy} includes the parameter $\theta_w$ which is the angular extension of the jet wings. For \emph{Stru1} we assume $\theta_w=15$ deg, while for \emph{Stru2} we assume $\theta_w=30$ deg. The impact of the choice of $\theta_w$ on the detectability of X-ray emission is discussed later in the text.
The model depends also on the micro-physical parameters $n_0,p, \epsilon_{e}$ and $ \epsilon_{B}$ which are the ISM density, the slope of the electron energy distribution, the fraction of energy carried by electrons and the fraction of energy carried by the magnetic field, respectively. In the following, we perform our simulation using two setups of parameters:
\begin{enumerate}
    \item $n_0=2.5\times 10^{-4} \text{ cm}^{-3}$, $p=2.2$, $\epsilon_e=0.1$ and $\epsilon_B=1.3\times 10^{-4}$, which are the best fit values obtained from the multi-wavelength (X-ray, optical, radio) modeling of the X-ray afterglow of GRB 170817A (\citealt{Ghirlanda2019}). These values are also consistent with the confidence intervals reported by \cite{Wu2019}. We call this configuration \emph{FS-GW17}.
    \item For the ISM density we take the fiducial interval derived by \cite{Fong2015} $n_0\in(3-15)\times 10^{-3} \text{ cm}^{-3}$. The other parameters are fixed to $p=2.2$, $\epsilon_e=0.1$ and $\epsilon_B \in [0.01-0.1]$. $n_0$ and $\epsilon_B$ are uniformly extracted from the confidence intervals reported above. This configuration is more representative for SGRB population with respect to \emph{FS-GW17}, and we call it \emph{FS-SGRB}.
\end{enumerate}
If we call $\eta$ the fraction of kinetic energy which is transformed into radiation, then $E_{\rm rad}/E_{\rm kin}=(1-\eta)/\eta=\tilde{\eta}$ and we assume $\tilde{\eta}$ randomly distributed in the interval $[0.01-0.1]$. Fig.~\ref{lc_sample} shows a collection of X-ray light curves produced with our model at different viewing angles, including both HLE and FS (configuration \emph{FS-SGRB}). The curves are obtained with a random extraction from the probability distributions of the prompt parameters described in sec.~\ref{sec_prompt} and the parameter of each distribution is extracted from the Monte Carlo posterior described in sec.~\ref{calib}.

\section{Results}
\label{results}
\subsection{Joint detection of GWs and the prompt emission}
\label{gw+pro}

\begin{figure}
    \centering
    \includegraphics[width=1.0\columnwidth]{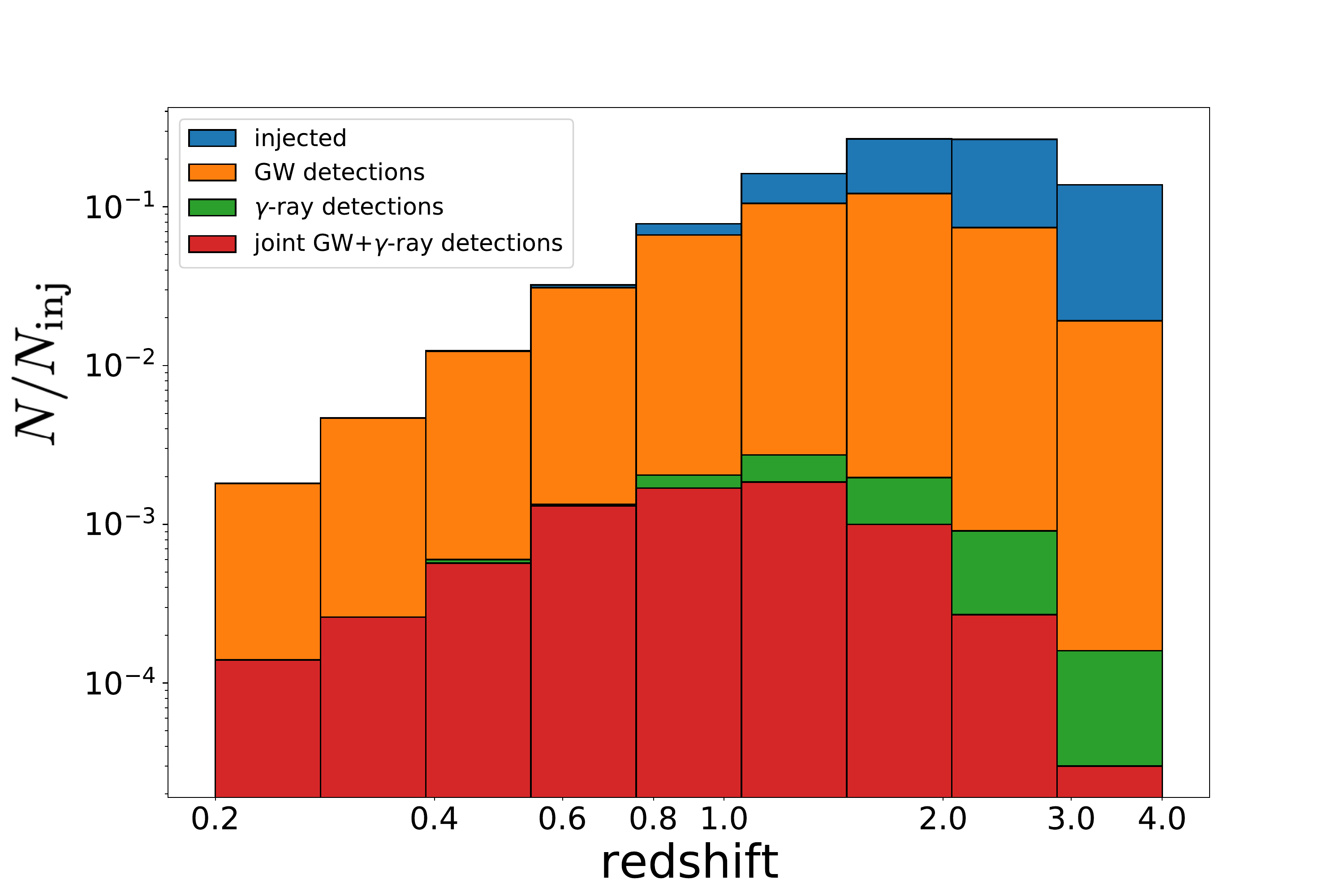}
    \caption{Histogram of the joint ET+\emph{Fermi}-GBM detections (red), together with the distribution of injected BNS (blue), the mergers detected by ET (orange), the mergers detected by \emph{Fermi}-GBM (green). The histogram is normalised to $N_{inj} =10^5$, which is the number of injected BNS mergers in the angle range $0^{\circ}<\theta_v<15^{\circ}$ and in the redshift range $0<z<4$. For visualization purposes, the fraction of BNS producing a jet, $f_j$, has been assumed to be one.}
    \label{hist_JD}
\end{figure}

\begin{figure}
    \centering
    \includegraphics[width=1.0\columnwidth]{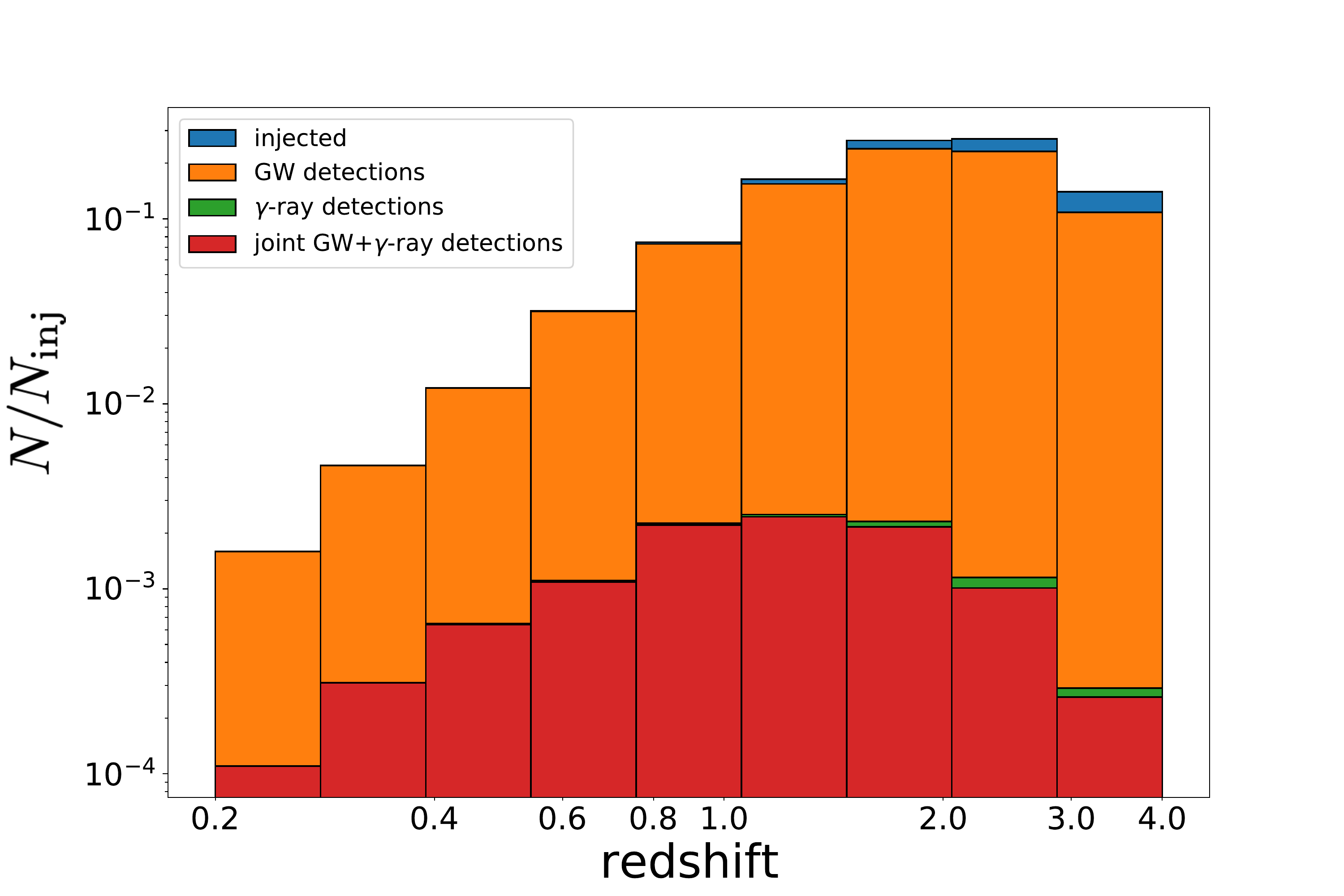}
    \caption{Same as Fig. \ref{hist_JD}, but considering \emph{Fermi}-GBM+(ET+CE).}
    \label{hist_JD_ETCE}
\end{figure}

\begin{table*}[t]
\centering
\begin{tabular}{|l|c|c|c|c|c|c|c|c|}
\hline
\multirow{2}{*}{INSTRUMENT}&band &$F_{\rm lim}$&\multirow{2}{*}{FOV/$4\pi$}&\multirow{2}{*}{loc. acc.}&Joint ET& \multirow{2}{*}{$N_{JD}/N_{\gamma}$} &Joint (ET+CE)& \multirow{2}{*}{$N_{JD}/N_{\gamma} $}\\ 
&MeV&erg cm$^{-2}$ s$^{-1}$&&&+$\gamma$-ray&&+$\gamma$-ray&\\ \hline
\emph{Fermi}-GBM  & 0.01 - 25 &0.5$(^*)$& 0.75 & 5 deg $(^{a})$  & $33^{+14}_{-11}$          & $68_{-18}^{+13} \% $   & $47_{-14}^{+14}$ & $95_{-7}^{+5}\%$            \\\hline

\emph{Swift}-BAT        & 0.015 - 0.15 &$2\times 10^{-8}$& 0.11 & 1-3  arcmin  & $10^{+3}_{-3}$          & $62_{-14}^{+11}\%$  & $13_{-4}^{+5}$ & $94_{-7}^{+6}\% $         \\\hline

GECAM      & 0.006 - 5 &$2\times 10^{-8}$& 1.0 & 1 deg  & $121_{-48}^{+84}$           & $57_{- 10}^{+8}\%$  & $205_{-72}^{+145}$ & $92_{-5}^{+4}\%$        \\\hline

SVOM-ECLAIRs    &0.004 - 0.250 &1.792(*)& 0.16 & $<10$ arcmin &$3_{-1}^{+1}$           & $69_{-9}^{+10}\%$  & $4_{-1}^{+1}$ & $95_{-4}^{+5}    \%$  \\
\hline
SVOM-GRM    &0.03 - 5 & 0.23(*)& 0.16 & $\sim 5$ deg &$9_{-3}^{+4}$           & $59_{-6}^{+6}\%$  & $14_{-4}^{+6}$ & $92_{-3}^{+3}    \%$  \\
\hline

THESEUS-XGIS       & 0.002 - 10 &$3\times 10^{-8}$& 0.16 & $<15$ arcmin  &$10_{-4}^{+5}$           & $63_{-13}^{+13}\%$  & $15_{-4}^{+6}$ & $94_{-7}^{+6}\%$       \\\hline

HERMES     & 0.05 - 0.3 &0.2$(^*)$& 1.0 & 1 deg  &$84_{-30}^{+42}$          & $61_{-11}^{+10}\%$    & $139_{-36}^{+54}$ & $94_{-6}^{+6}\%$     \\\hline
TAP-GTM    &0.01 - 1 &1$(^*)$& 1.0 & 20 deg &$60_{-24}^{+24}$           & $67_{-14}^{+13}\%$  & $84_{-24}^{+30}$ & $95_{-6}^{+5} \% $     \\\hline

\end{tabular}
\flushleft
\footnotesize{$(^a)$ The value indicates the 50$\%$ percentile of the localization error of the fourth \emph{Fermi}-GBM catalog \citep{Kien2020}. 90$\%$ of the GRBs detected by Fermi have a localization error below 15 deg.\\
$(^*)$ expressed in ph cm$^{-2}$ s$^{-1}$}
\rule{\textwidth}{0.4pt}
    \caption{In this table we report the number of joint GW+$\gamma$-ray detections for different combinations of $\gamma$-ray instruments. Columns 5 and 7 give the number of joint GW+$\gamma$-rays detections during one year of observation with ET alone, and the network of ET+CE, respectively. Column 6 and 8 give $N_{JD}/N_{\gamma}$, the fraction of $\gamma$-ray detections which have also a GW counterpart for ET and ET+CE, respectively. The jet structure \emph{Stru1} is assumed.}
    \label{tab_joint}
\end{table*}

\begin{table*}[t]
\centering
\begin{tabular}{|l|c|c|c|c|c|c|c|c|}
\hline
\multirow{2}{*}{INSTRUMENT}&band &$F_{\rm lim}$&\multirow{2}{*}{FOV/$4\pi$}&\multirow{2}{*}{loc. acc.}&Joint ET& \multirow{2}{*}{$N_{JD}/N_{\gamma}$} &Joint (ET+CE)& \multirow{2}{*}{$N_{JD}/N_{\gamma} $}\\ 
&MeV&erg cm$^{-2}$ s$^{-1}$&&&+$\gamma$-ray&&+$\gamma$-ray&\\ \hline
\emph{Fermi}-GBM  & 0.01 - 25 &0.5$(^*)$& 0.75 & 5 deg $(^{a})$& $36^{+14}_{-14}$          & $70_{-14}^{+16} \% $   & $47_{-14}^{+18}$ & $95_{-7}^{+5} \% $          \\\hline

\emph{Swift}-BAT        & 0.015 - 0.15 &$2\times 10^{-8}$& 0.11 & 1-3  arcmin  & $9^{+6}_{-3}$          & $64_{-13}^{+11}\%$  & $14_{-4}^{+7}$ & $94_{-7}^{+6}  \% $        \\\hline

GECAM      & 0.006 - 5 &$2\times 10^{-8}$& 1.0 & 1 deg  & $163_{-60}^{+175}$           & $60_{-8}^{+7}\%$  & $253_{-96}^{+271}$ & $92_{-4}^{+3}  \%  $  \\\hline

SVOM-ECLAIRs    &0.004 - 0.250 &1.792(*)& 0.16 & $<10$ arcmin &$3_{-1}^{+1}$           & $71_{-10}^{+15}\%$  & $4_{-1}^{+1}$ & $95_{-4}^{+5}    \%$  \\
\hline
SVOM-GRM    &0.03 - 5 & 0.23(*)& 0.16 & $\sim 5$ deg &$11_{-4}^{+6}$           & $60_{-6}^{+13}\%$  & $15_{-5}^{+10}$ & $92_{-3}^{+3}    \%$  \\
\hline

THESEUS-XGIS       & 0.002 - 10 &$3\times 10^{-8}$& 0.16 & $<15$ arcmin    &$11_{-4}^{+7}$           & $66_{-16}^{+12}\%$  & $16_{-5}^{+7}$ & $94_{-7}^{+6} \% $     \\\hline

HERMES     & 0.05 - 0.3 &0.2$(^*)$& 1.0 & 1 deg  &$96_{-31}^{+60}$          & $64_{-12}^{+12}\%$    & $151_{-48}^{+66}$ & $94_{-6}^{+4}  \% $   \\\hline
TAP-GTM    &0.01 - 1 &1$(^*)$& 1.0 & 20 deg &$66_{-24}^{+24}$           & $69_{-14}^{+15}\%$  & $90_{-24}^{+30}$ & $95_{-7}^{+5}    \% $ \\
\hline

\end{tabular}
    \caption{As in Tab. \ref{tab_joint}, assuming the jet structure \emph{Stru2}.}
    \label{tab_joint_stru2}
\end{table*}

\begin{figure}
    \centering
    \includegraphics[width=1.0\columnwidth]{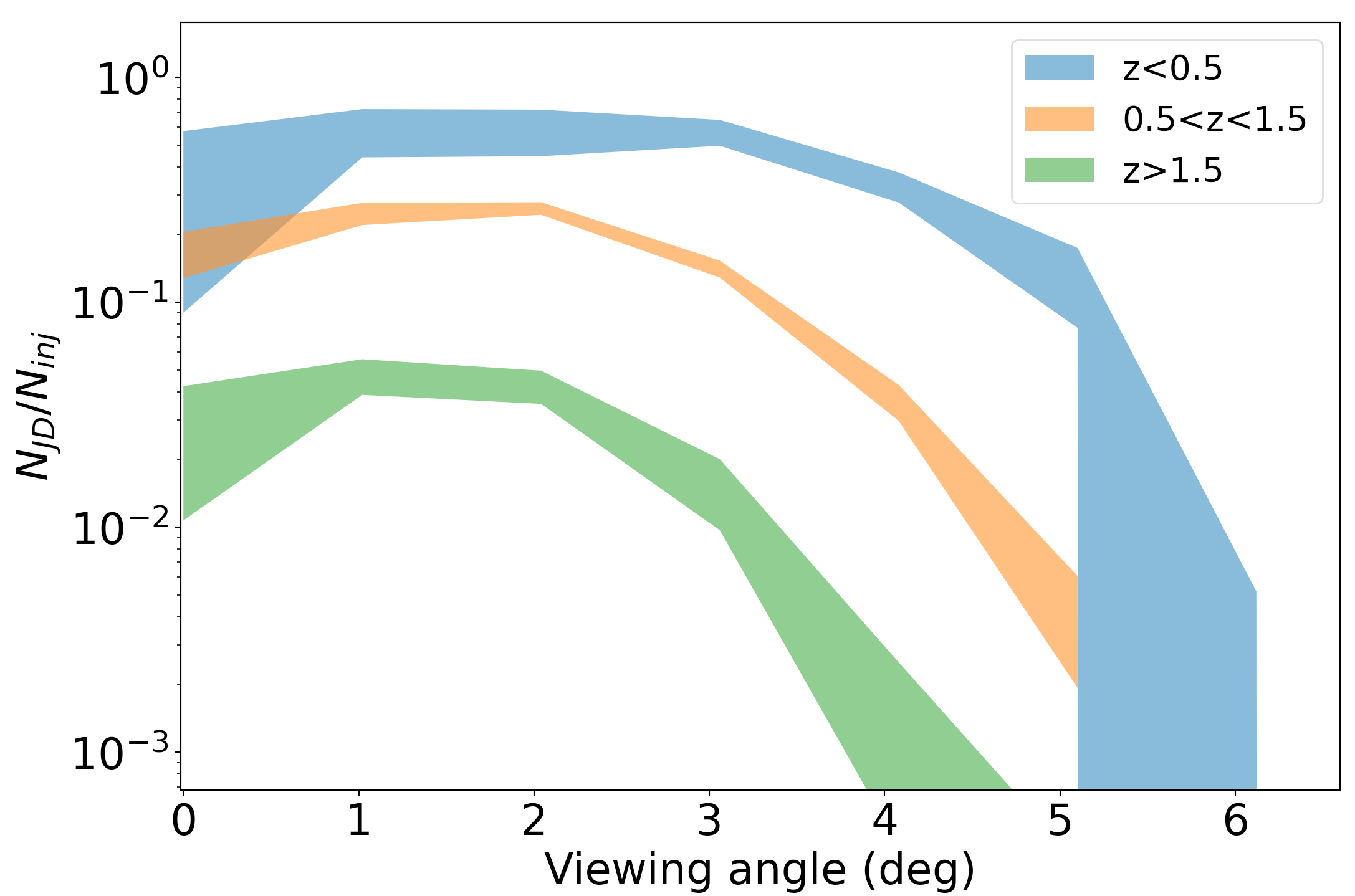}
    \caption{Angle dependency of the joint GW+$\gamma$-ray detections for three redshift bins, considering ET+\emph{Fermi}-GBM. The bands are obtained with random realizations extracted from the MCMC posterior samples. For each redshift bin, the y-axis gives the number of joint detections normalised to $5\times 10^4$ BNS injections with viewing angle $0^{\circ}<\theta_v<15^{\circ}$. We assume that all BNS produce a jet ($f_j=1$).}
    \label{JD_vs_angle}
\end{figure}

For the detection of the prompt emission from our SGRBs population, we consider the following $\gamma$-ray instruments: the Gamma-ray Burst Monitor on board of \emph{Fermi} (\emph{Fermi}-GBM, \citealt{GBM}), the \emph{Swift}-Burst Alert Telescope (\emph{Swift}-BAT, \citealt{BAT}), the Gravitational wave high-energy Electromagnetic Counterpart All-sky Monitor (GECAM, \citealt{gecam}), the ECLAIRs telescope and the Gamma-Ray burst Monitor (GRM) on board of the Space-based multi-band astronomical Variable Objects Monitor (SVOM, \citealt{svom2015,svom2012}), the X/Gamma-ray Imaging Spectrometer (XGIS) on board of the Transient High-Energy Sky and Early Universe Surveyor (THESEUS, \citealt{theseus,Amati2021,Ciolfi2021,Rosati2021}), the Gamma-ray Transient Monitor (GTM) on board of the Transient Astrophysical Probe (TAP, \citealt{TAP}), and the High Energy Rapid Modular Ensemble of Satellites (HERMES, \citealt{hermes}).
The first three are currently operating missions; \emph{Swift} and \emph{Fermi} have been observing for more than 10 years, and GECAM was launched in 2020. SVOM is expected to be launched in 2023. Nominally they are not expected to be operative in mid '30s and '40s when ET and CE will start observations, but we take them as reference instruments.  
THESEUS, TAP, and HERMES 
are mission concepts. The HERMES-Pathfinder project consisting in the deployment of six cube-satellites is expected to be operative in the next few years \citep{Fiore2021}. In this work we consider the final HERMES as a full constellation of cube-satellites. In order to establish whether the prompt emission is detected, we compute the peak flux during 1 second exposure and we compare this value with the detection threshold of the instrument. The viewing angle of each injection is assumed to be exactly the same used for the evaluation of the GW signal, and thus extracted from the randomly oriented BNS population (see Sect.~\ref{population}).\\
In order to estimate the number of joint detections, we simulate the high-energy emission (keV-MeV) according to the model described in Sect.~\ref{sec_prompt}. The results relative to one year of joint GW+$\gamma$-ray detections are reported in Tabs.~\ref{tab_joint} and \ref{tab_joint_stru2} for \emph{Stru1} and \emph{Stru2}, respectively, where  we considered ET and ET+CE as GW networks. The uncertainty intervals are computed simulating 1000 years of observations and for each year the prompt emission parameters are randomly extracted from the Monte Carlo posterior distribution.  In the computation of the absolute numbers, we took into account the FOV of each instrument, assuming that the probability of detection is FOV$/4\pi$, which implicitly assumes an optimistic duty cycle of 100$\%$. In the case of \emph{Fermi}-GBM, \emph{Swift}-BAT and SVOM, we adopt a more realistic value of sky-averaged FOV, which takes into account Sun and Moon occultations \citep{Burns2016}. In the case of THESEUS-XGIS the FOV we assumed ($\sim 2$ sr) is relative to the energy band 2-150 keV. Since the FOV in the energy band above 150 keV is 2$\pi$ sr,  the reported number of joint detections are a conservative estimate. Our results show that, already with ET alone, $\sim 60-70 \%$ of all the SGRBs will have a detectable GW counterpart, and this fraction approaches 100$\%$ if we consider ET operating with CE. The less steep structure profile of \emph{Stru2}
increases the number of joint detections for the majority of the satellites. 
Depending on the properties of each satellite, Tabs.~\ref{tab_joint} and \ref{tab_joint_stru2} show instruments giving a few tens of detections per year and other hundreds of detections per year. While instruments like GECAM are optimal for statistical studies by giving a large number of detections, instruments with a smaller number of detections but able to localize the source, such as THESEUS-XGIS, are crucial for science cases requiring the identification of the host galaxy, the knowledge of the source redshift, and the complete multi-wavelength characterization of  the source emission. Furthermore, the sensitivity of the instruments considered for the detection of the prompt emission maximizes in different energy bands. While instruments like GECAM are appropriate for detecting GRBs with harder spectrum, GRBs less energetic (intrinsically or because viewed off-axis) will peak at lower energies (soft/hard X-rays) and for them instruments such as XGIS are more suitable. The presence of multiple instruments will be extremely valuable to cover the entire energy range typical of the prompt emission of short GRBs.\\

Figs.~\ref{hist_JD} and \ref{hist_JD_ETCE} show the distribution in redshift of the joint detections, in the specific case of \emph{Fermi}-GBM in synergy with ET and ET+CE, respectively. These figures are produced injecting $10^5$ BNS mergers (extracted from the BNS population, and thus following the astrophysical merger rate evolution with z), with the viewing angle uniformly distributed in the range $0^{\circ}-15^{\circ}$. Even if we distribute the injections over a wider range of angles, the number of joint detections (given as fraction of $\gamma$-ray detections having an associated GW counterpart) does not change, since, while there are GW sources detectable at $\theta_v>15^{\circ}$, there are no $\gamma$-ray detections for $\theta_v>15^{\circ}$ (see Fig~\ref{JD_vs_angle} and  Fig~\ref{JD_vs_angle_s2}).
The $\gamma$-ray detections are computed considering the best fit configuration of parameters derived from MCMC posterior distribution, except for $f_j$, which has been assumed to be one, for better visibility. The value of $f_j$ shifts the green and red histograms vertically but does not change their relative ratio. In the case of ET alone, the GW detector is so sensitive that up to $z\simeq 0.8-1.0$ the probability that a SGRB has a detectable GW counterpart is close to 100$\%$. Adding CE to the network, the GW detection efficiency remains close to 100$\%$ for redshifts above the BNS merger peak. In this section we do not include the results for ET+2CE, since the addition of another CE would not further increase the joint detection efficiency in the redshift range accessible to $\gamma$-ray instruments for SGRBs. 
The $\gamma$-ray missions that are more suited to maximize the joint detection rates are those with large FOV and best sensitivity around MeV energies. However, another parameter to take into account is the localization accuracy by the $\gamma$-ray detectors (given in Tables \ref{tab_joint} and \ref{tab_joint_stru2}), which defines what are the instruments that are able to drive the follow-up observations by ground-based telescopes, crucial for obtaining the source redshift and to completely characterize the multi-wavelength emission of the source.  \\

Fig.~\ref{JD_vs_angle} shows the distribution of the ET+$\gamma$-ray joint detections as a function of the viewing angle, assuming \emph{Stru1}, for three redshift bins: $z<0.5,0.5<z<1.5,z>1.5$. Fig.~\ref{JD_vs_angle_s2} in Appendix shows the same, but assuming \emph{Stru2}. Again, as example, we consider \emph{Fermi}-GBM as $\gamma$-ray detector, but we obtain consistent results for the other $\gamma$-ray satellites considered in the present paper. On the y axis we report the ratio $N_{\rm JD}/N_{\rm inj}$ between the joint detections over the total injected BNS (assuming $f_j=1$) per redshift bin and viewing angle bin. For the viewing angle we consider a linearly spaced grid, where the width of the single bin is $\theta_c/3$. The uncertainty bands are obtained with 15 random extractions of the prompt parameters and each realization considers $5\times 10^4$ BNS injections with viewing angle $0^{\circ}<\theta_v<15^{\circ}$. As before, a different assumed value of $f_j$ just shifts the bands vertically. The plot shows a mild decrease of the ratio $N_{\rm JD}/N_{\rm inj}$ for very small viewing angles ($\theta_v\lesssim \theta_c/3$). Such effect is due to the fact that SGRB viewed on-axis have very high values of peak energy, meaning that the bulk of the flux is above the \emph{Fermi}-GBM band. Therefore, considering two identical SGRBs at the same redshift, the one viewed at $\theta_v\sim\theta_c$ appears slightly brighter than the one viewed at $\theta_v\sim 0$. Moreover, the number of sources per unit angle scales as $\sin(\theta_v)$, therefore the small decrease of $N_{\rm JD}/N_{\rm inj}$ around $\theta_v\sim 0$ is related to the paucity of sources contained in the first angle bin.

\subsubsection{Joint detection of GWs and $\gamma$-rays from cocoon shock break-out}
\begin{table}[t]
\centering
\begin{tabular}{|c|c|c|}
\hline
instrument & $D_{\rm L,max}$ (Mpc) & $N_{\rm det}$  \\\hline
\emph{Fermi}-GBM & 76.4     & $<1$ \\\hline
\emph{Swift}-BAT & 123 & 3 \\\hline
GECAM & 210 & 12 \\\hline
THESEUS-XGIS & 114 & 2 \\\hline
HERMES & 121 & 3 \\\hline
TAP-GTM & 80 &1 \\\hline

\end{tabular}
    \caption{Maximum luminosity distance $D_{\rm L,max}$ and number of joint GW+$\gamma$-ray detections of a SBO with the same properties of the $\gamma$-ray flash observed in coincidence with GW 170817.}
    \label{sbo}
\end{table}

In the derivation of the joint GW+$\gamma$-ray detections, we have assumed that all the SGRBs share a common jet structure and that the $\gamma$-ray emission is given by the dissipation of the internal energy of the jet. However, for GRBs observed at large viewing angles, i.e. $\theta_v\gg \theta_c$, the $\gamma$-ray emission from the shock break-out (SBO) of the jet should be taken into consideration. Indeed, even if the formation and successful break-out of a relativistic jet through the post-merger ejecta is not guaranteed, the formation of a cocoon at very wide angles is commonly expected in the BNS mergers \citep{Ramirez2002,Nakar2017,Gott2018a}. The shocks driven by the mildly relativistic expansion of the cocoon can produce $\gamma$-ray emission which is potentially detectable at least in the local Universe. Moreover, several studies found a remarkable agreement among the properties of the gamma-ray emission of GW 170817 and the shock-breakout model \citep{Kasliwal2017,Gott2018b,Nakar2018,Brom2018,Pozanenko2018}.
The $\gamma$-ray emission detected is not emitted by the wings of a structured jet, but rather by the shock breakout of the cocoon produced by the interaction of the jet with the NS merger ejecta \citep{Gott2018b,Brom2018}. 
In order to investigate this scenario, as a reference, we take the spectrum (cut-off power law with photon index $\alpha=0.62\pm0.40$ and cut-off energy $E_c=185\pm62$ keV) and the \emph{Fermi}-GBM peak flux ($(3.1\pm0.7)\times 10^{-7}$ erg cm$^{-2}$ s$^{-1}$) of the $\gamma$-ray flash associated to GW 170817 \citep{Gold2017} and we compute the maximum distance $D_{\rm L,max}$ at which this emission can be detected by the $\gamma$-ray instruments considered in the present work. In a first order approximation, the SBO emission can be considered isotropic, therefore we do not include any angle dependency in our treatment. In order to evaluate the number of $\gamma$-ray detections associated to SBO, we compute the number of BNS at a luminosity distance $D_L<D_{\rm L,max}$. In our derivation, we assume that each BNS produces a SBO. Since the detection efficiency of any 3G GW detector is $\sim 100\%$ for distances $D_L<D_{\rm L,max}$, the number of $\gamma$-ray detections per year that we report corresponds also to the number of joint GW+$\gamma$-ray detections associated to SBO from BNS mergers. The results are reported in Tab.~\ref{sbo}.

\begin{figure*}[]
     \centering
     \begin{subfigure}[h]{0.45\textwidth}
         \centering
         \includegraphics[width=\textwidth]{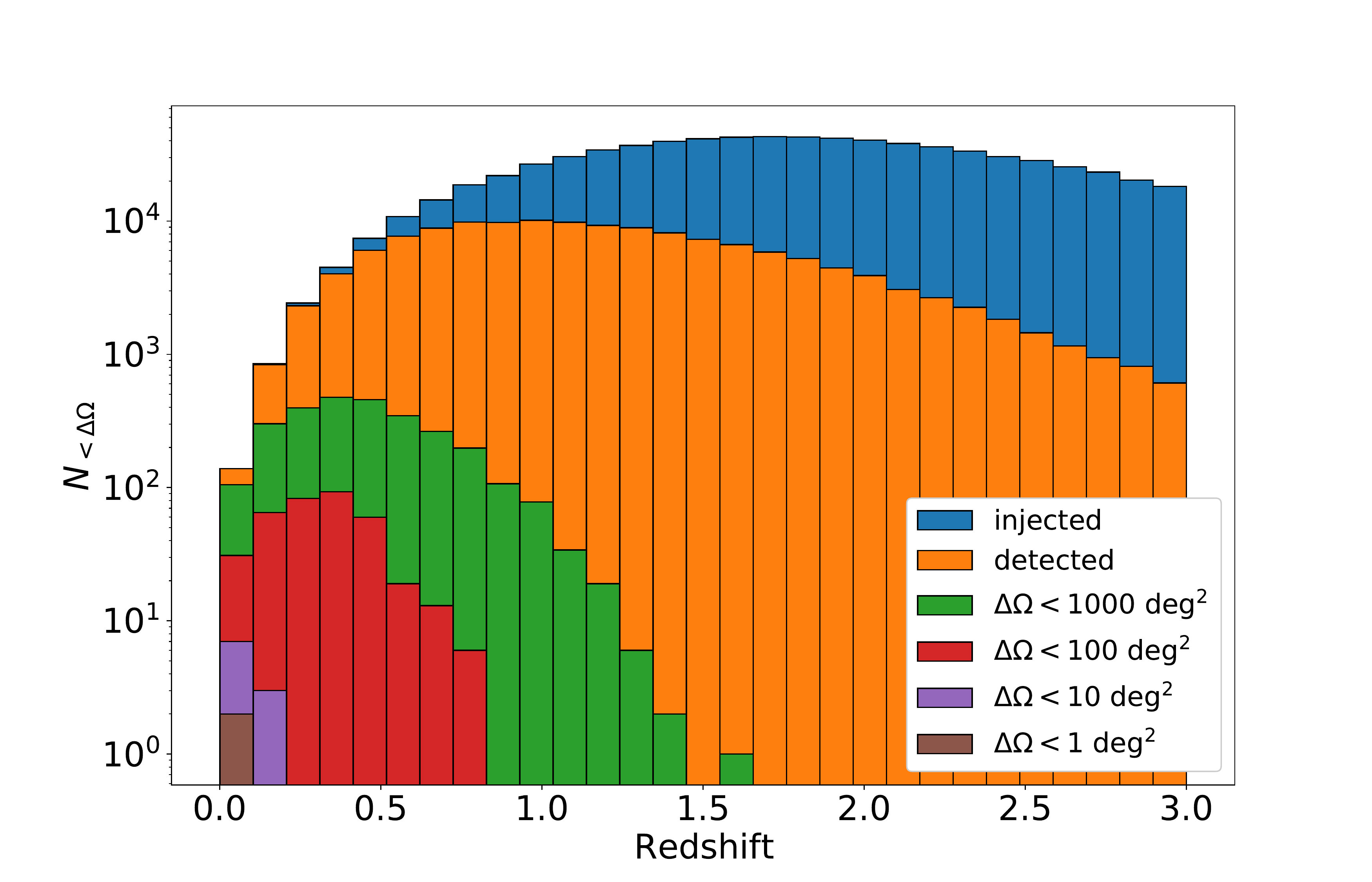}
         \caption{ET, all angles}
         \label{fig:y equals x}
     \end{subfigure}
     \begin{subfigure}[h]{0.45\textwidth}
         \centering
         \includegraphics[width=\textwidth]{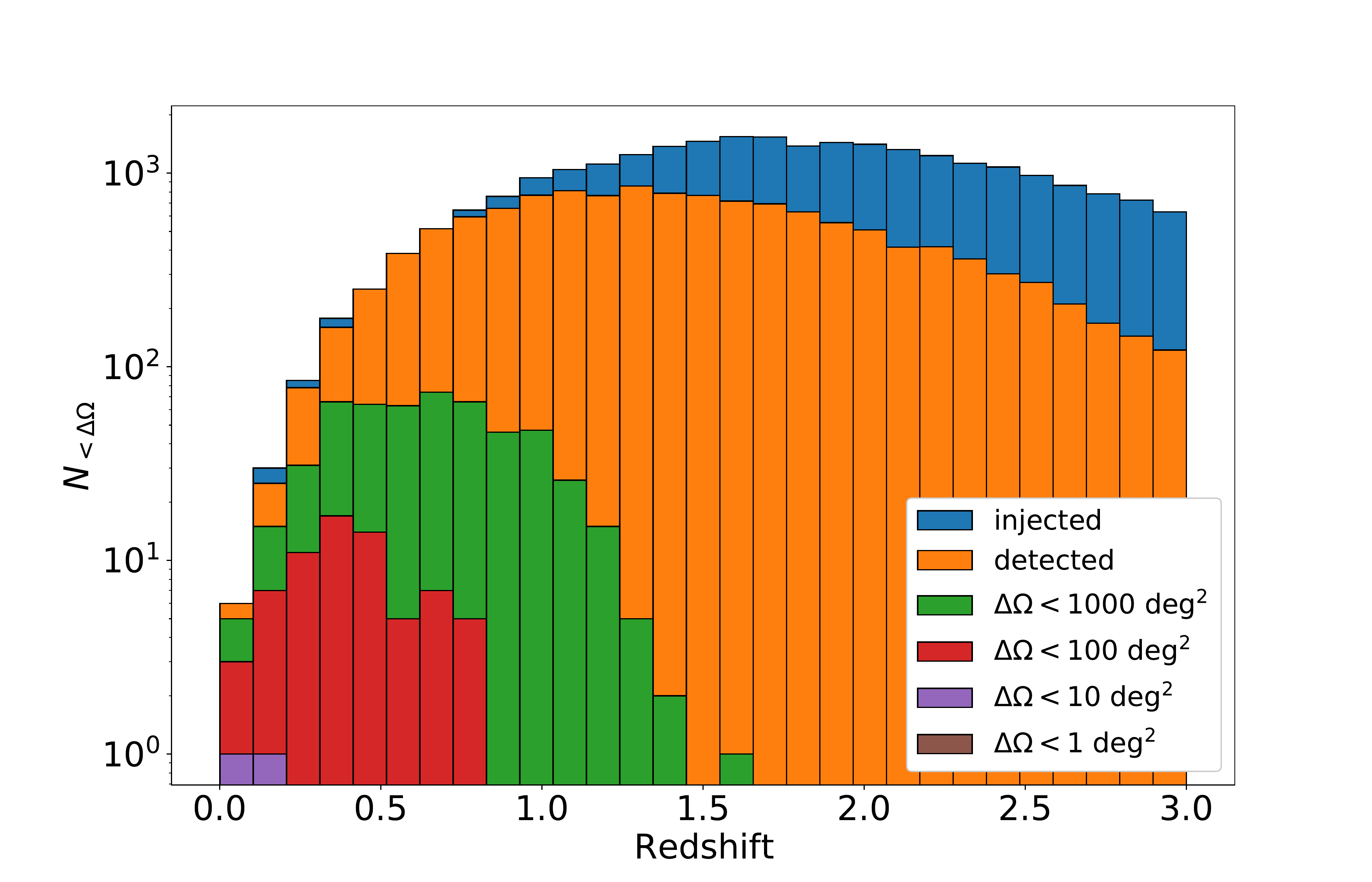}
         \caption{ET, $\theta_v<15^{\circ}$}
         \label{fig:three sin x}
     \end{subfigure}
     \\
     \begin{subfigure}[h]{0.45\textwidth}
         \centering
         \includegraphics[width=\textwidth]{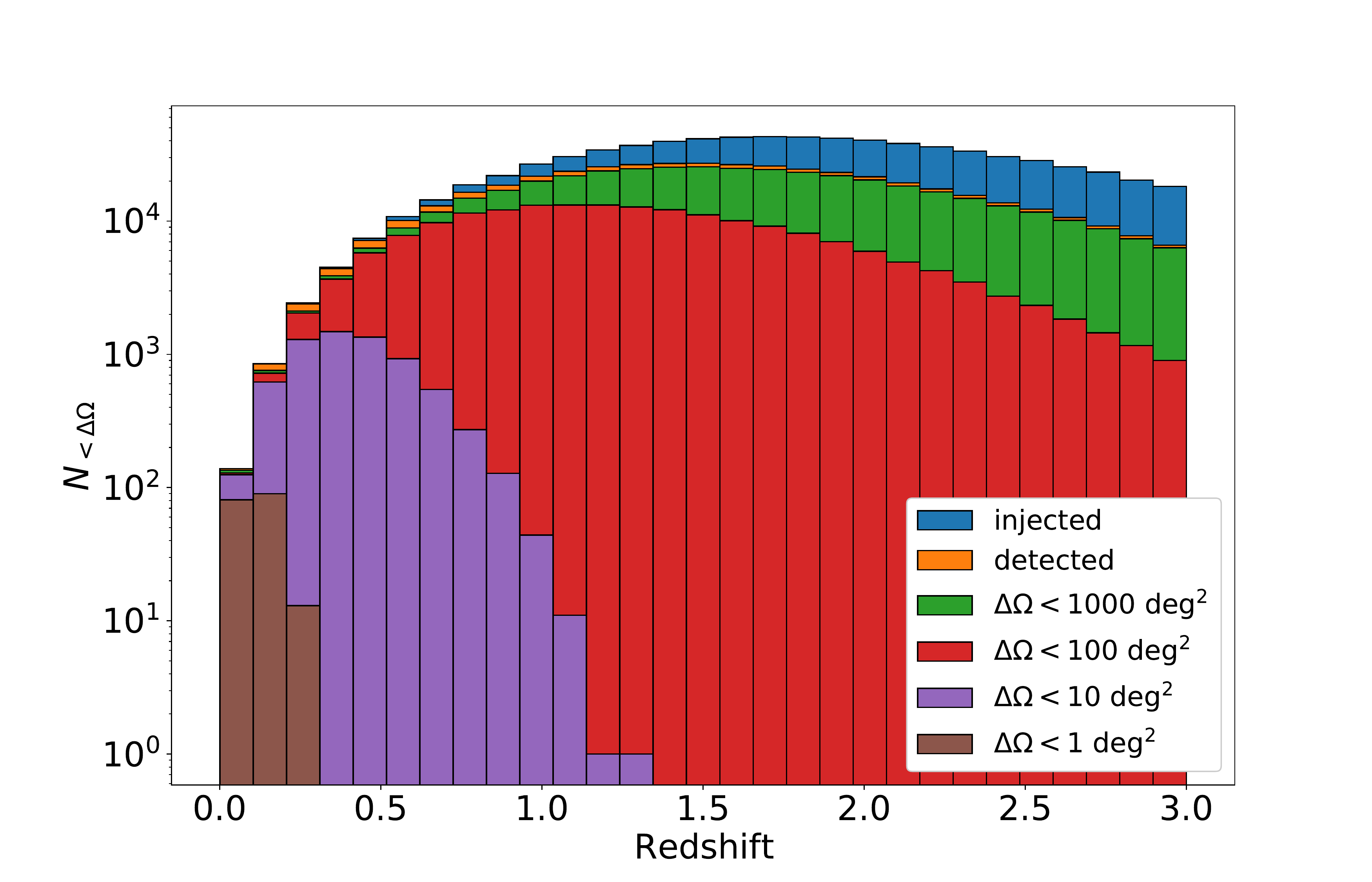}
         \caption{ET+CE, all angles}
         \label{fig:five over x}
     \end{subfigure}
     \begin{subfigure}[h]{0.45\textwidth}
         \centering
         \includegraphics[width=\textwidth]{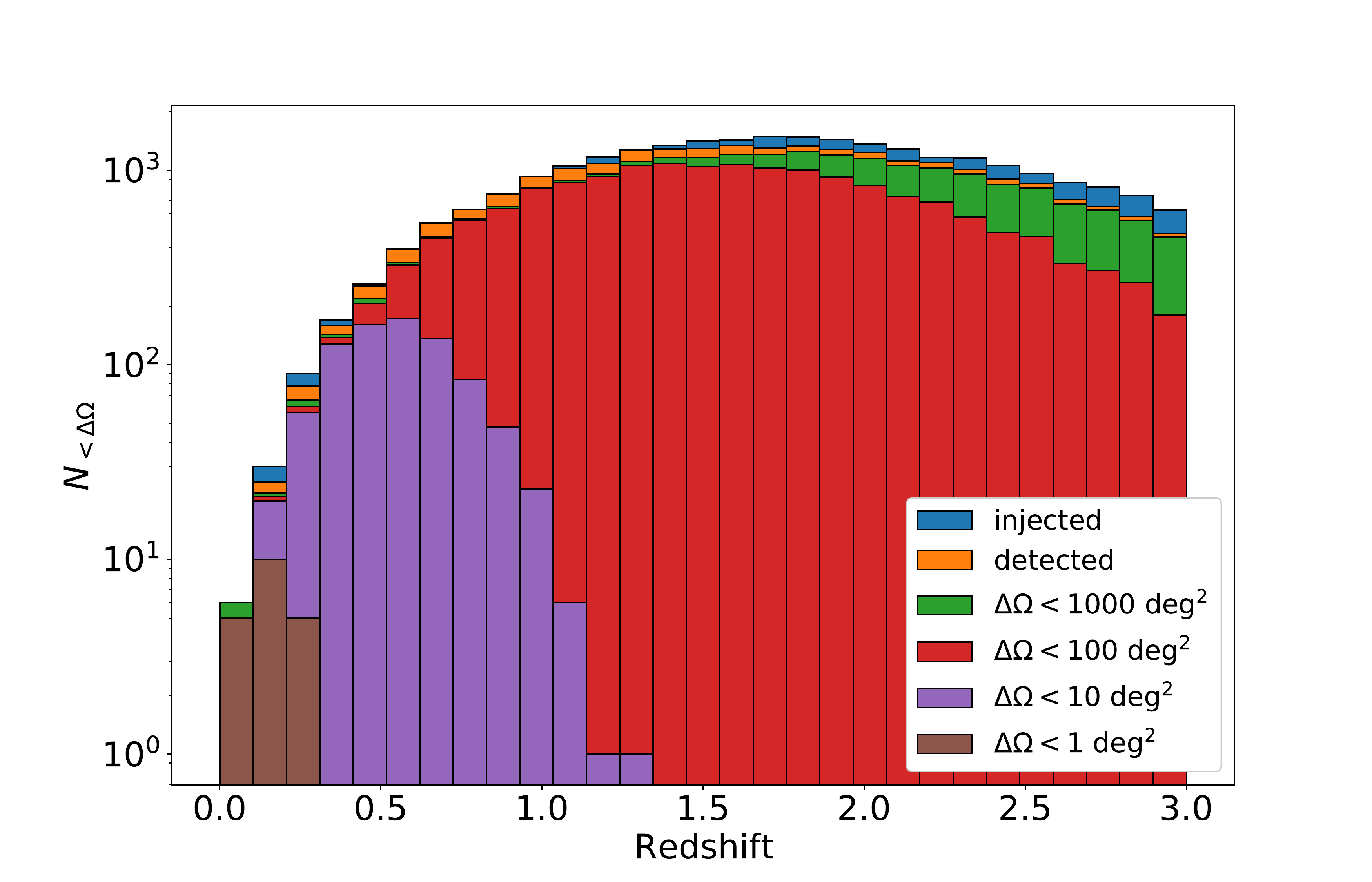}
         \caption{ET+CE, $\theta_v<15^{\circ}$}
         \label{fig:five over x}
     \end{subfigure}
     \\
     \begin{subfigure}[h]{0.45\textwidth}
         \centering
         \includegraphics[width=\textwidth]{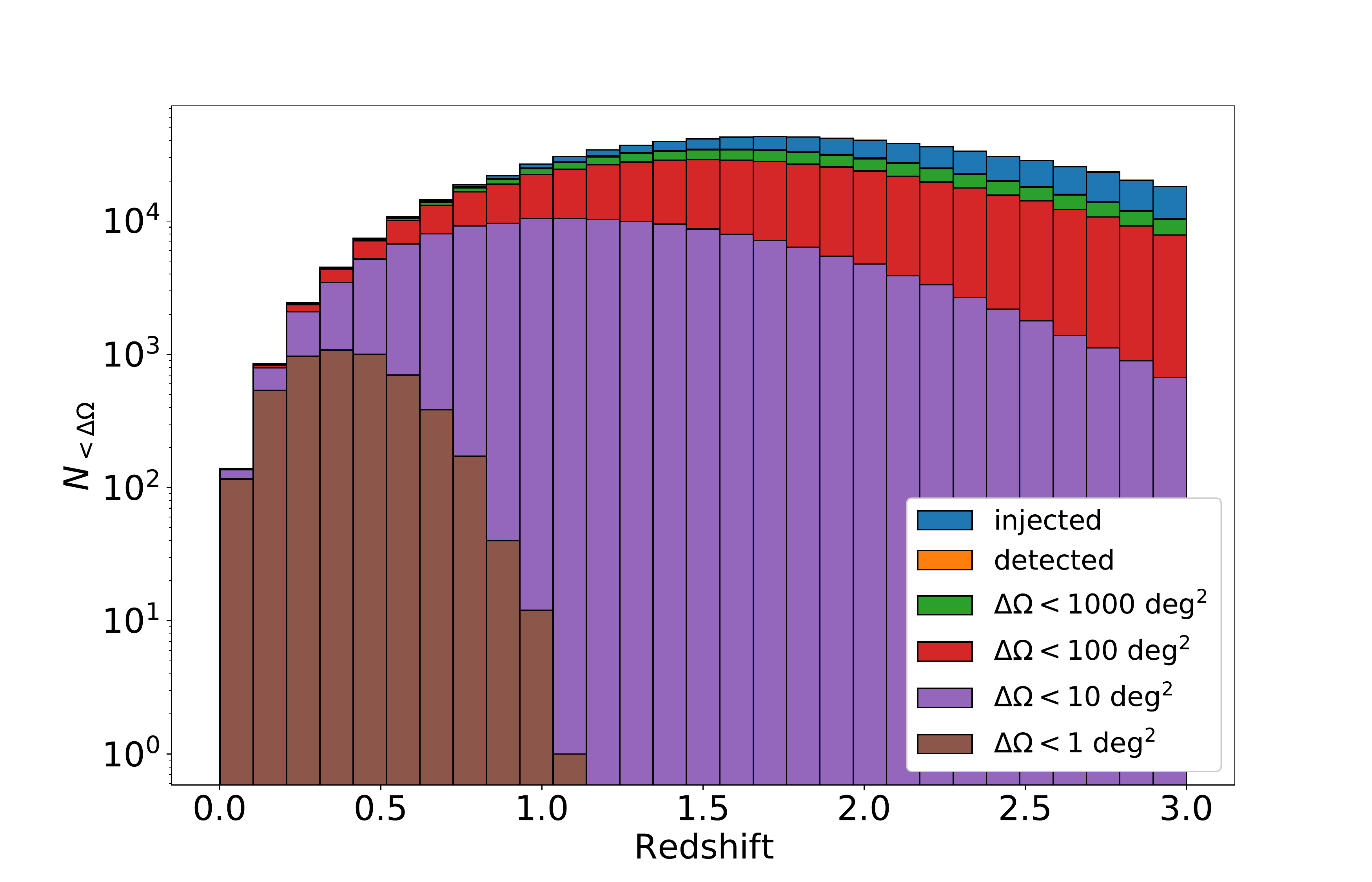}
         \caption{ET+2CE, all angles}
         \label{fig:five over x}
     \end{subfigure}
     \begin{subfigure}[h]{0.45\textwidth}
         \centering
         \includegraphics[width=\textwidth]{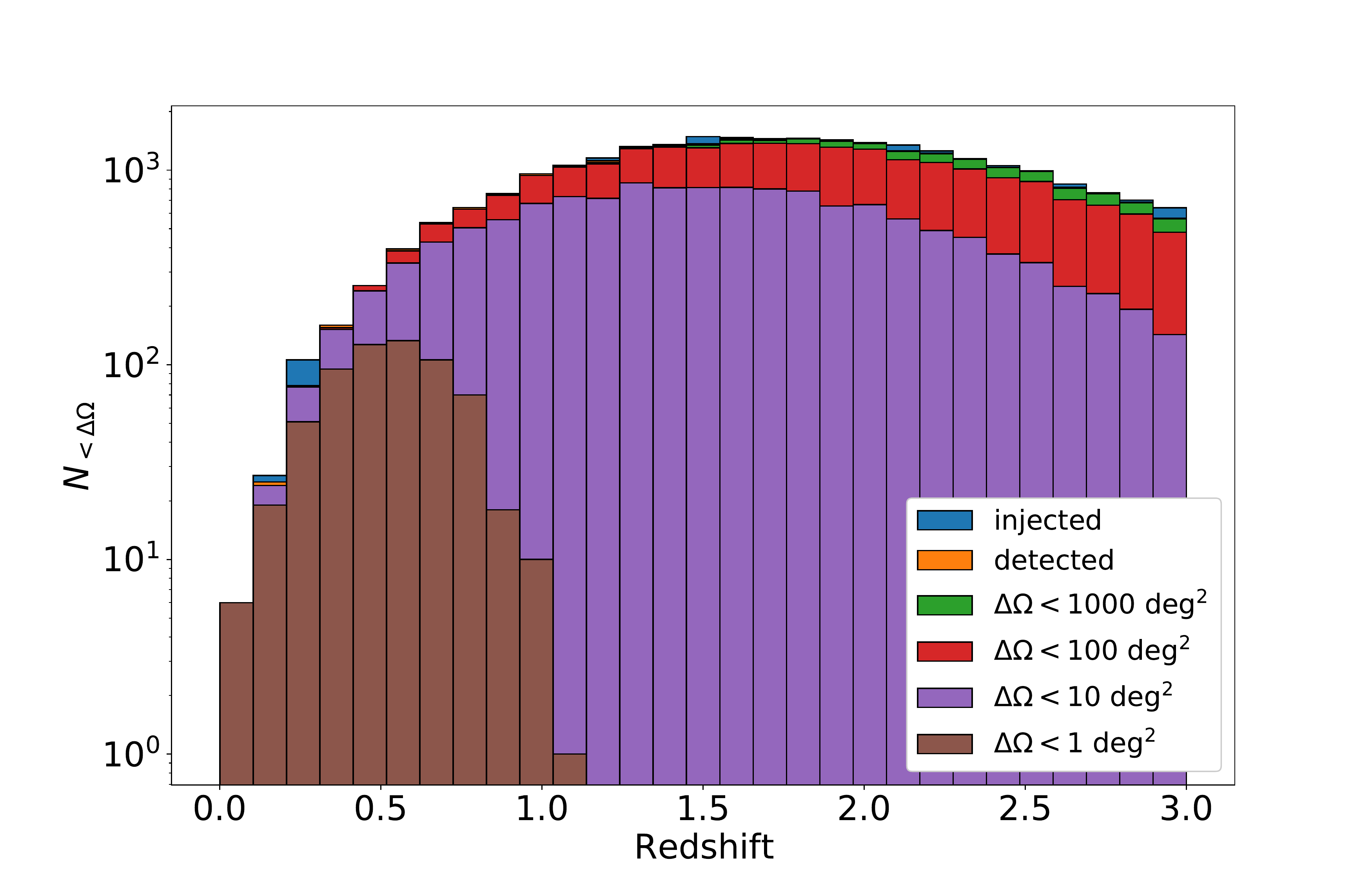}
         \caption{ET+2CE, $\theta_v<15^{\circ}$}
         \label{fig:five over x}
     \end{subfigure}

        \caption{Redshift distribution of the sky-localization uncertainty (given as 90$\%$ credible region) for three detector configurations: ET, ET+CE and ET+2CE. The absolute numbers are relative to one year of observation and assuming a duty cycle of 0.85 as described in the text. Plots on the right show BNS systems with a viewing angle $0<\theta_v<15^{\circ}$.}
        \label{sky_loc_hist}
\end{figure*}

\begin{table}[t]
\centering
\begin{tabular}{|l|c|c|c|}
\hline
    & ET & ET+CE & ET+2CE \\\hline 
$N_{\rm det}$ & 12970 & 23600 & 25668\\ \hline 
$N_{\text{det}}(\Delta \Omega<1 \text{ deg}^2)$ & 0  & 20    & 636     \\\hline
$N_{\text{det}}(\Delta \Omega<10 \text{ deg}^2)$ & 2  & 845    & 13673     \\\hline
$N_{\text{det}}(\Delta \Omega<100 \text{ deg}^2)$ & 69  & 17049    & 23935    \\\hline
$N_{\text{det}}(\Delta \Omega<1000 \text{ deg}^2)$ & 526  & 21564   & 25367    \\\hline
\end{tabular}
    \caption{Number of GW detections per year with sky localization better than 1, 10, 100 and 1000 $\text{deg}^2$. The reported numbers are relative to BNS mergers with $\theta_v<15^{\circ}$. The numbers of GW detections per year are obtained assuming a duty cycle of 0.85 as described in the text.}
    \label{tab_sky_loc}
\end{table}

\begin{table}[t]
\centering
\begin{tabular}{|l|c|c|c|}
\hline
    & ET & ET+CE & ET+2CE \\\hline 
$N_{\rm det}$ & 143970 & 458801 & 592565\\ \hline 
$N_{\text{det}}(\Delta \Omega<1 \text{ deg}^2)$ & 2  & 184    & 5009     \\\hline
$N_{\text{det}}(\Delta \Omega<10 \text{ deg}^2)$ & 10  & 6797    & 154167     \\\hline
$N_{\text{det}}(\Delta \Omega<100 \text{ deg}^2)$ & 370  & 192468    & 493819     \\\hline
$N_{\text{det}}(\Delta \Omega<1000 \text{ deg}^2)$ & 2791  & 428484   & 585317    \\\hline
\end{tabular}
    \caption{Same as Tab.~\ref{tab_sky_loc}, but without any selection on $\theta_v$}
    \label{tab_sky_loc_allth}
\end{table}

\subsection{Joint detection of GWs and X-ray emission}
\label{gw+aft}
In this section, we evaluate the detectability of the afterglow emission with future X-ray telescopes. Specifically, we show the expected rate of BNS mergers which will have both a detectable GW signal and an X-ray emission associated to the afterglow phase of the SGRB. For the X-ray emission, we consider two components: 1) the standard forward shock (FS) emission, 2) the HLE associated to the last photons emitted during the prompt phase. Under the assumption of a structured jet, the HLE produces an X-ray flux which can be comparable or even dominant with respect to the FS emission \citep{Oga2020, Pan2020}. We evaluate the joint X-ray and GW detections considering satellites in survey mode as well as satellites pointing to the GW source. For the pointing strategy, we use the localization capabilities of ET alone and in the network of 3G observatories. Since the number of sources suitable for pointing observations result to be high (especially for network of 3G detectors), we also evaluate other source parameters estimated from the GW signals, such as viewing angle and distance, which can be used to down-select sources to be observed.

\subsubsection{GW sky localization}

Fig.~\ref{sky_loc_hist} shows the sky localization of ET, ET+CE and ET+2CE. The figures show the error on sky position, expressed in deg$^{2}$, as a function of redshift for one year of BNS injections. The injections are extracted from the BNS population described in Sect.~\ref{population}, thus following the same redshift evolution of the merger rate. We show the sky localization both for sources with $0<\theta_v<15^{\circ}$ and for sources with no selection on $\theta_v$.
We chose $\theta_v$ maximum equal to $15^{\circ}$ as it is consistent with the largest viewing angle up to which the X-ray emission from a \emph{Stru1} jet is observable by the analysed WFX-ray satellites (see Fig.~\ref{th_vs_z_s1}).
For the combination ET+CE (ET+2CE), a considerable fraction of sources detected at $z<1$ ($z<3$) has a sky localization smaller than $10$ deg$^{2}$, which is small enough to enable prompt and efficient multi-wavelength search for EM counterparts.\\

In Tab.~\ref{tab_sky_loc} and \ref{tab_sky_loc_allth} we summarize the sky localization capabilities of ET, ET+CE and ET+2CE for systems observed with $\theta_v<15^{\circ}$ and for systems observed at all angles, respectively. The tables list the number of detections per year with sky localization uncertainty $\Delta \Omega<1,10, 100, 1000$ deg$^2$. Fig.~\ref{sky_loc_et} in Appendix shows the redshift distribution of $\Delta \Omega$ as scatter plot with a color code indicating the SNR for ET, ET+CE and ET+2CE, considering cases with $\theta_v<15^{\circ}$ and cases with no selection on $\theta_v$.\\

Fig.~\ref{SNR} in Appendix shows the cumulative distribution of the SNR for BNS detected with ET, ET+CE and ET+2CE with a sky uncertainty $\Delta \Omega<100$ deg$^2$, for $\theta_v<15^{\circ}$ (panel a) and without selection on $\theta_v$ (panel b). The SNR distribution shows that, even adopting a larger SNR threshold, e.g. SNR$>12$, the number of sources with $8<\rm SNR <12$ is negligible and therefore not relevant for the evaluation of the joint GW+X-ray detection rate at least for well localized sources. 

\subsubsection{Viewing angle and luminosity distance from GWs}
\label{gw+angledistance}
\begin{figure}[h!]
     \centering
     \begin{subfigure}[h]{0.5\textwidth}
         \centering
         \includegraphics[width=\textwidth]{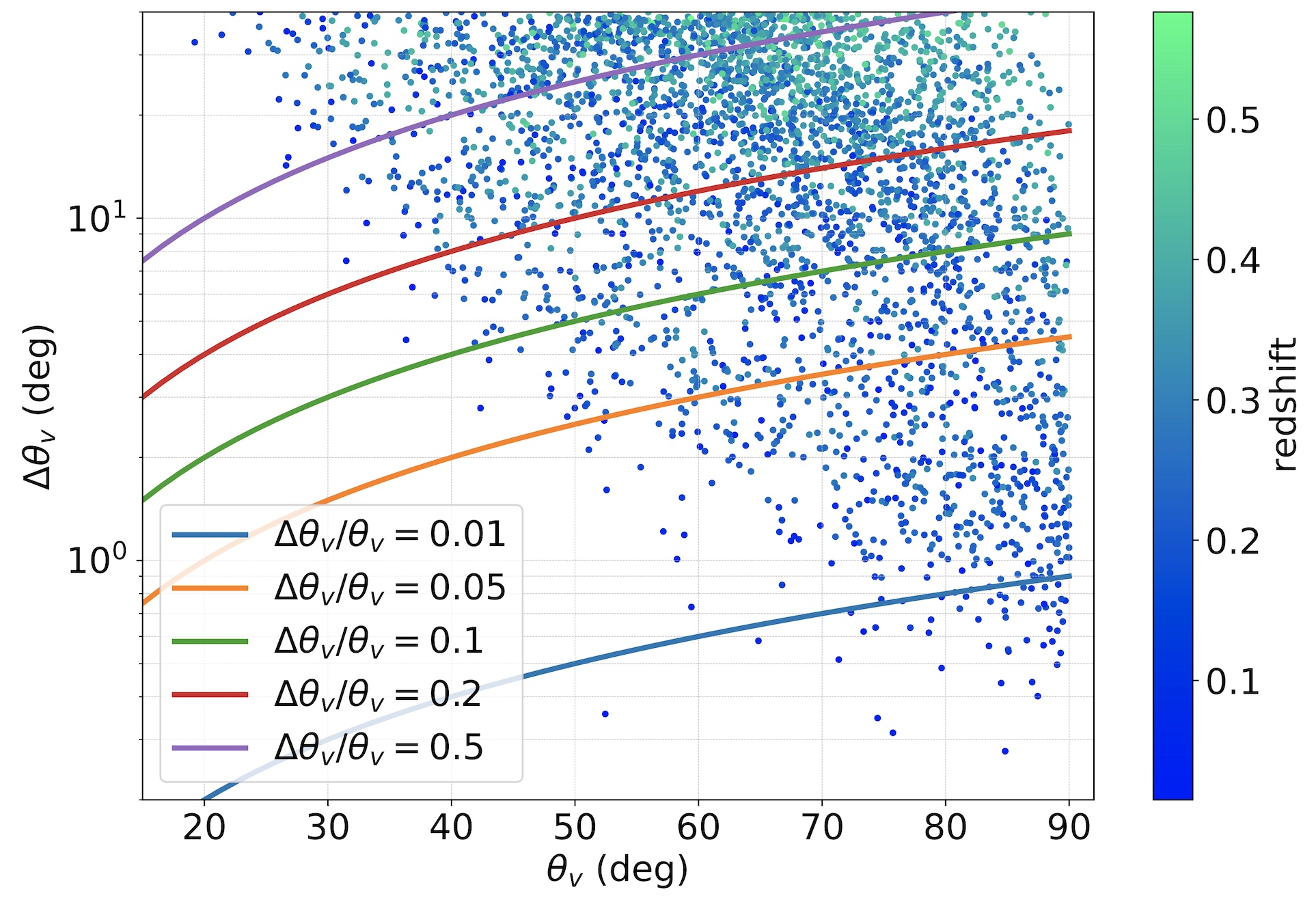}
         \caption{}
         \label{fig:y equals x}
     \end{subfigure}
     \hfill
     \begin{subfigure}[h]{0.5\textwidth}
         \centering
         \includegraphics[width=\textwidth]{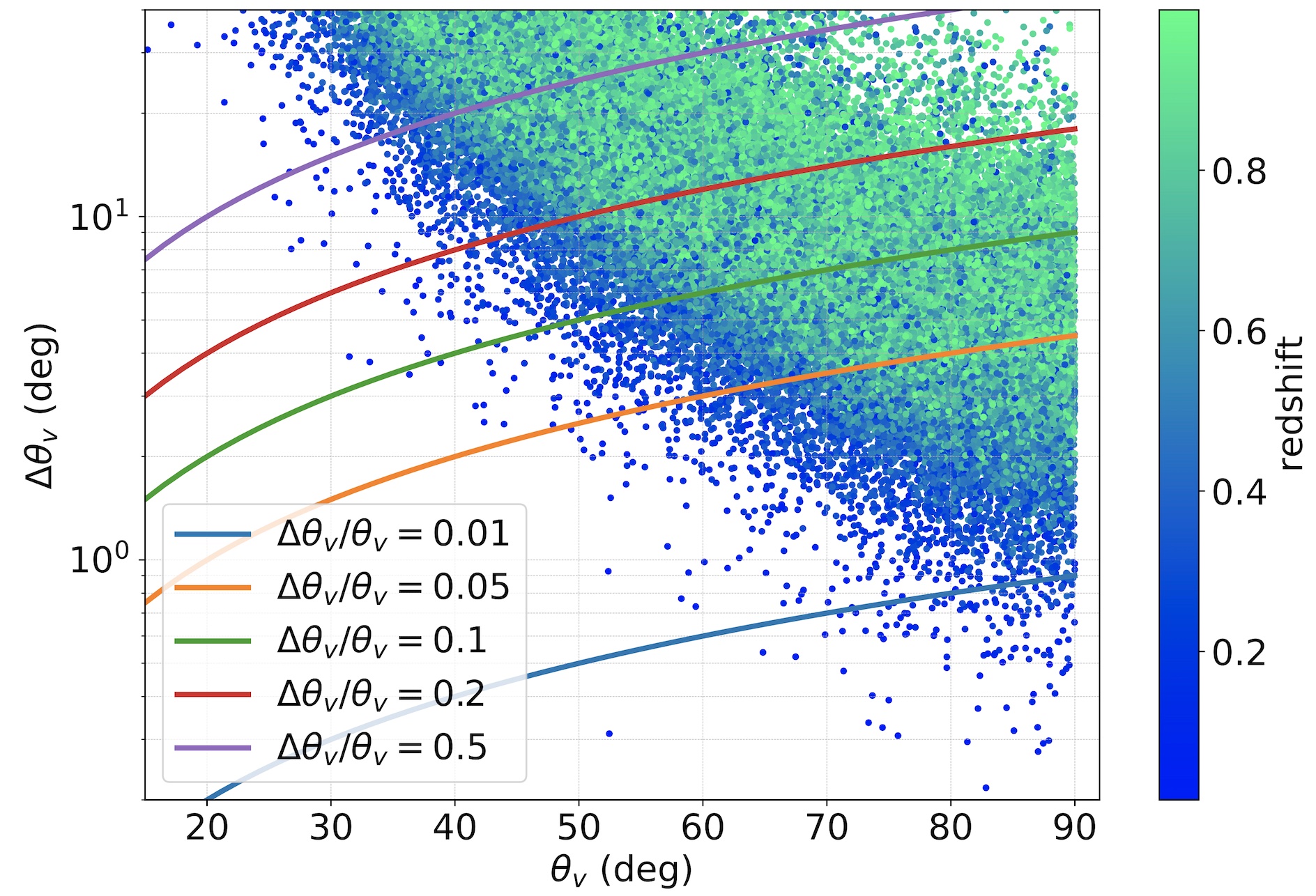}
         \caption{}
         \label{fig:three sin x}
     \end{subfigure}
     \hfill
     \begin{subfigure}[h]{0.5\textwidth}
         \centering
         \includegraphics[width=\textwidth]{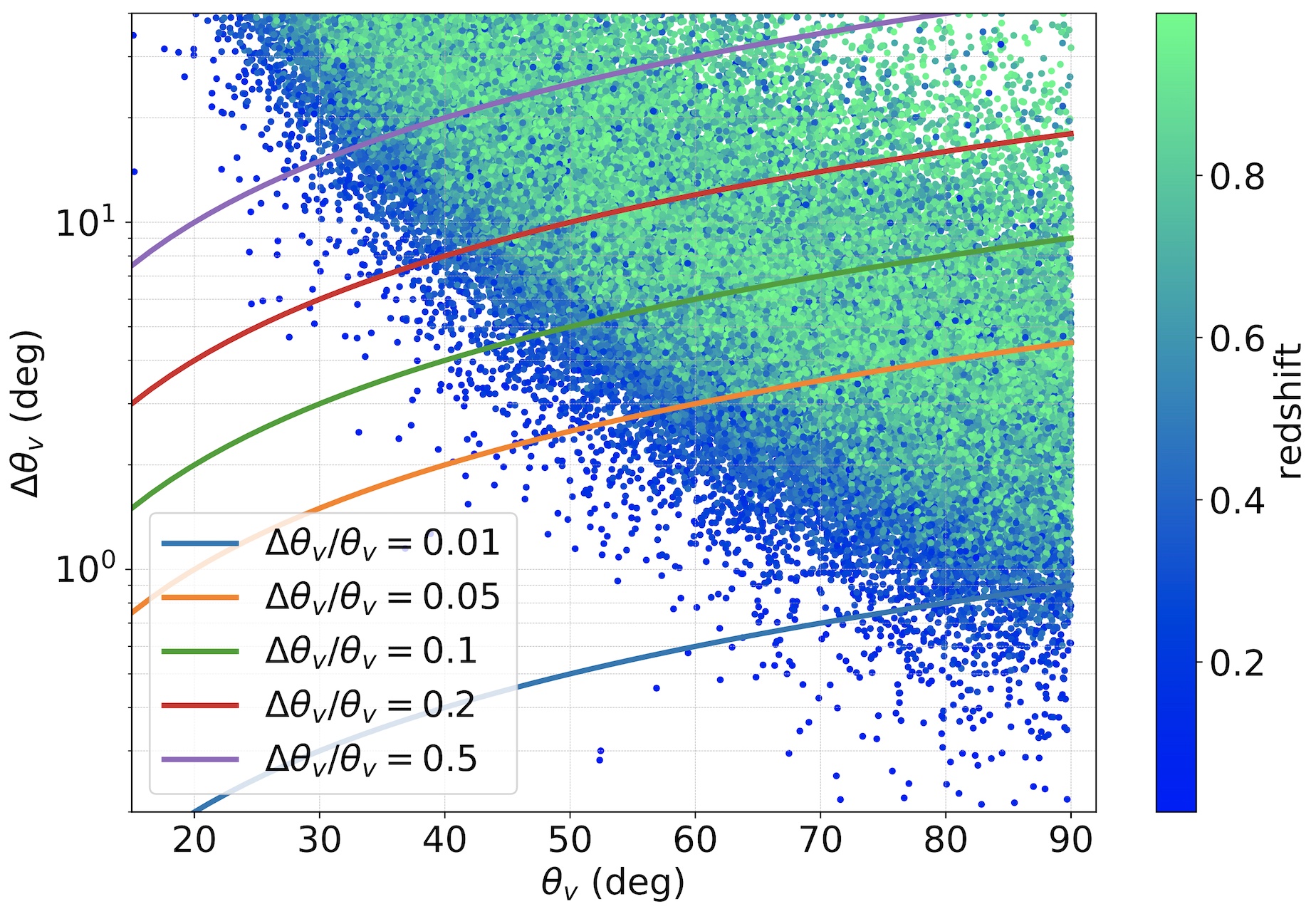}
         \caption{}
         \label{fig:five over x}
     \end{subfigure}
        \caption{Distribution of $\Delta \theta_v$ vs $\theta_v$ relative to one year of observation, in the case of ET (a), ET+CE (b) and ET+2CE (c). The color bar indicates the redshift. We report only BNS mergers detected at $z<1$ and with a $\Delta \theta_v<40^{\circ}$. A duty cycle of 0.85 has been assumed for the GW detectors as described in the text.}
        \label{th_v}
\end{figure}

\begin{figure}[h!]
     \centering
     \begin{subfigure}[h]{0.5\textwidth}
         \centering
         \includegraphics[width=\textwidth]{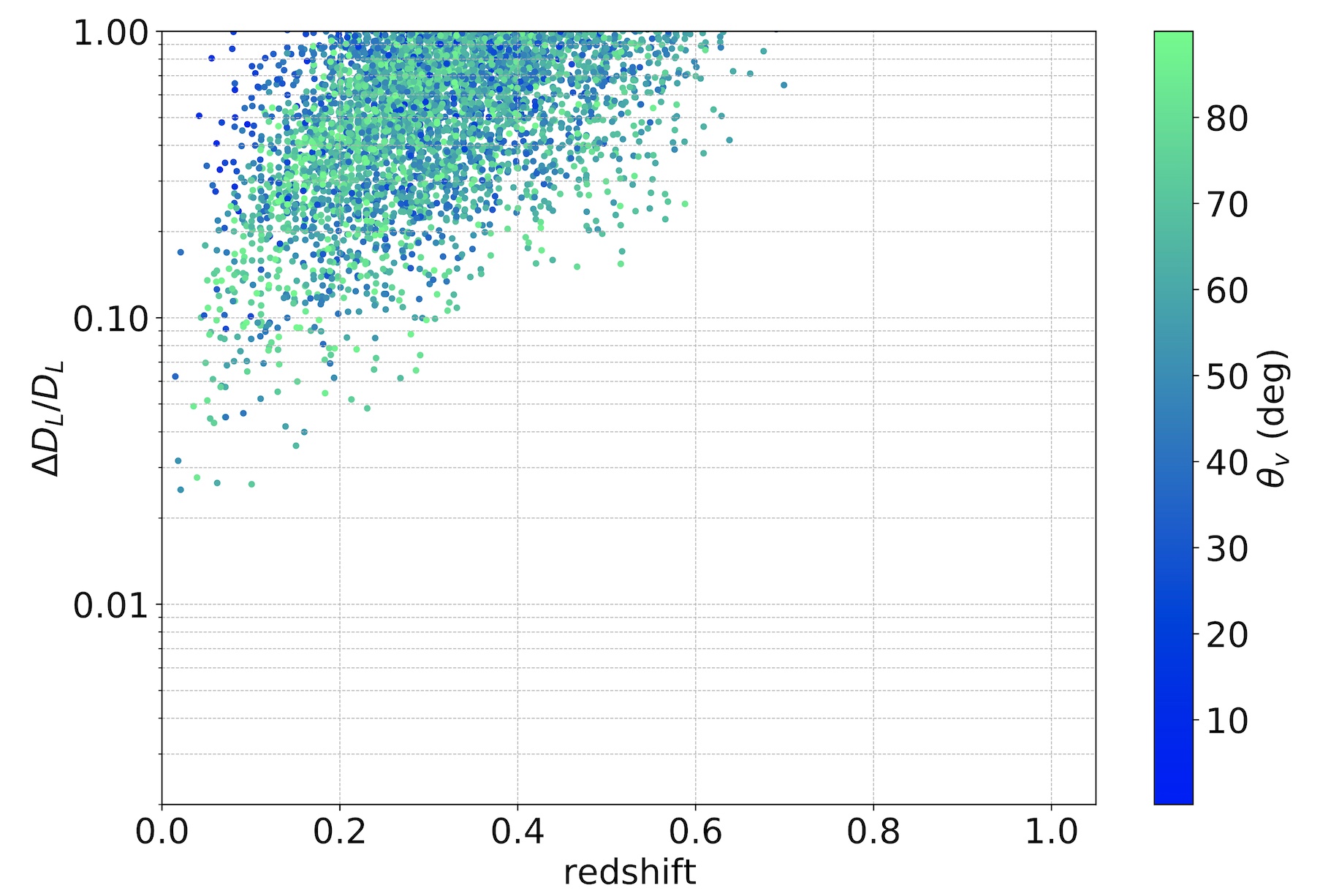}
         \caption{}
         \label{fig:y equals x}
     \end{subfigure}
     \hfill
     \begin{subfigure}[h]{0.5\textwidth}
         \centering
         \includegraphics[width=\textwidth]{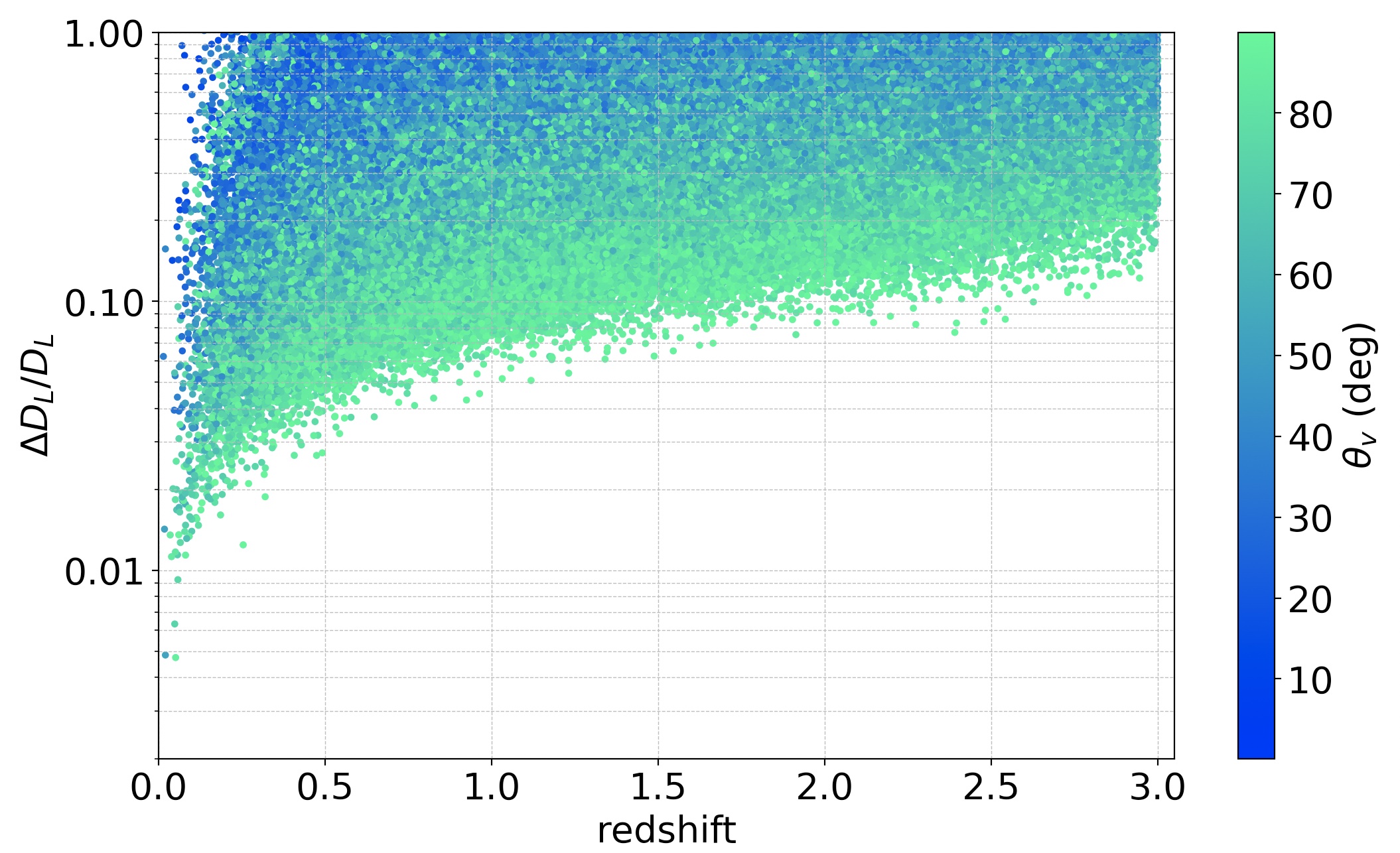}
         \caption{}
         \label{fig:three sin x}
     \end{subfigure}
     \hfill
     \begin{subfigure}[h]{0.5\textwidth}
         \centering
         \includegraphics[width=\textwidth]{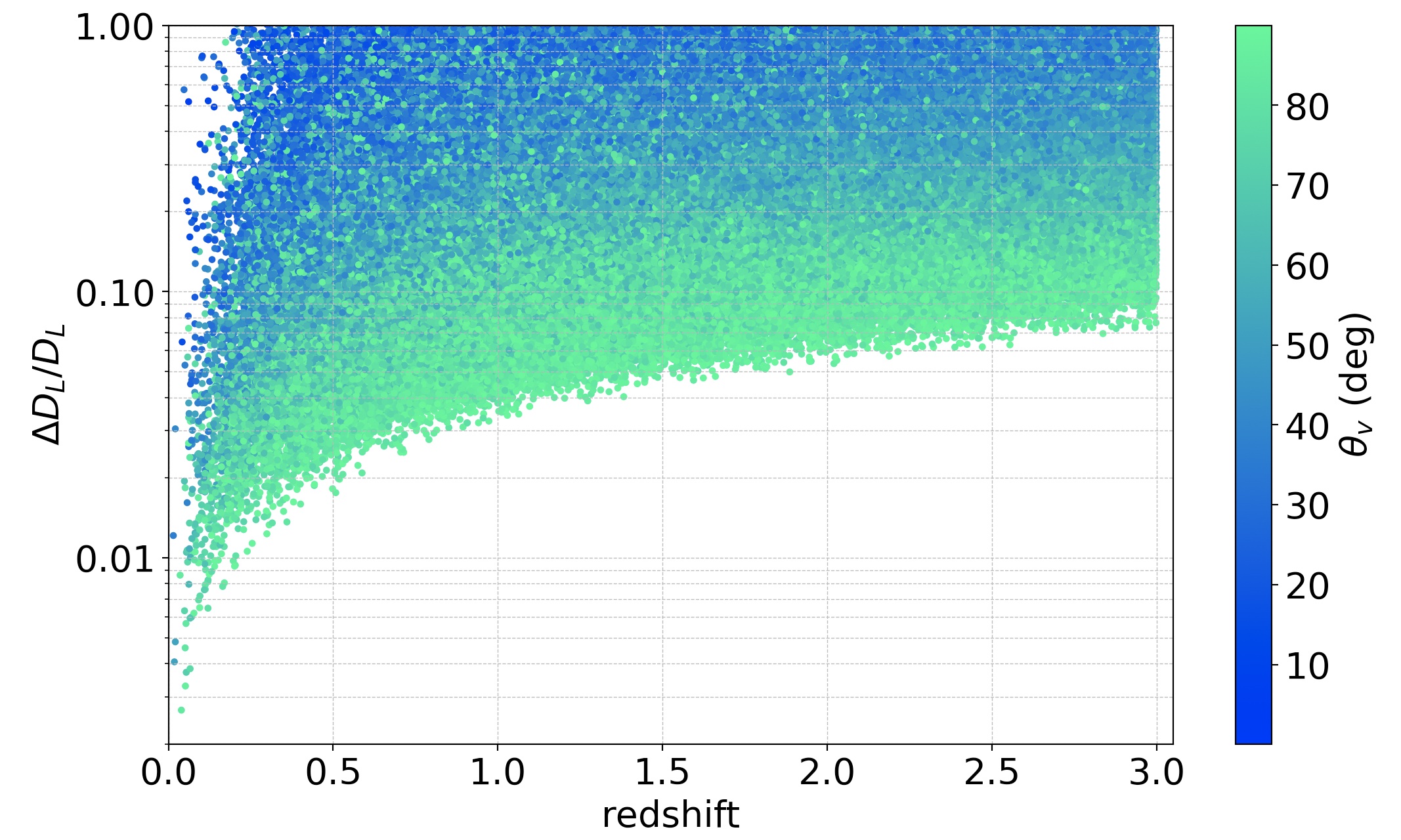}
         \caption{}
         \label{fig:five over x}
     \end{subfigure}
        \caption{Distribution of the relative error on the luminosity distance $D_L$ as a function of redshift relative to one year of observation, in the case of ET (a), ET+CE (b) and ET+2CE (c). The color bar indicates the viewing angle. Only cases with $\Delta D_L/D_L<1$ and at $z<3$ are shown. A duty cycle of 0.85 has been assumed for the GW detectors as described in the text.}
        \label{dl}
\end{figure}

Here, we explore how the GW observations can give constraints about the viewing angle with respect to the axis perpendicular to the BNS orbital plane (assumed to be parallel to the  
jet axis) and the luminosity distance of the system.
Fig.~\ref{th_v} shows how the detected sources are distributed in the plane $\Delta\theta_v-\theta_v$, considering ET, ET+CE and ET+2CE. $\Delta\theta_v$ is the 1$\sigma$ error on the viewing angle. 
Only sources at $z<1$ and with $\Delta\theta_v<40$ deg are shown. For all the networks, a considerable fraction of sources have $\Delta\theta_v/\theta_v<0.1$. Moreover, since the gradient of the GW amplitude is maximum at $\theta_v=90 $ deg, the $\theta_v$ parameter is better constrained when the source is seen edge-on. This result is particularly important to prioritize the BNS events to be followed-up. 
Indeed, in the era of 3G GW detectors we expect hundreds of BNS detections per day. Due to the limited observational time at each observatory, with a pointing strategy only a limited fraction of GW-detected signals can be followed up. For example, aiming at detecting the high-energy emission, one can exclude all the sources edge-on for which the viewing angle estimate is more precise from the GW parameter estimation.\\

Fig.~\ref{dl} shows the distribution in redshift of $\Delta D_L/D_L$, where $\Delta D_L$ is 1$\sigma$ error on the luminosity distance $D_L$. We consider ET, ET+CE and ET+2CE. Only sources with $\Delta D_L/D_L<1$ are shown. For ET operating alone the relative error can be below $10\%$ only for sources with $z \lesssim 0.2$, while for ET+CE and ET+2CE the same is true for $z \lesssim 1.5$ and $z \lesssim 3$, respectively. The plots also show that the constraint on distance is tighter for sources seen edge on. This is due to the strong degeneracy between luminosity distance and viewing angle and therefore the error on the first is significantly reduced when the second is well constrained.

\subsubsection{Wide-FOV X-ray telescopes: survey and pointing observations}

For the detection of the afterglow emission from our SGRBs population, we consider the following X-ray instruments: the Einstein Probe (EP, \citealt{ep}) which is scheduled for launch by the end of 2022, the ECLAIRs instrument on board of SVOM \citep{svom2015} which is scheduled for launch in 2023, and three mission concepts, THESEUS (\citealt{theseus,Amati2021}), the Wide Field Imager on board of TAP (\citealt{TAP}), and the \textit{Gamow Explorer} \citep{Gamow2021}. The launch date of this last is expected to be in 2028-2032.\\
If the instrument operates in survey mode, the probability of detecting the source (if the flux is above the sensitivity limit) is given by $\sim$FOV/$4\pi$.
Instead, if the instrument operates only in pointing mode, the light curve is monitored only after a time $t_{\rm resp}$ from the trigger time, which is the sum of the time to respond to a trigger from ground and the time to re-point the instrument.
In order to evaluate how each instrument will sample the afterglow temporal evolution, we use the sensitivity curve (The sensitivity curve of EP is taken here\footnote{\url{https://sci.esa.int/documents/34375/36249/1567258027270-ESA-CAS-workshop1_20140225_7__Einstein_Probe-exploring_the_dynamic_X-ray_Universe__Yuan.pdf}}, while for all the other X-ray missions, we use the sensitivity curves reported in the references cited at the beginning of this section), which relates the minimum detectable flux as a function of the exposure time. The temporal sampling is computed using an iterative approach. Starting from the beginning of the light curve, we determine the length $\Delta T$ of each bin at time $t_i$ such that
\begin{equation}\label{sens}
\frac{1}{\Delta T}\int_{t_i}^{t_i+\Delta T} F(t)dt > F_{\rm lim}(\Delta T)
\end{equation}
where $F_{\rm lim}(\Delta T)$ is derived from the sensitivity curve. The afterglow light curve is classified as undetectable if, even increasing $\Delta T$, the eq.~\ref{sens} is never satisfied. For \emph{Gamow}, THESEUS and EP we consider sensitivity curves including a median Galactic absorption $N_H=5\times 10^{20}\rm cm^{-2}$. For TAP the published sensitivity curve does not include this information. \\

For all the results reported in this section, we simulate 25 realizations of one year of BNS mergers, each with parameters extracted randomly from the MCMC posterior distributions. For each realization we consider a number of injections $N_{\rm inj}=\min(N_{\rm req},10^4)$, where $N_{\rm req}$ is the number of BNS in one year that satisfy the selection criteria (e.g. about SNR, $\theta_v$ and $\Delta \Omega$). The choice $N_{\rm inj}=\min(N_{\rm req},10^4)$ is made in order to have enough statistics and such that the full parameter space is correctly covered. Then the number of detections correspondent to one year is obtained re-scaling by the factor $N_{\rm req}/N_{\rm inj}$.  The parameters of the prompt model influence the afterglow light curve because both the FS and HLE brightness depend on the $E_{\rm iso}$ which is predicted by our model.
First we evaluate the expected rate of joint GW+X-ray detections, considering the telescopes operating in survey mode. The results are reported in Tab.~\ref{tab_jd_x_survey}. Due to the good sensitivity down to 2 keV and the large FOV, THESEUS-XGIS can contribute substantially to the total amount of X-ray detections.  Thanks to the larger FOV (1.1 sr) and the better sensitivity, EP gives a larger number of detections compared to THESEUS-SXI and TAP.\\

The probability of detection can be increased considering the possibility to point GW sources localized with enough precision through the GW signal.
In such case, wide field X-ray (WFX-ray) telescopes can point to the sky error box and start the search of the X-ray counterpart. We explore this scenario simulating one year of BNS mergers and selecting only those that are detected with an error on sky position better than 100 deg$^2$. The typical FOV of WFX-ray telescopes is larger than 1 sr, so an error region of 100 deg$^2$ can be well contained inside the FOV of the instrument. A sky region of 100 deg$^2$ is around 1/10 of a typical FOV of a WFX-ray telescope. If, instead, we select sources localized better than 1000 $\text{deg}^2$ and considering that the bulk of sources detectable in the X-rays are at $z\lesssim 1$, we expect, with respect to the cases with $\Delta \Omega<100 $ deg$^2:$
\begin{enumerate}
    \item a factor $\sim$ 10 more joint GW+X-ray detections with ET
    \item no substantial variation for ET+CE, since sources with $100\text{ deg}^2<\Delta \Omega<1000\text{ deg}^2$ are located at $z\gtrsim 1$
    \item no substantial variation for ET+2CE, since $N(\Delta \Omega<100$ deg$^2) \sim N(\Delta \Omega<1000$ deg$^2$) for $z\lesssim 1$.
\end{enumerate}
For the selected sub-sample of well localized sources, we predict the X-ray emission and we check if it is detectable with the telescopes listed before. The numbers of THESEUS-SXI and \textit{Gamow} are identical since we consider for them the same sensitivity, and the slightly different FOV does not influence pointing observations. In Tab.~\ref{tab_jd_x} we show the number of expected joint GW+X-ray detections for one year of observation with ET, ET+CE and ET+2CE. We consider the configuration \emph{FS-SGRB} for the FS parameters and we assume that the X-ray telescopes are able to point towards the sky position provided by GW detectors within 100 s from the BNS merger. A response time of 100 s is a short amount of time to communicate the trigger from the ground to the satellites, but the possibility of accessing lower frequencies by the next generation GW detectors will make it possible pre-merger alerts. We evaluated that, in the case of ET, a fraction of $\sim$30\% and $\sim$60\% of sources localized better than 100 deg$^{2}$ at the merger are above the detection threshold (SNR=8) and with a sky-localization better than 1000 deg$^{2}$, which is well within the FOV of the satellites analyzed here, 20 and 10 minutes before the merger, respectively (see tab.~\ref{pre_merger}). For ET+CE (ET+2CE) this fraction is $\sim$5\% (6\%) 10 minutes before the merger, leaving the absolute number of triggers to be followed up of order of several hundreds. The typical exposure time for detection in X-rays is $t_{\rm exp}=66_{-42}^{+249}$ s for SXI and \emph{Gamow}, $t_{\rm exp}=96_{-29}^{+219}$ s for TAP, $t_{\rm exp}=239_{-173}^{+171}$ s for EP, where the uncertainties are reported at $1\sigma$ level of confidence and they are computed considering a random sampling of the MCMC posterior distribution.
\begin{table}[t]

 \centering
\begin{tabular}{|c|c|c|}
\hline
    & FOV (sr) & loc. accuracy (arcmin) \\\hline 
Einstein Probe  & 1.1   & 5   \\\hline
\textit{Gamow} & 0.4     & 1-2    \\\hline
THESEUS-SXI & 0.5    & 1-2     \\\hline
TAP-WFI & 0.4    & 1     \\\hline

\end{tabular}
    \caption{Field of view and localization accuracy of the WFX-ray telescopes considered in this work.}
    \label{tab_instr}
\end{table}

\begin{table}[t]
 \centering
\begin{tabular}{|c|c|c|}
\hline
                    & ET                 & ET+2CE            \\\hline 
SVOM-ECLAIRs        & $4\pm2$            & $5\pm2$          \\ \hline                   
Einstein Probe                  & $50^{+15}_{-16}$   & $64^{+12}_{-20}$  \\\hline
\textit{Gamow}      & $9^{+2}_{-2}$      & $10^{+3}_{-3}$     \\\hline
THESEUS-SXI         & $11^{+3}_{-3}$     & $13^{+4}_{-3}$     \\\hline
THESEUS-(SXI+XGIS)  & $23^{+6}_{-5}$     & $27^{+7}_{-5}$     \\\hline
TAP-WFI             & $16^{+3}_{-4}$     & $17^{+6}_{-3}$     \\\hline

\end{tabular}
    \caption{Expected number of joint GW+X-ray detections in one year of observations, for different X-ray instruments operating in survey mode. The assumed structure is \emph{Stru1} and the FS configuration is \emph{FS-SGRB}.}
    \label{tab_jd_x_survey}
\end{table}

\begin{table}[t]
\centering

\begin{tabular}{|c|c|c|c|}
\hline
    & ET & ET+CE & ET+2CE \\\hline
Einstein Probe  & $9^{+5}_{-3}$  & $294^{+80}_{-59}$   & $359^{+168}_{-110}$    \\\hline
THESEUS-SXI/ & \multirow{2}{*}{$7^{+5}_{-3}$}  & \multirow{2}{*}{$95^{+43}_{-14}$}    & \multirow{2}{*}{$122^{+41}_{-23}$  }   \\
\textit{Gamow} &&&    \\\hline
TAP-WFI & $8^{+5}_{-3}$  & $182^{+43}_{-31}$    & $225^{+76}_{-72}$     \\\hline

\end{tabular}
    \caption{Expected number of joint GW+X-ray detections in one year of observations, for different X-ray instruments operating in pointing mode. The reported numbers correspond to BNS mergers with sky-localization uncertainty $\Delta \Omega<100$ deg$^2$ and with a detectable X-ray emission, assuming a $t_{\rm resp}=100$ s. The assumed structure is \emph{Stru1} and the FS configuration is \emph{FS-SGRB}.}
    \label{tab_jd_x}
\end{table}

\begin{table}[t]
\centering
\begin{tabular}{|c|c|c|c|}
\hline
    & 100 s                & 1 hr              & 4 hr \\\hline 
Einstein Probe  & $359^{+168}_{-110}$    & $48^{+24}_{-15}$ & $17_{-10}^{+15}$ \\\hline
THESEUS-SXI/ & \multirow{2}{*}{$122^{+41}_{-23}$}     & \multirow{2}{*}{$12\pm 7$}             & \multirow{2}{*}{$<9$}  \\
\textit{Gamow} &&&  \\\hline  
TAP-WFI & $225^{+76}_{-72}$      & $50^{+20}_{-10}$  & $17_{-5}^{+10}$  \\\hline
\end{tabular}
    \caption{As in Tab. \ref{tab_jd_x}, but considering different delay times between the merger and the beginning of X-ray observation. In this table, we consider ET+2CE as GW network.}
    \label{tab_jd_x_delay}
\end{table}
\begin{table}[t]
\centering
\begin{tabular}{|c|c|c|c|}
\hline
    & HLE       & FS-SGRB &HLE\\
    & +FS-SGRB  &         &+FS-GW17 
\\\hline 
Einstein Probe  & $359^{+168}_{-110}$    & $344^{+106}_{-95}$ & $67^{+14}_{-17}$ \\\hline
THESEUS-SXI/ & \multirow{2}{*}{$122^{+41}_{-23}$}     & \multirow{2}{*}{$98^{+34}_{-31}$}  & \multirow{2}{*}{$22^{+7}_{-6}$}  \\
\textit{Gamow} &&&  \\\hline
TAP-WFI & $225^{+76}_{-72}$    & $174^{+94}_{-27}$  & $33^{+12}_{-8}$  \\\hline

\end{tabular}
    \caption{As in Tab. \ref{tab_jd_x}, but considering different assumptions for the computation of the X-ray light curve. In this table we consider ET+2CE as GW network and $t_{\rm resp}=100$ s.}
    \label{tab_jd_x_par}
\end{table}

\begin{table}[t]
\centering
\begin{tabular}{|c|c|c|}
\hline
    & \emph{Stru1}             & \emph{Stru2}   \\\hline 
Einstein Probe  & $359^{+168}_{-110}$    & $383^{+131}_{-112}$  \\\hline
THESEUS-SXI/ & \multirow{2}{*}{$122^{+41}_{-23}$}     & \multirow{2}{*}{$128^{+73}_{-61}$}  \\
\textit{Gamow} &&\\\hline
TAP-WFI & $225^{+76}_{-72}$     & $219^{+128}_{-73}$    \\\hline

\end{tabular}
    \caption{Comparison between \emph{Stru1} and \emph{Stru2} for the prediction of joint GW+X-ray detections in pointing mode, considering ET+2CE. Both HLE and FS are included. The FS configuration is \emph{FS-SGRB} and we assume $t_{\rm resp}=100$ s.}
    \label{tab_jd_x_stru}
\end{table}

\begin{figure}[h!]
     \centering
     \begin{subfigure}[h]{0.45\textwidth}
         \centering
         \includegraphics[width=\textwidth]{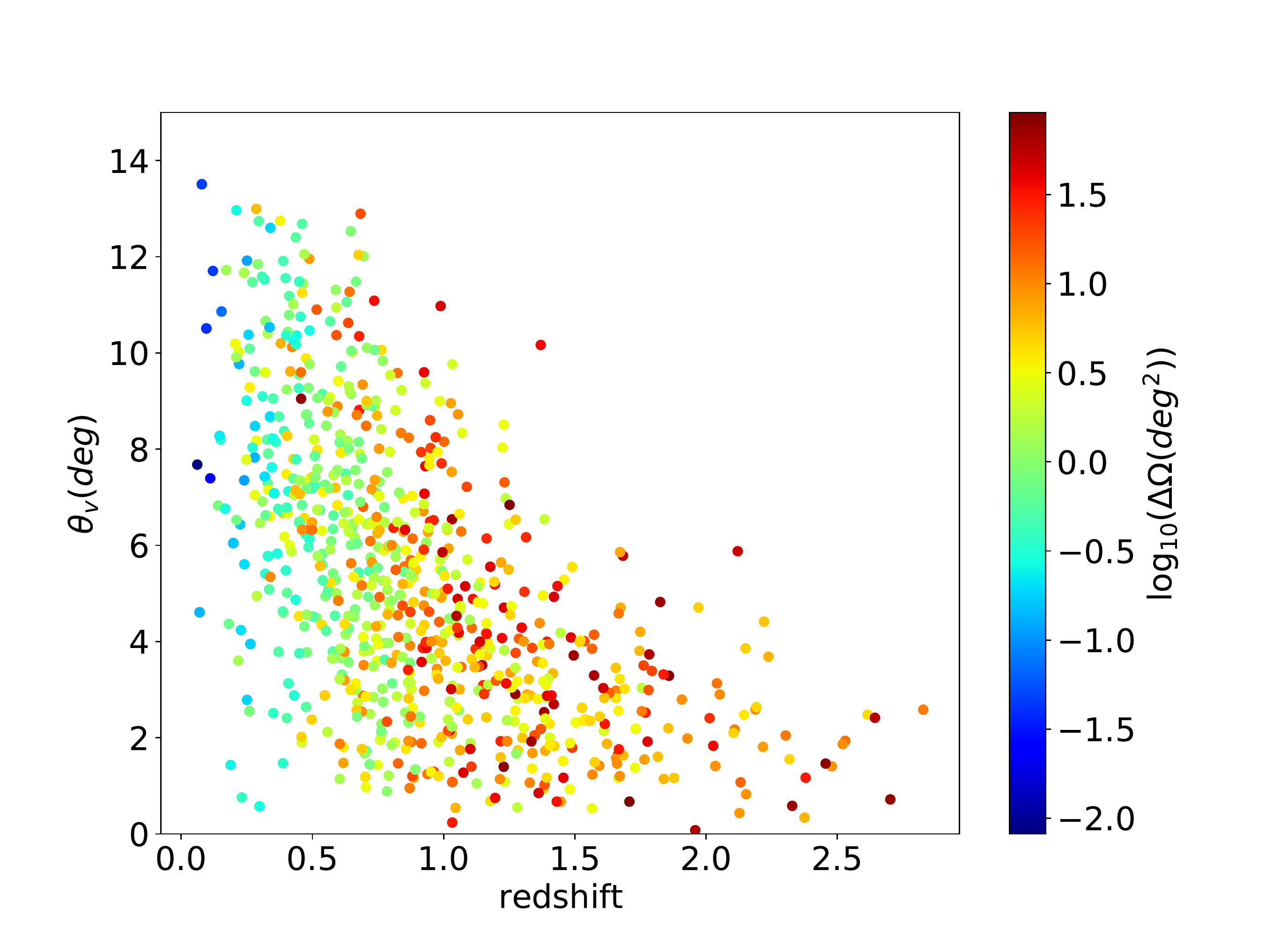}
         \caption{}
         \label{fig:y equals x}
     \end{subfigure}
     \hfill
     \begin{subfigure}[h]{0.48\textwidth}
         \centering
         \includegraphics[width=\textwidth]{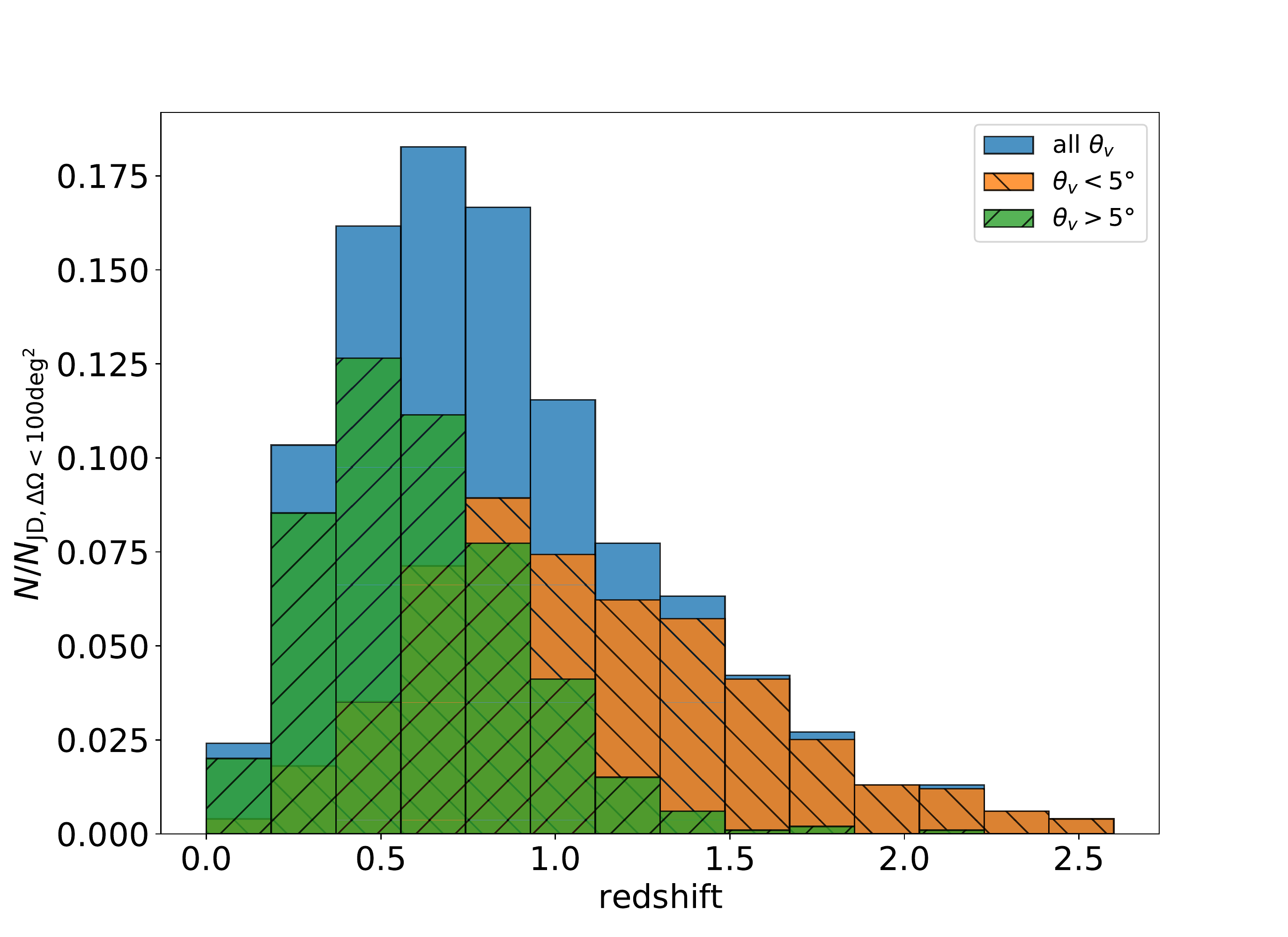}
         \caption{}
         \label{fig:three sin x}
     \end{subfigure}
        \caption{Panel (a): Distribution of the viewing angle as a function of redshift for a sample of 1000 joint GW+X-ray detections, considering ET+2CE and SXI and under the assumption \emph{Stru1}. Only BNS detections with GW sky localization $\Delta \Omega<100$ deg$^2$ are selected. The assumed jet core angle is $\theta_c = 3.4$ deg. The color bar indicates the GW sky localization uncertainty of each detection. Panel (b): distribution in redshift of the same sample, where we distinguish between detections with $\theta_v<5^{\circ}$ and $\theta_v>5^{\circ}$. The histogram is normalized to the number of total joint detections with $\Delta \Omega<100$ deg$^2$.}
        \label{th_vs_z_s1}
\end{figure}

\begin{figure}[h!]
     \centering
     \begin{subfigure}[h]{0.5\textwidth}
         \centering
         \includegraphics[width=\textwidth]{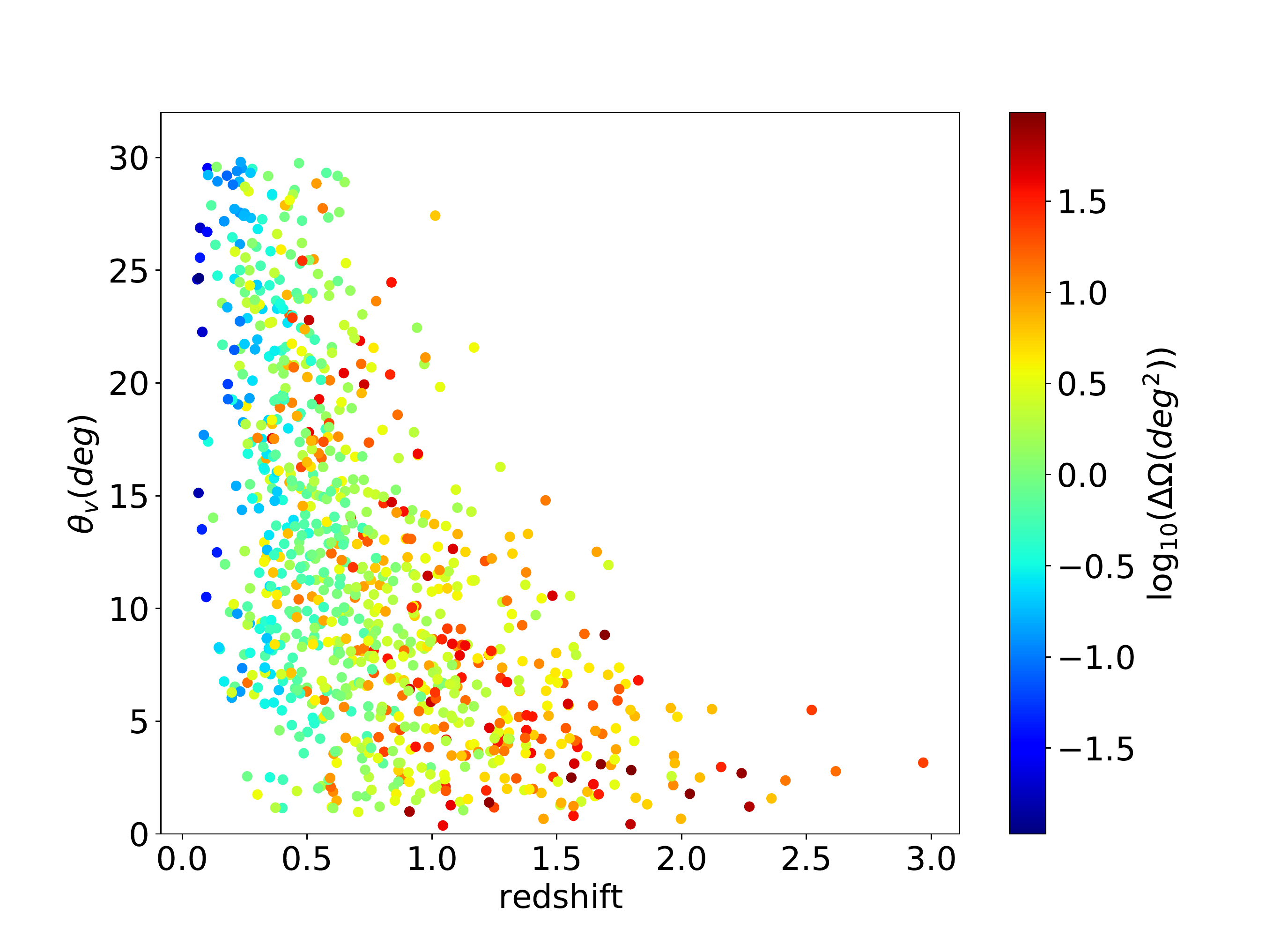}
         \caption{}
         \label{fig:y equals x}
     \end{subfigure}
     \hfill
     \begin{subfigure}[h]{0.48\textwidth}
         \centering
         \includegraphics[width=\textwidth]{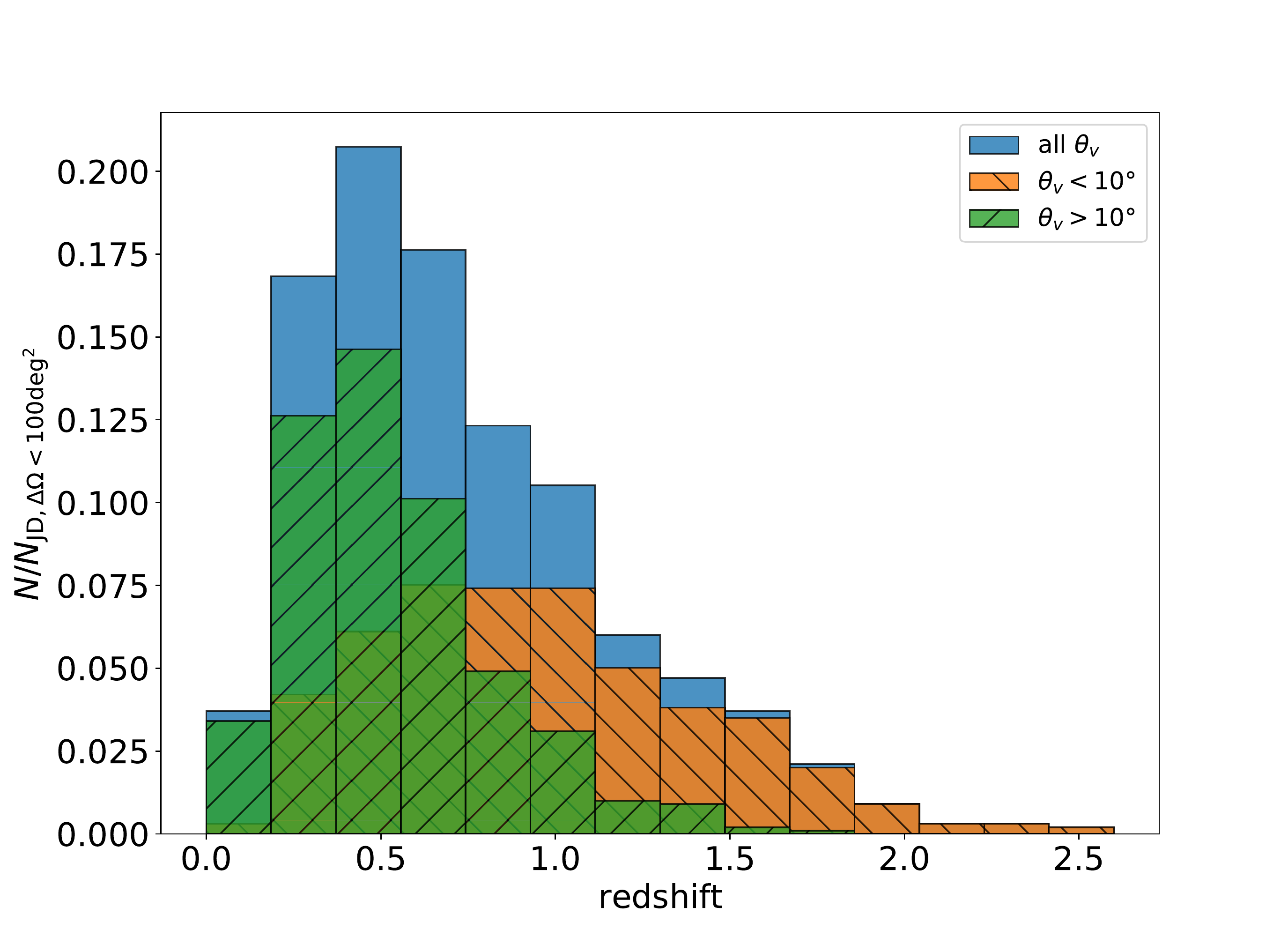}
         \caption{}
         \label{fig:three sin x}
     \end{subfigure}
        \caption{Same as in Fig.~\ref{th_vs_z_s1}, but for \emph{Stru2}.}
        \label{th_vs_z_s2}
\end{figure}

Considering ET alone and FS configuration (\emph{FS-SGRB}), Tab.~\ref{tab_jd_x_survey} and \ref{tab_jd_x} show that the survey mode is more promising than the pointing mode in terms of number of GW+X-ray detections per year. For ET+CE and ET+2CE, due to the larger number of well localized sources, the number of joint detections is considerably larger than the survey mode case.
Hence, the exploitation of sky localization provided by GW detectors can enhance substantially the probability of identifying the EM counterpart of BNS mergers. We point out that the reported numbers for pointing mode do not take into account the duty factor of the instruments, which is assumed to be 100$\%$. Moreover, each X-ray mission, depending on the scientific objectives, can dedicate only a limited amount of time for the follow-up of GW triggers. Therefore, if we call $\epsilon_{\rm  GW}$ the fraction of observational time dedicated to GW follow-up and $\epsilon_{\rm DC}$ the duty cycle, a more realistic estimate of the number of joint detections would be $N_{\rm JD, real}=\epsilon_{\rm GW}\epsilon_{\rm DC}N_{\rm JD}$, where $N_{\rm JD}$ are the numbers reported in this section about the pointing mode. Regarding the survey mode, instead we just have  $N_{\rm JD, real}=\epsilon_{\rm DC}N_{\rm JD}$. Another pivotal point of the future pointing strategy, which can strongly influence the real joint detection efficiency, is the trigger selection. As shown in Table ~\ref{tab_sky_loc}, the number of sources localized better than 100 deg$^2$ are of order of 100 for ET and of 10$^5$ for ET+CE and ET+2CE. In order to not lose observational time, the triggers to be observed will require to be prioritized. A strategy based on the knowledge of the satellite detection efficiency, the distance and viewing angle from the GW signals (see Sect.~\ref{gw+angledistance}) will be mandatory to select the triggers with a higher chance of joint detection. Therefore the reported numbers for pointing mode should be considered as potential detections, which can be achieved only provided that an efficient selection of the GW triggers is performed to restrict the follow-up to cases with highest probability of detection.\\

In Tab.~\ref{tab_jd_x_delay} we consider longer response time (relaxing the assumption of $t_{\rm resp}=100$ s) and we show how the detection rate significantly decreases if the telescope needs more time to respond to the trigger and point the GW localization. The significant decrease of the chance to detect the afterglow emission pointing with one hour delay is due to a combination of the rapid decline of the GRB X-ray flux and the sensitivity of wide field X-ray instruments. Among the population of detected BNS, nearby events similar to GW170817 are a tiny fraction, while the large majority of joint detections comes from on-axis (or slightly off-axis) short GRBs distributed at cosmological distances, with an afterglow light curve which peaks at early times and followed by rapid drop of the flux. 
For these cases a rapid response is fundamental to catch the first decaying part of the X-ray emission. In Tab.~\ref{tab_jd_x_par}, we show how the prediction of joint GW+X-ray detections depends on the assumption of FS parameters and on the inclusion of HLE. In this case, we assume ET+2CE as GW network and $t_{\rm resp}=100$ s. In the specific, we test three possible setups, where we compute the X-ray light curve:\\
\begin{enumerate}
    \item including HLE + FS emission, with \emph{FS-SGRB} configuration
    \item including HLE + FS emission, with \emph{FS-GW17} configuration
    \item including only FS emission, with \emph{FS-SGRB} configuration.
\end{enumerate}
Our results show that the assumption of FS parameters like GW 170817 leads to less bright FS emission, mainly due to lower value of $\epsilon_{B}$. Moreover, the tables show that the inclusion of HLE is non negligible in the estimation of detectability of the X-ray afterglow. The reasons are mainly twofold: 1) the HLE produces an initial steep decay phase which is well visible in X-rays and exceeds the FS contribution at least in the first tens of seconds, 2) the HLE associated to a structured jet typically produces a plateau phase for sources observed at $\theta_v\lesssim \theta_c$ and this contribution can be as relevant as the FS emission itself.\\

Finally, we test how the expected joint detection rate of GW+X-rays from BNS mergers depends on the assumption of the jet structure. 
The specific structure \emph{Stru1} assumed so far, namely the one derived from GW 170817, has a very steep off-core profile, meaning that the probability of detecting prompt and/or afterglow emission from SGBRs viewed at $\theta_v>\theta_c$ is lower than the one of detecting a shallower off-core jet profile, such as \emph{Stru2}. Tab.~\ref{tab_jd_x_stru} shows the difference in joint GW+X-ray detections in pointing mode assuming \emph{Stru1} and \emph{Stru2}. The overall rate of joint GW+X-ray detections for \emph{Stru2} is, within the uncertainties, comparable with the one of \emph{Stru1}. This is due to the fact that, though the X-ray emission is detectable at larger viewing angles for \emph{Stru2}, the best fit value for the fraction of BNS producing a jet, $f_j$, derived from the Monte Carlo is smaller for \emph{Stru2} than the one of \emph{Stru1}.
Fig.~\ref{th_vs_z_s1} and \ref{th_vs_z_s2} show the distribution of viewing angle of GW+X-rays joint detections as a function of redshift, adopting the jet structure \emph{Stru1} and \emph{Stru2}, respectively. For each detection, we use a color code for the sky-localization uncertainty. The plot is realized simulating 1000 GW+X-rays detections (ET+2CE + SXI) which satisfy the requirement $\Delta \Omega<100$ deg$^2$, assuming $t_{\rm resp}=100$ s and \emph{FS-SGRB} configuration. Below $z\sim 0.5-1$, for both the structures, the X-ray emission is detectable up to viewing angles much larger than $\theta_c = 3.4$ deg. In the bottom panel of Fig.~\ref{th_vs_z_s1}, we show how the GW+X-rays joint detections are distributed in redshift and we distinguish between cases detected at $\theta_v<5^{\circ}$ and $\theta_v>5^{\circ}$. We adopt $\theta_v=5^{\circ}$ as a threshold because, according to Fig.~\ref{JD_vs_angle}, this is the maximum angle up to which prompt emission is detectable by \emph{Fermi}-GBM (similarly for XGIS onboard of THESEUS), considering an average on redshift.
This implies that all the sources detected in X-rays at $\theta_v>5^{\circ}$ have a non-detectable prompt emission in $\gamma$-ray. In the case of \emph{Stru1}, sources with $\theta_v>5^{\circ}$ represent $\sim40\%$ of all the GW+X-rays joint detections. Similarly, for \emph{Stru2} the threshold angle for detectability of prompt emission is $\theta\sim10^{\circ}$ and $\sim50\%$ of all the GW+X-rays joint detections is above this limit.
Independently on the assumption on the jet structure, this result means that there is a significant number of off-axis events concentrated below $z\sim 0.5-1$ which are only detectable with WFX-ray instruments. In terms of absolute numbers of off-axis GW+X-rays joint detections, a combination of ET+2CE + SXI would observe few a tens of such events per year. For TAP and Einstein Probe a similar fraction ($\sim 40-50 \%$) of events are expected to be observed off-axis. This result demonstrates the fundamental role of WFX-ray telescopes in synergy with GW detectors for the detection of off-axis emission from BNS mergers that would be otherwise undetactable for $\gamma$-ray instruments. We finally evaluated how the choice of a larger value of the angular extension of the jet wings, $\theta_w$ (used in \emph{Afterglowpy}), could increase the number of detections. Increasing $\theta_w$ also the maximum viewing angle at which the source is detectable increases. We estimated that increasing $\theta_w$ from 15 deg (30 deg) to 45 deg, the number of joint detections increases of a factor $<5\%$ ($\sim 10 \%$) for \emph{Stru1} (\emph{Stru2}).

\subsubsection{Medium and small-FOV sensitive X-ray observatories }
When the GW sources are localized at the level of arcminutes the X-ray afterglow can be directly detected by pointing small-FOV instruments. However, 
there are no BNSs localized at arcminute precision through the GW signal when ET is operating as a single detector, and the number is also negligible when ET is in a network of 3G detectors.
When the GW localization is of order of 1-10 deg$^2$, medium-FOV X-ray instruments can scan the entire localization region by performing mosaic observations. Considering the GW sources localized better than 10 deg$^2$ from Table~\ref{tab_sky_loc_allth},
the sources that can be followed-up by medium-FOV instruments 
are around 10 when ET is operating as a single detector, but several thousands (hundred of thousands) when the ET+CE (ET+2CE) network is observing. However, in the case of ET+CE, the distances of these sources are limited to $z<1.2$. For larger $z$ the way to detect the X-ray counterpart remains through the use of wide-FOV $\gamma$-ray and/or X-ray  satellites. In this case, the more sensitive small-FOV satellites can operate at a later time by pointing the precise localization provided by the wide-FOV satellites, following-up and characterizing the detected counterparts.\\ 

During the activity period of 3G GW detectors, the mission Athena is expected to be operative \citep{Nandra2013}.
The exceptional sensitivity of Athena X-ray telescope and its focal plane instruments will allow us to detect the faint afterglow emission from SGRBs and GW counterparts hours to years (for closer events) after the merger and to make high resolution spectra \citep{Piro2021}. Athena will host the Wide Field Imager (WFI) with a FOV of 0.4 deg$^2$ and the X-IFU high resolution spectrometer with a FOV of 5 arcmin equivalent diameter. The WFI is able to directly point GW signals with sky localization uncertainties smaller than its FOV (a few tens per year for ET+CE and thousands for ET+2CE considering all the orientations of the binary systems), and in addition is capable to carry out a mosaic of sky regions extended up to 10 deg$^2$. 
On the other hand, X-IFU requires a precise localization ($\simeq$2 arcmin) to directly point the source. Even considering a GW network with ET+2CE, there are no sources with such a small uncertainty on the GW sky location. Therefore, instead of being directly triggered by the GW detection, X-IFU can be used either for following up precise localizations delivered by WFX-ray telescopes or by the Athena WFI itself.
Taking into account that Athena will likely have a Time of Opportunity (TOO) response time of 4 hours, we find that the totality of the sources jointly detected by ET+2CE and the WFX-ray telescopes can be detected and followed-up by Athena X-IFU.\\

In order to evaluate the Athena-WFI capabilities to detect GW counterparts, 
we consider the sources localized by the GW detectors within a region of 10 deg$^2$.
We assume an average number of exposures to fully cover the GW sky region equal to $N_{\rm exp}\sim \Delta \Omega_{\rm GW}$/FOV$_{\rm WFI}$, where FOV$_{\rm WFI}$ = 0.4 deg$^2$, and an exposure time $T_{\rm exp}= 10^4$ s for each mosaic observation.  
For simplicity, we work in the approximation that the source is randomly located within the 90$\%$ GW error region and that each position is equi-probable. This implies that, if the GW region is tiled in $N_{\rm exp}$ sub-regions, the probability that the source is located inside the $k^{th}$ sub-region is $P(k)=1/N_{\rm exp}$ and is the same for all the sub-regions. We consider a TOO response time $T_{\rm TOO}$ of 4 hr and we include a dead time interval $T_d$ of 30 minutes corresponding to the time necessary to slew from one sub-region of the mosaic to the adjacent one\footnote{\url{https://www.cosmos.esa.int/documents/400752/400864/Athena+Mission+Proposal/18b4a058-5d43-4065-b135-7fe651307c46}}. Hence, the time necessary to point and detect the source is $ T(k)=T_{\rm TOO}+k\times(T_{\rm exp}+T_d)$. Operatively, for each source we extract a random number $k$ in the range $[0,N_{\rm exp}]$ and we check if at time $T(k)$ after the GW trigger the X-ray afterglow emission is above the sensitivity of Athena-WFI corresponding to $10^4$ s of exposure.\\

Given the several science goals of Athena, only a fraction of its observational time can be realistically dedicated to the follow-up of GWs; specifically we assume one month over one year of observations. While for ET alone the number of sources with $\Delta \Omega < 10 $ deg$^2$ is around 10, and all of them can in principle be followed up by Athena in this total amount of time, the number of the sources detected by ET+CE with $\Delta \Omega < 10 $ deg$^2$ is of several thousands. In order to maximize the effectiveness of the search and not lose time on sources without any chance of detection, a further selection of the sources to be followed-up results to be necessary. We select the sources by excluding all those with $\theta_v > 50^{\circ} $ (for which the error on $\theta_v$ is relatively small, see Fig.~\ref{th_v}, and thus can be reliably considered off-axis) and $z>0.5$.
We select randomly among these sources ($\sim130$) up to reach a total amount of time of one month of observation, and find that Athena-WFI is able to detect $N_{\rm GW+ Athena}=\epsilon_{\rm FOR}\times (6_{-3}^{+2})$ events per year, where $\epsilon_{\rm FOR}$ is a factor that takes into account the TOO efficiency and it corresponds to a field of regard of $\sim50\%$ (i.e. only a fraction $\epsilon_{\rm FOR}$ of the events can be successfully followed-up by Athena). Including another CE in the GW network does not increase significantly the number of Athena detections ($N_{\rm GW+ Athena}=\epsilon_{\rm FOR}\times 8_{-2}^{+3}$ detections per year), because even increasing the number of sources well localized, the maximum number of Athena detections is anyway limited by the time dedicated to TOO. If we exclude more sources by lowering the threshold on the viewing angle to $\theta_v < 30^{\circ} $ instead of $\theta_v < 50^{\circ}$, the number of joint detection increases to $\epsilon_{\rm FOR}\times14^{+4}_{-3}$/yr  ($\epsilon_{\rm FOR}\times27^{+3}_{-3}$/yr) for ET+CE (ET+2CE). The knowledge about the luminosity distance and the viewing angle coming from GW analysis can considerably help for the selection of a golden sample of events that have a larger probability to be detected, giving the possibility to maximize the GW+X-ray joint detection efficiency.

\section{Discussion and conclusions}
Our study evaluates the perspectives for multi-messenger astronomy in the ET era focusing on the search of high-energy signals, which is a unique way to detect EM counterparts of GW signals at high redshift. 
We estimate the expected rate of joint detections of GWs and $\gamma$-ray and/or X-ray signals from BNS mergers considering prompt and afterglow emissions, different GW detector configurations (ET, ET+CE, ET+2CE), and different observational strategies (survey and pointing mode) for several satellites. Our theoretical framework starts from an astrophysically-motivated population of BNS mergers and predicts the high-energy signals associated with these mergers making the BNS population able to reproduce all the statistical properties of currently observed SGRBs. This approach makes it possibile to reproduce not only the observational features of SGRBs, such as luminosity, duration and spectral properties, but also the average detection rate with current $\gamma$-ray instruments, such as \emph{Fermi}-GBM. This enables us to normalize the number of joint detections, evaluate the fraction of BNS producing a jet, and be less dependent of the BNS merger absolute number, which is still largely uncertain. \\

Our main results are summarized as follows:
\begin{itemize}
    \item Regarding the joint detection of GWs and $\gamma$-ray emission, 
we find that already considering ET alone, more than $60\%$ of all the SGRBs having a detectable prompt emission will also have a detectable GW counterpart. This percentage approaches $100\%$ if we include CE as an additional interferometer in the network. From a few tens to hundred joint detections are expected per year, mainly depending on the FOV of the high energy-satellites. However, although increasing the number of joint detections is important for statistical studies, characterizing the sources and identifying the host galaxies is of primary importance. As shown by \emph{Swift} in many years of observations, this is possible and effective when the source is localized at arcmin level enabling to drive the follow-up of ground-based telescopes. 
Mission concepts like THESEUS-XGIS, will be essential to detect and localize high-energy counterpart of BNS mergers with enough precision (arcmin uncertainty). This is particularly important when the sources are located at large redshift, namely $z>1$ where also in the optimistic scenario of a network of 3G GW detectors, the GW sky localization will be larger than 1 deg$^2$ from the GW signals. Instruments such as HERMES (full constellation of cube-satellites) can also be a good compromise of number of detections and sky localization uncertainties which are however order of a few deg$^2$.
\item Considering the possible shock breakout of the cocoon produced by the interaction of the jet with the NS merger ejecta, we show that sensitive (more than the current ones) and wide FOV $\gamma$-ray are required.
\item Regarding the joint detection of GWs and X-ray emission, we demonstrate  the important role of WFX-ray missions, such as SVOM, Einstein Probe, \textit{Gamow}, THESEUS, and TAP. We predict tens of detections per year  when these instruments operates in survey mode. The joint detection rate in survey mode is limited by the chance of having the source inside the FOV at the moment of the merger. We propose an additional observational strategy, the pointing strategy, which could in principle enhance the probability of joint detection exploiting the information about sky localization provided by GW instruments (see below).
\item ET operating as a single observatory can detect several hundreds of BNS merger per year with $\Delta \Omega <100$ deg$^2$, among them about $\sim70$ BNS/yr have a viewing angle $\theta_v<15^{\circ}$, i.e. sources with potentially detectable high-energy emission. The inclusion of CE in the network significantly improves the localization capabilities by reaching several thousands (hundreds with $\theta_v<15^{\circ}$) of BNS detections per year with $\Delta \Omega <10$ deg$^2$. The network of CE+ET is able to localize a few hundreds (a few tens with $\theta_v<15^{\circ}$) with $\Delta \Omega <1$ deg$^2$ up to redshift 0.3. This number increases to thousands (hundreds on-axis) per year up to redshift 1 with the network of ET+2CE. 
\item Considering WFX-ray telescopes slewing to the sky position provided by GW detectors for sources with a good sky localization (better than 100 deg$^2$), we evaluate that the X-ray afterglow emission from ten to hundreds SGRBs could be detected during one year of observation assuming a response time $t_{\rm resp}=$100 s.
\item A rapid communication of the GW sky location and a rapid repointing of the high-energy satellites is crucial to maximize the efficiency of the poiting strategy. Indeed, due to the typical rapid decline of the X-ray flux of SGRBs, we showed that a longer delay time between the merger and the beginning of X-ray monitoring would decrease the probability of detection by a factor $\sim$ 10 or more going from $t_{\rm resp}=$100 s to $t_{\rm resp}=$ 1 hr. The low-frequency sensitivity of ET will make it possible 
to provide a good sky localization even several minutes before the merger \citep[see][Banarjee et al. in prep.]{Chan2018,Li2021,Tito2021}. This will allow us to send information about the sky location of the source well before the merger and X-ray telescope can realistically start earlier the slewing procedure. We evaluate that a not negligible fraction of GW events, localized within 100 deg$^2$ at merger, are detectable tens of minutes before the BNS merger with a sky-localization well within the FOV of the WFX-ray satellites.
\item While the number of detections for WFX-ray satellites in survey mode is reliable, the number of detections in pointing is optimistic and relies on a perfect prioritization of the trigger to be followed (see below).
\item We show that 
using WFX-ray instruments makes it possible to detect a significant fraction of BNS merger which are too off-axis to have a detectable $\gamma$-ray emission and that would be otherwise missed in absence of these instruments. 
The presence of WFX-ray monitors is also fundamental 
to provide the spectral coverage at smaller energies than $\gamma$-rays and to precisely localize the source. The WFX-ray telescopes can provide sky location accuracy of $\sim$ arcmin, which enables to trigger follow-up observations by ground-based optical telescopes, radio arrays and exceptional sensitive X-ray instruments, such as the X-IFU instrument onboard of Athena. We evaluate that the totality of the sources jointly detected by the GW detectors and the WFX-ray telescopes can be detected and followed-up by the Athena X-IFU.
\item The network of ET+CE and ET+2CE is expected to localize a large number of BNS signals with $\Delta \Omega <10$ deg$^2$. An instrument such as the WFI (FOV of $0.4$ deg$^2$) on-board of Athena is able to carry out a mosaic of these sky regions by detecting from a few to a few dozen signals by using 1 month of Athena observations. The joint detection number comes from an observational strategy which maximizes the chance of detections by removing the BNS signals with smaller probability to have observable jet. In particular, this is done by removing BNS with larger viewing angles and distances.
\item Due to the high number of BNS events expected to be released by the next generation GW detectors,  for the pointing observational strategy, it will be crucial to select and prioritize the ones to be followed-up. The selection based on sky-localization will not be enough to avoid to lose observational time of the high-energy satellites when the network ET+CE and ET+2CE will operate. On the basis of the different scientific science goals, specific selection based on the GW source parameters such as distance and viewing angles, but also pre-merger sky-localization, will be mandatory to maximize the joint detection efficiency. This shows the importance in the ET era to send in low-latency (updating) information about GW parameter estimation.
\end{itemize}

Our work shows that to maximize the science return of 3G GW detectors in the multi-messenger context, it is necessary to develop sensitive wide-FOV high-energy instruments able to detect but also localize a large number of short GRBs.  Instruments, such as THESEUS, combining $\gamma$-ray  and WFX-ray telescopes, are of primary importance to guarantee the observation of $\gamma$-ray and X-ray counterparts, by detecting $\sim 10-100$ well localized SGRBs up to high redshifts, for which also the host galaxy can be identified. This is of primary importance for evaluating the cosmological parameters and testing general relativity at cosmological scales. WFX-ray telescopes are required to detect GRB off-axis and  make it possible to explore the jet structure, its interconnection with the ejecta, and the nature of low-luminosity SGRBs with a unique level of detail.
Medium-size X-ray instruments, such-as Athena WFI, will become effective (operating mosaic of the GW sky-localization) for the X-ray counterpart search when a network of GW detector will observe. Mosaic observations with deep exposure by Athena WFI can enhance the possibility to detect off-axis X-ray emission. Sensitive instruments such as Athena X-IFU will be crucial to follow-up and characterize the X-ray emission at later times with respect to the merger.

\section{Acknowledgements}
 We thank Yann Bouffanais, Boris Goncharov, Andrea Maselli, Om Sharan Salafia, Giulia Stratta for valuable discussions. We thank Michele Maggiore, Francesco Iacovelli, Michele Mancarella, and Stefano Foffa for checking the gravitional-wave results and for very helpful discussions. We acknowledge Stefano Bagnasco, Federica Legger, Sara Vallero, and the INFN Computing Center of Turin for providing support and computational resources. We thank Ssohrab Borhanian, Bangalore Sathyaprakash,  Man Leong Chan, Yufeng Li, Ik Siong Heng for useful information and cross-check on the gravitational-wave data analysis. We thank Lorenzo Amati, Fabrizio Fiore, Paul O'Brien and Nicholas White for useful information about some of the X-ray instruments. We thank Maria Grazia Bernardini for the information about SVOM. MB and SR acknowledge financial support from MIUR (PRIN 2020 grant 2020KB33TP$\_$001). MB and GO acknowledge financial support from the AHEAD2020 project (grant agreement n. 871158). BB, GG and MB acknowledge financial support from MIUR (PRIN 2017 grant 20179ZF5KS). GG acknowledges ASI-INAF agreement n. 2018-29-HH. MM and FS acknowledge financial support from the European Research Council for the ERC Consolidator grant DEMOBLACK, under contract no. 770017. The research leading to these results has been conceived and developed within the Einstein Telescope Observational Science Board.

\section*{Code availability}
The latest public version of {\sc mobse} can be downloaded from \href{https://gitlab.com/micmap/mobse_open}{this repository}. {\it GWFish} is publicly available at this repository\footnote{\url{https://github.com/janosch314/GWFish}}

\bibliographystyle{aa}
\bibliography{references}

\begin{appendix}

\section{Computation of the observables in the structured jet scenario}

\begin{figure}[t]
     \centering
     \begin{subfigure}[h]{0.5\textwidth}
         \centering
         \includegraphics[width=\textwidth]{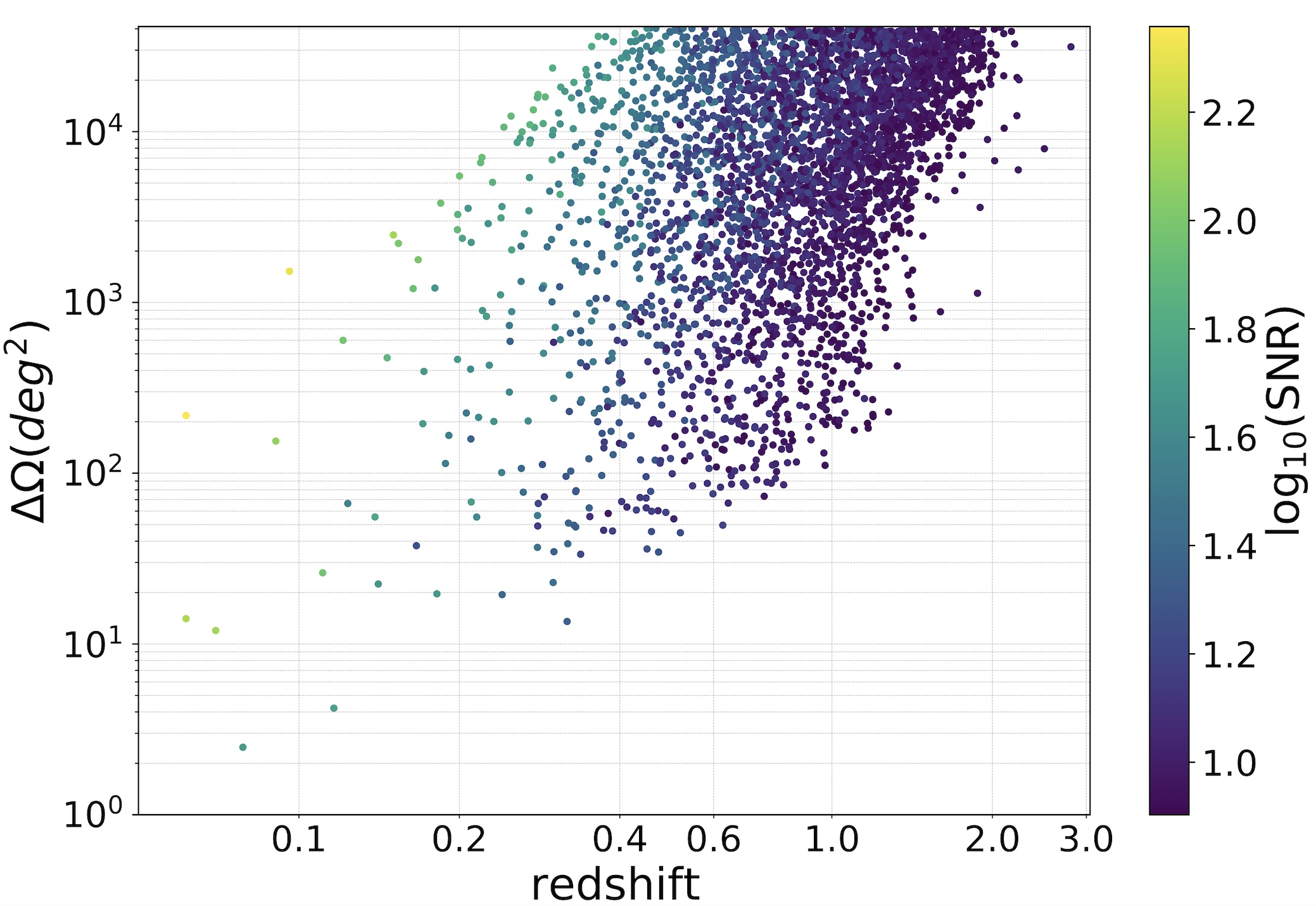}
         \caption{}
         \label{fig:y equals x}
     \end{subfigure}
     \hfill
     \begin{subfigure}[h]{0.5\textwidth}
         \centering
         \includegraphics[width=\textwidth]{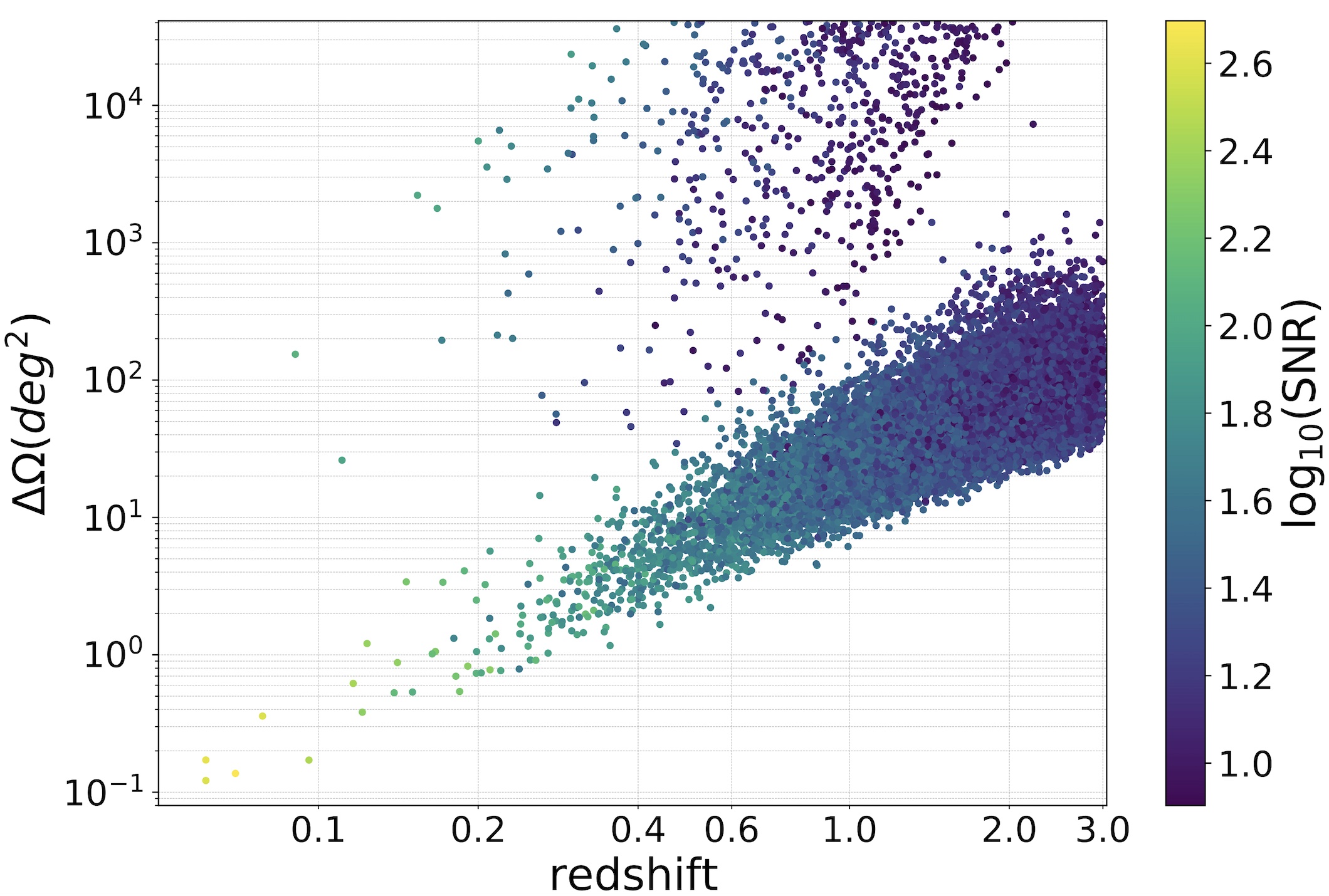}
         \caption{}
         \label{fig:three sin x}
     \end{subfigure}
     \hfill
     \begin{subfigure}[h]{0.5\textwidth}
         \centering
         \includegraphics[width=\textwidth]{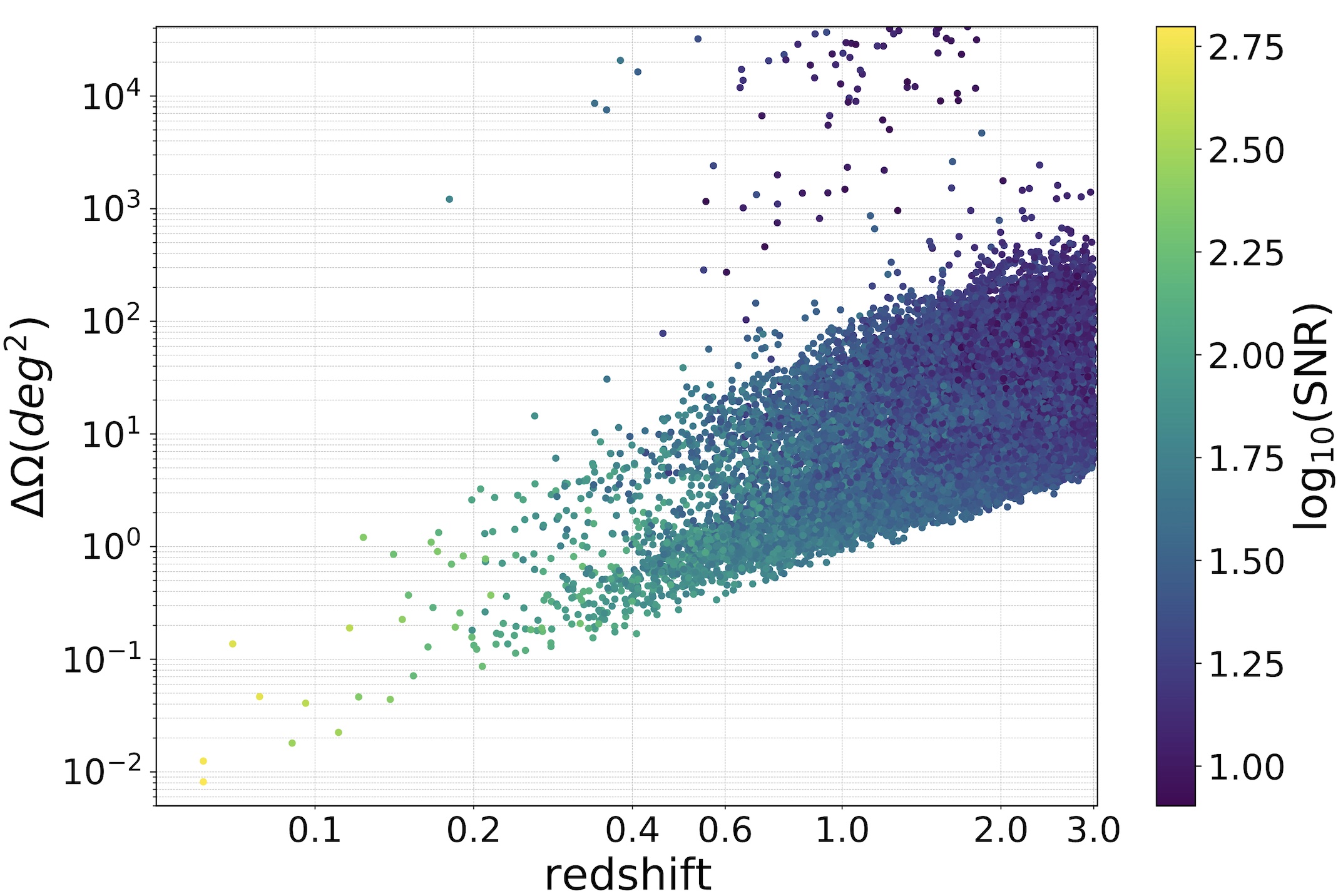}
         \caption{}
         \label{fig:five over x}
     \end{subfigure}
        \caption{Sky localization (90$\%$ credible region in deg$^{2}$) as a function of the redshift for one year of BNS injections with viewing angle in the range $0^{\circ}<\theta_v<15^{\circ}$ and  considering ET (a), ET+CE (b) and ET+2CE (c). The color bar indicates the SNR of each detection. A duty cycle of 0.85 has been assumed for the GW detectors as described in the text.}
        \label{sky_loc_et}
\end{figure}

\begin{figure}
    \centering
    \includegraphics[width=1.0\columnwidth]{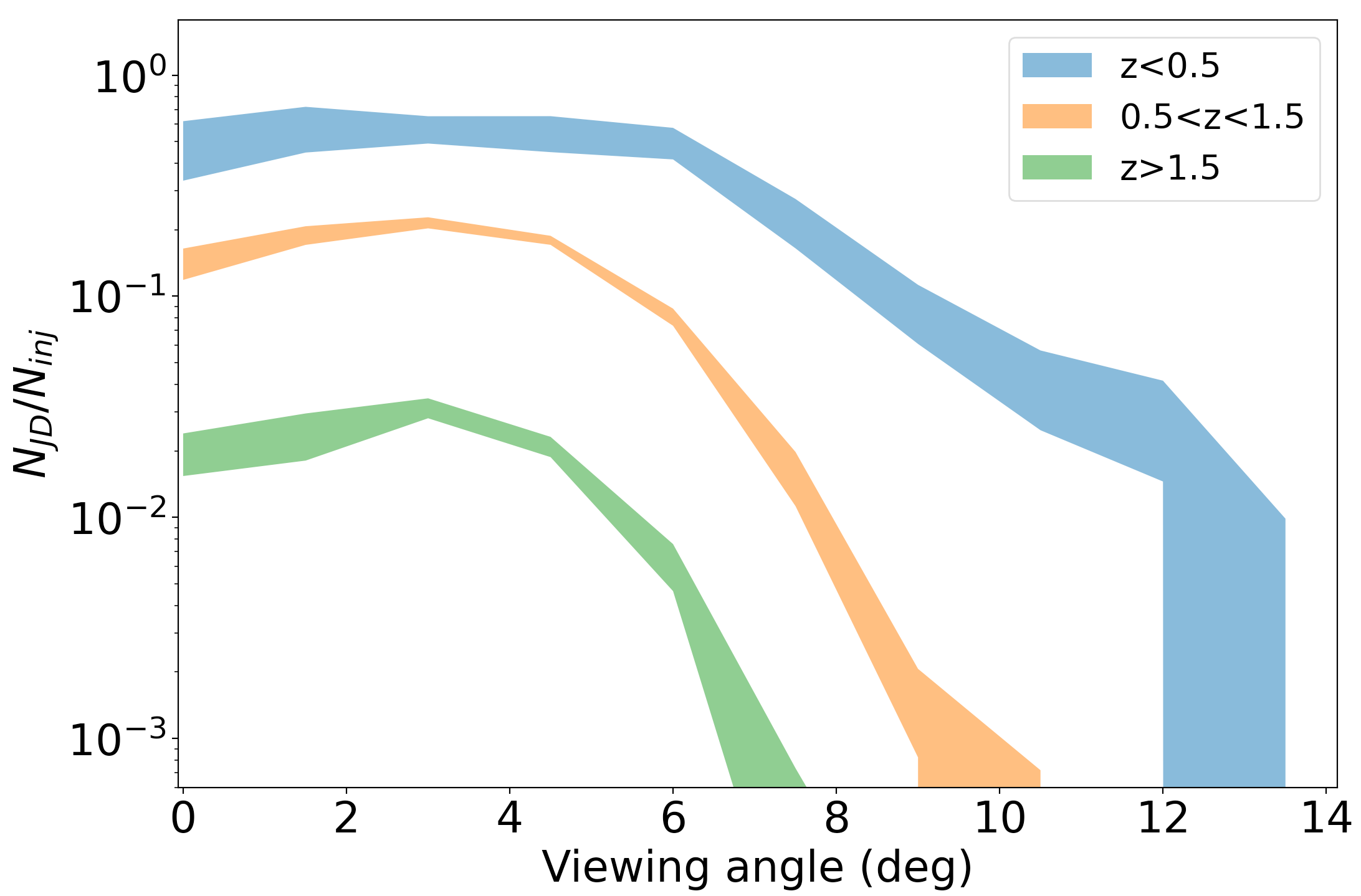}
    \caption{Same as Fig. \ref{JD_vs_angle}, but for \emph{Stru2}.}
    \label{JD_vs_angle_s2}
\end{figure}

\begin{figure}[]
     \centering
     \begin{subfigure}[h]{0.45\textwidth}
         \centering
         \includegraphics[width=\textwidth]{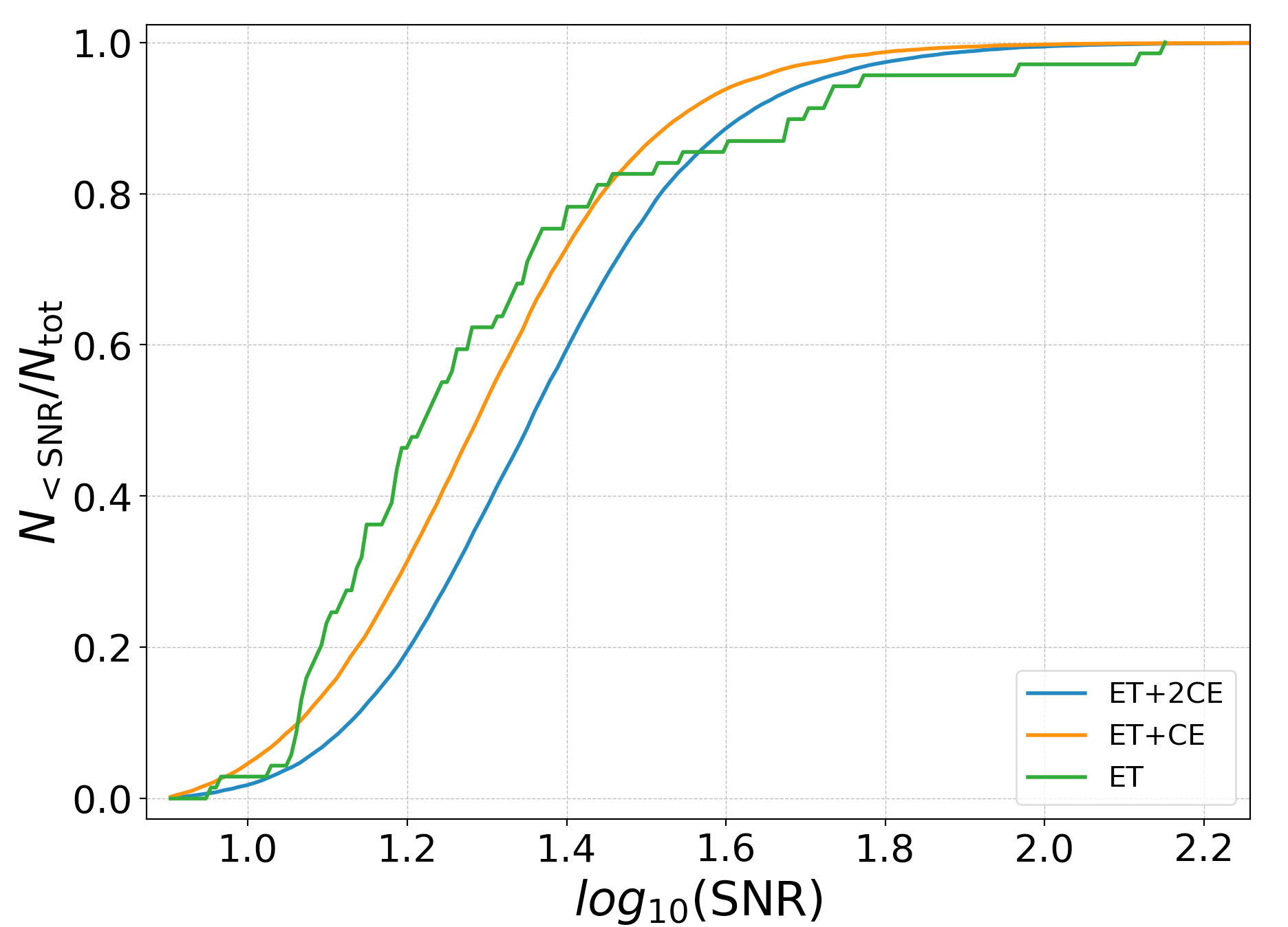}
         \caption{}
         \label{fig:y equals x}
     \end{subfigure}
     \hfill
     \begin{subfigure}[h]{0.45\textwidth}
         \centering
         \includegraphics[width=\textwidth]{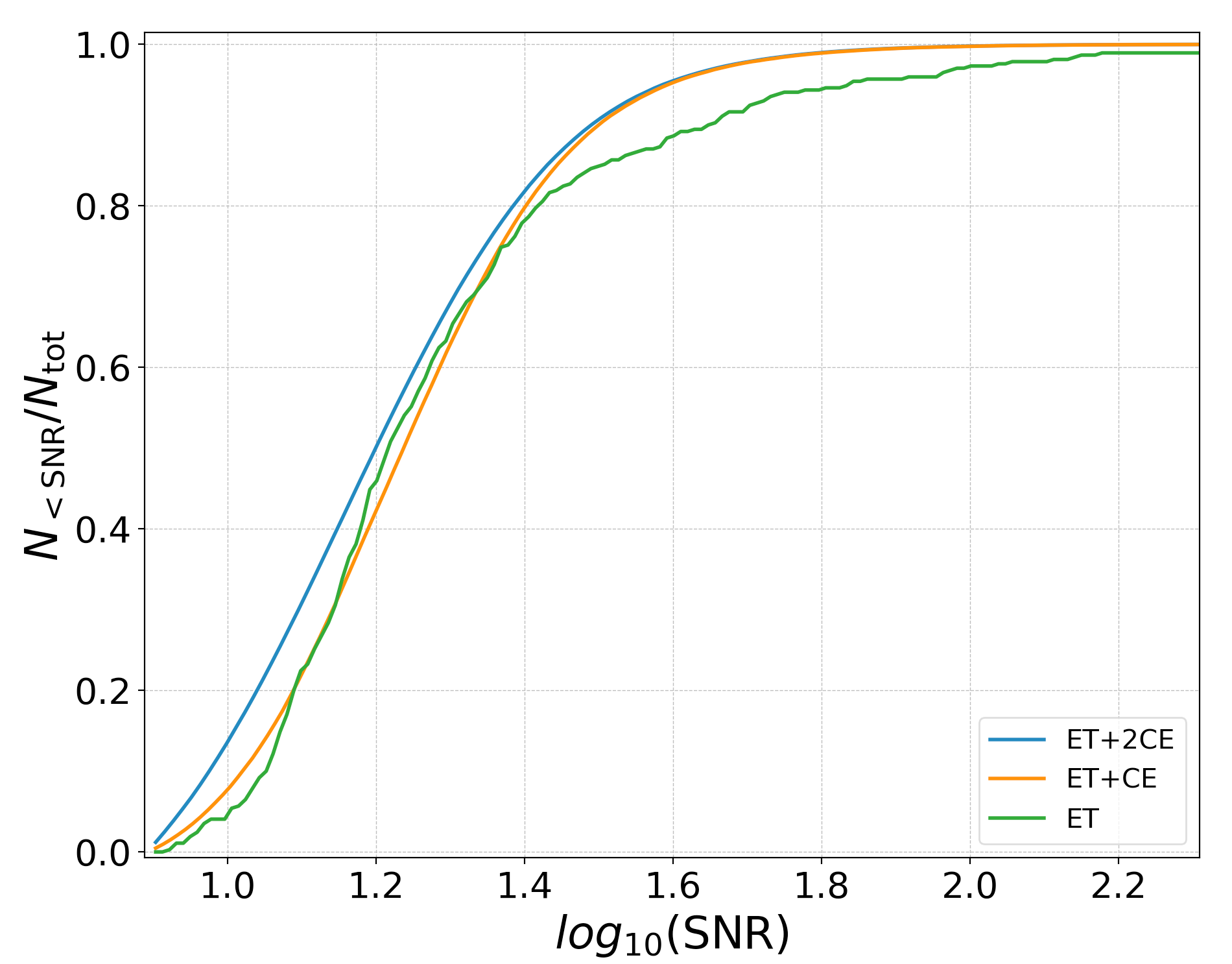}
         \caption{}
         \label{fig:three sin x}
     \end{subfigure}
     
        \caption{Cumulative distribution of the SNR for BNS mergers localized better than $100$ deg$^2$ for the three GW networks considered in this work. Panel (a) shows cases with $\theta_v<15^{\circ}$, while in panel (b) all angles are considered.}
        \label{SNR}
\end{figure}

\begin{figure}
    \centering
\includegraphics[width=1.0\columnwidth]{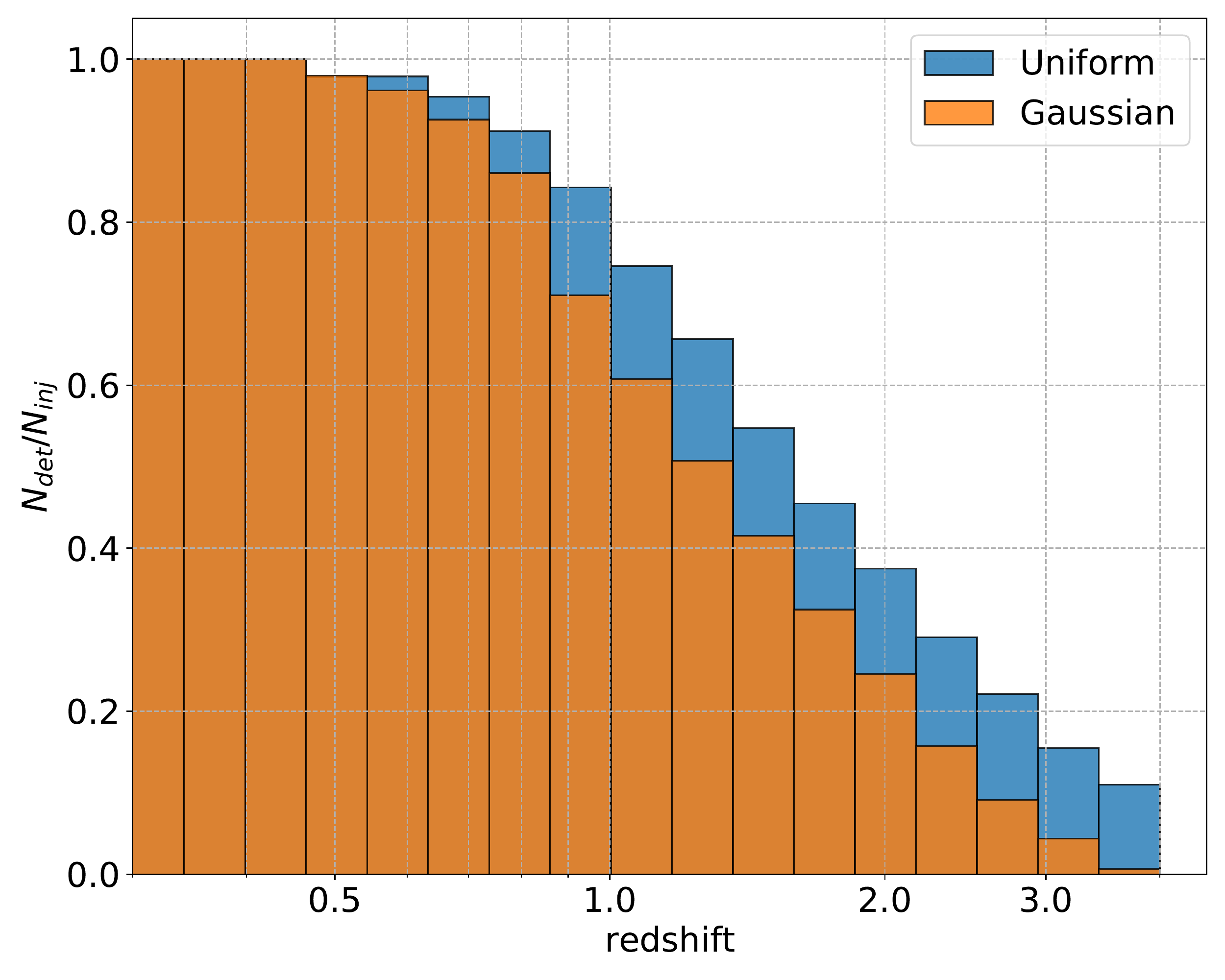}    
\caption{The plot shows the detection efficiency of ET as a function of redshift assuming a Gaussian distribution ($\expval{M}=1.33 M_{\odot}$, $\sigma_{M}=0.1 M_{\odot}$) and a uniform distribution ($M \in [1.0-2.5] M_{\odot}$). Only cases with $\theta_v<15^{\circ}$ are considered. A duty cycle of 0.85 has been assumed for the GW detectors as described in the text. The ratio between the orange bars and the blue bars corresponds to the function $\phi(z)$ defined in the text.}
    \label{de}
\end{figure}

\begin{table}[t]
\centering
\begin{tabular}{|c|c|}
\hline
parameter & prior interval\\\hline
$\lambda_E$             & [1,12]    \\\hline
$\log_{10}{(E_t/10^{49} \text{erg/s})}$         &  [-2,$\log_{10}(50)$] \\\hline
$\log_{10}{(\mu_E/\text{keV})}$         & $[\log_{10}(700),\log_{10}(3000)]$             \\\hline
$\sigma_E$              & [0.05,1]    \\\hline
$\log_{10}{(\mu_{\tau}/s)}$     &  $[-2,\log_{10}(4)]$  \\\hline
$\sigma_{\tau}$             & [0.05,1]    \\\hline
$\log_{10}{f_j}$        & [-3,0]  \\\hline

\end{tabular}
    \caption{Prior intervals for the prompt emission model. For $\lambda_E$, $\sigma_E$ and $\sigma_{\tau}$ we consider a flat prior distribution, while for the others the distribution is flat in logarithm.}
    \label{priors}
\end{table}

\begin{table}[t]
\centering
\begin{tabular}{|c|c|c|c|c|}
\hline
& \multicolumn{4}{c|}{$N_{\Delta \Omega <1000 \rm \,deg^2, t_{\rm pre}}/N_{\Delta \Omega <100 \rm \,deg^2, t_{\rm merger}}$}\\\hline
     & \multicolumn{2}{c|}{SNR$>8$} & \multicolumn{2}{c|}{SNR$>7$}\\\hline
   $t_{\rm pre}$  & 10 min & 20 min& 10 min & 20 min \\\hline
ET &    $63 \%$ &  $29 \%$  & $68\%$  & $37\%$   \\\hline

ET+CE &   $5\%$ &  $0.5\%$ & $5\%$   &  $0.7\%$  \\\hline

ET+2CE &   $6 \%$ &   $0.5\%$& $6\%$  & $0.6\%$   \\\hline

\end{tabular}
    \caption{We show the fraction $N_{\Delta \Omega <1000 \rm \,deg^2, t_{\rm pre}}/N_{\Delta \Omega <100 \rm \,deg^2, t_{\rm merger}}$ of BNS localized better than 100 deg$^2$ at the merger that are localized better than 1000 deg$^2$ at a pre-merger time $t_{\rm pre}$, for ET, ET+CE and ET+2CE. We show both cases that are detected with SNR$>8$ and SNR$>7$ at the pre-merger time.  We consider only BNS observed at a viewing angle $\theta_v<15^{\circ}$.}
    \label{pre_merger}
\end{table}

In this section we show how the peak flux, the fluence and the isotropic energy are derived in the structured jet framework. The derivation is done assuming a unique jet structure, with fixed values of $R_0$ and $\Gamma$, defined in the main text. For a given viewing angle $\theta$, we write the rest frame  photon spectrum as
\begin{equation}
\label{n_p}
    N(E,\theta,t)=N_0(\theta,t) f(E/E_0(\theta,t))
\end{equation}
where we included the time dependency only in the normalisation, assuming that there is not a strong spectral evolution during the GRB pulse. $f(E/E_0(\theta))$ is the spectral shape and it is assumed to be a smoothly broken power law, whose spectral slopes are assumed to be angle independent. $f$ is normalised in such a way that $f(1)=1$. $E_0(\theta,t)$ is the peak energy and we express it as
\begin{equation}
    E_0(\theta,t)=E_0(\theta,t=t_p)P_E(t/t_p,\theta)
\end{equation}
where $t_p$ is the peak time of the pulse and $P(t/t_p\theta)$ describes the temporal profile, in principle angle-dependent. Analogously, we express the time behaviour of the photon spectrum normalisation as:
\begin{equation}
    N_0(\theta,t)=N_0^p(\theta) P_N(t/t_p,\theta)
\end{equation}
In both cases, we approximate the temporal profiles of peak energy and normalisation as
\begin{equation}
P_X(t)=\begin{cases}
t/t_p & 0<t<t_p \\
(t/t_p)^{-\alpha_X(\theta)} & t>t_p
\end{cases}
\quad X=E,N
\end{equation}
At the peak time we write the photon spectrum as
\begin{equation}
\label{n_p}
    N^p(E,\theta)=N_0^p(\theta) f(E/E_0(\theta))
\end{equation}
We define also
\begin{equation}
   R_E(\theta)=E_0(\theta)/E_0(\theta=0)=E_0(\theta)/\hat{E}_0
\end{equation}
and 
\begin{equation}
\label{r_f}
   R_F(\theta)=\frac{N^p(\theta)|_{E=E_0(\theta)}}{\hat{N}^p|_{E=\hat{E}_0}}
\end{equation}
The functions $\alpha_E(\theta)$,  $\alpha_N(\theta)$, $R_E(\theta)$ and $R_F(\theta)$ are computed numerically using the structured jet model. Hereafter the notation $\hat{x}$ indicates that the quantity $x$ is evaluated at viewing angle $\theta=0$. In the observer frame, the peak photon spectrum is:
\begin{equation}
    N^p(E,\theta)=(1+z)^2N_0^p(\theta)f\left(\frac{1}{R_E}\frac{(1+z)E}{\hat{E}_0}\right)
\end{equation}
and using eq. \ref{r_f}
\begin{equation}
    N^p(E,\theta)=(1+z)^2\hat{N}_0^p R_F f\left(\frac{1}{R_E}\frac{(1+z)E}{\hat{E}_0}\right)
\end{equation}
The value of $\hat{N}_0^p$ is derived from the rest frame on-axis isotropic energy $\hat{E}_{\rm iso}=E_t/(1-\cos{\theta_c})$, where $E_t$ is extracted from the assumed probability distribution defined in the main text. In the specific,
\begin{equation}
   \hat{E}_{\rm iso}=4 \pi D_L^2 \int dt \int_ {1\text{ keV}}^{10^4 \text{ keV}} F_{\nu}(t) dE
\end{equation}
where $D_L$ is the luminosity distance and $F_{\nu}$ is the flux density. Using eq.~\ref{n_p}:
\begin{equation}
   \hat{E}_{\rm iso}=4 \pi D_L^2 \int  dt \hat{N}_0(t)  \hat{E}_0^2(t)\int_ {1\text{ keV}/\hat{E}_0(t)}^{10^4 \text{ keV}/\hat{E}_0(t)} xf(x) dx 
\end{equation}
which can be written as
\begin{equation}
\label{e_iso}
   \hat{E}_{\rm iso}=4 \pi D_L^2  \hat{N}_0^p \int   \hat{P}_N(t/t_p) \hat{E}_0^2(t) I_E(t) dt
\end{equation}
where $I_E(t)=\int_ {1\text{ keV}/\hat{E}_0(t)}^{10^4 \text{ keV}/\hat{E}_0(t)} xf(x) dx$

This procedure allows us to write $\hat{N}_0^p$ in terms of known quantities, namely
\begin{equation}
   \hat{N}_0^p= \frac{\hat{E}_{\rm iso}}{4 \pi D_L^2 \int   \hat{P}_N(t/t_p) \hat{E}_0^2(t) I_E(t) dt}
\end{equation}
Once $\hat{N}_0^p$ is determined, the peak photon flux integrated in the energy band of a given instrument is
\begin{gather}
         F_p(\theta)  =\int_{(1+z)E_1}^{(1+z)E_2} N^p(E,\theta)dE \\
        = (1+z)\hat{N}_0^p R_F (R_E \hat{E}_0) \int _{(1+z)E_1/R_E\hat{E}_0}^{(1+z)E_2/R_E\hat{E}_0} f(x) dx
\end{gather}
The duration of the burst $t_{90}$ is computed imposing that the 10-1000 keV fluence accumulated up to $t_{90}$ is 90$\%$ of the total fluence, namely
\begin{gather}
   \int_0^{t_{90}} dt\int_{(1+z)\text{ keV}}^{(1+z)10^4 \text{ keV}} E N(E,\theta,t)dE\\
   =90\% \int_0^{\infty} dt\int_{(1+z)\text{ keV}}^{(1+z)10^4 \text{ keV}/} E N(E,\theta,t)dE
\end{gather}
For computational reasons, the last integral in time is performed up to a time $t_{max}$, which is the maximum time for which the $\gamma$-ray emission is still detectable by the instrument. Finally the isotropic luminosity is defined as $E_{\rm iso}/t_{90}$.

\section{Detection efficiency for different neutron-star mass distribution}

In Fig.~\ref{de} we show the detection efficiency (detected signals over injected ones) of ET as a function of redshift for a Gaussian mass distribution as observed in Galactic double neutron-star systems in radio \citep{Ozel2016} and a uniform distribution of NS mass consistent with gravitational-waves observations \citep{GWTC3pop}. For the Gaussian distribution we fix the mean value to $\expval{M}=1.33 M_{\odot}$ and the standard deviation to $\sigma_{m}=0.1M_{\odot}$. The detection efficiency obtained with the two distributions is consistent up to a redshift equal to 0.5, and then the Gaussian distribution leads to a lower number of GW detections accountable by the fact that the GW signal amplitude scales with the mass.

In order to evaluate the impact on joint detections, we estimate, as  example, the loss of joint GW+$\gamma$-rays detections with ET+\emph{Fermi-GBM}. 
We assume the probability of having a joint detection as
$$
P_{JD}(z)= P_{EM}(z) P_{GW}(z)
$$
where $P_{EM}(z)$ is the probability of detecting the EM counterpart, and $P_{GW}(z)$ the probability of detecting the associated GW signal. Each probability corresponds to the ratio between the number of detected and injected sources.
$$
P_i(z)= \frac{N_{det,i}(z)}{N_{inj}(z)}, \, i=GW,EM
$$
The assumption of NS mass distribution only affects $P_{GW}(z)$, since we specified in the main text that we neglect any dependence on the NS mass for the formation of a jet. We call $P_{GW, flat}(z)$ and $P_{GW, Gauss}(z)$ the GW detection efficiency for flat and Gaussian distributions, respectively.
The variation on the rate of joint detections can be computed as
$$
\frac{dN_{JD, Gauss}}{dz}(z)=\frac{dN_{inj}(z)}{dz} P_{JD, Gauss}(z) =
$$
$$
=\phi(z)\frac{dN_{inj}(z)}{dz} P_{EM}(z) P_{GW, flat}(z)=\phi(z)\frac{N_{JD, flat}}{dz}
$$
where we wrote $P_{GW, Gauss}(z)=\phi(z)P_{GW, flat}(z)$. The total variation, integrated in redshift, is
$$
N_{JD,Gauss}=\int \phi(z) \frac{N_{JD, flat}}{dz} dz
$$
In Fig.\ref{de} the function $\phi(z)$ corresponds to the ratio between the two histograms. Finally, knowing the function $\phi(z)$ and the redshift distribution of joint detections, we derive the variation in joint detection rate for a Gaussian mass distribution. We obtain that the joint GW+$\gamma$-rays detections with ET+\emph{Fermi-GBM}
decreases of a factor $\sim 27\%$ (from 33/yr to 24/yr), assuming a Gaussian mass distribution.

\clearpage
\begin{figure*}
    \centering
    \includegraphics[width=1.0\textwidth]{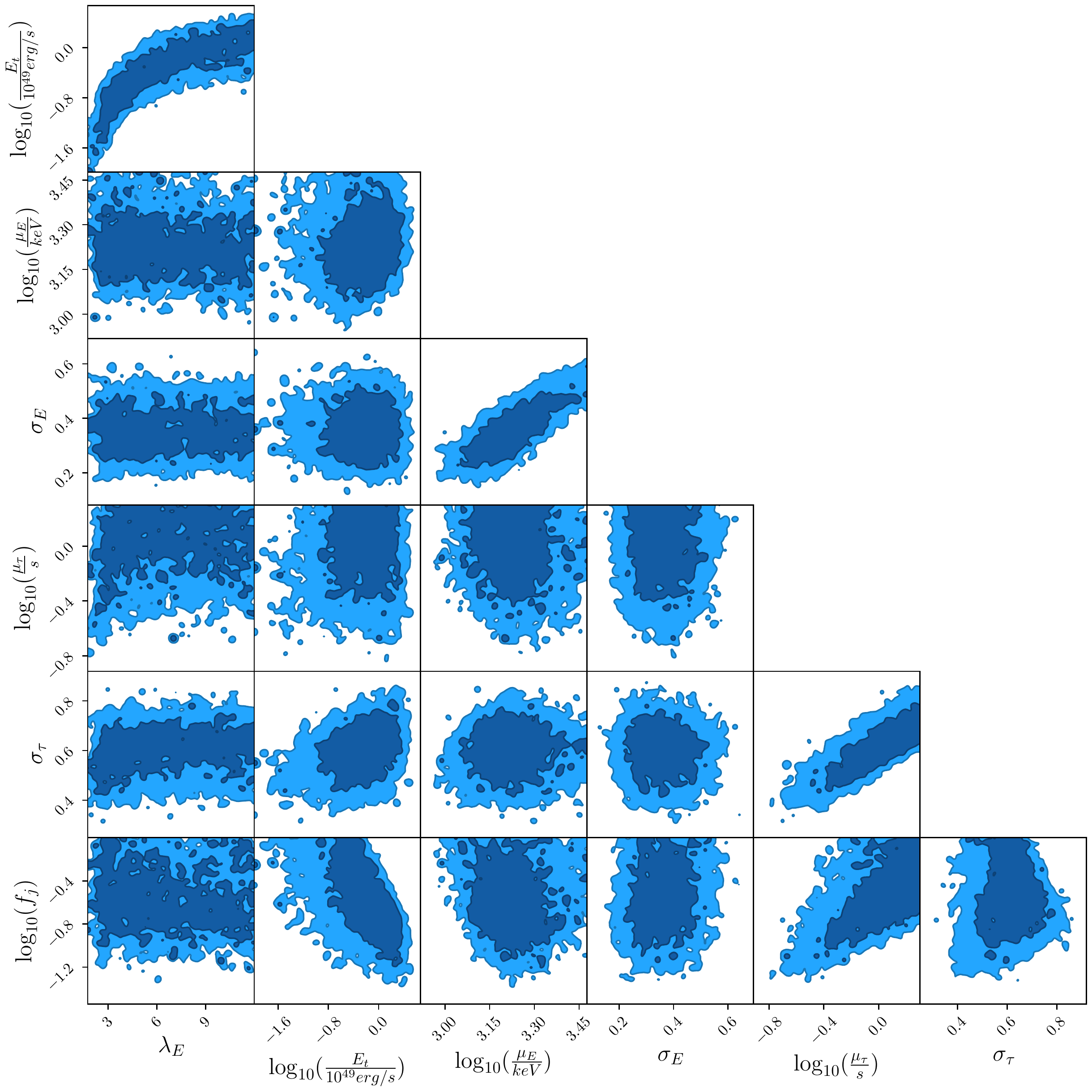}
    \caption{Corner plot of the posterior distribution of the parameters adopted for the prompt emission model. The assumed structure is \emph{Stru1}. The number of MCMC steps is chosen such that the auto-correlation time reaches a plateau, as described in \cite{emcee}.}
    \label{corner_s1}
\end{figure*}

\begin{figure*}
    \centering
    \includegraphics[width=1.0\textwidth]{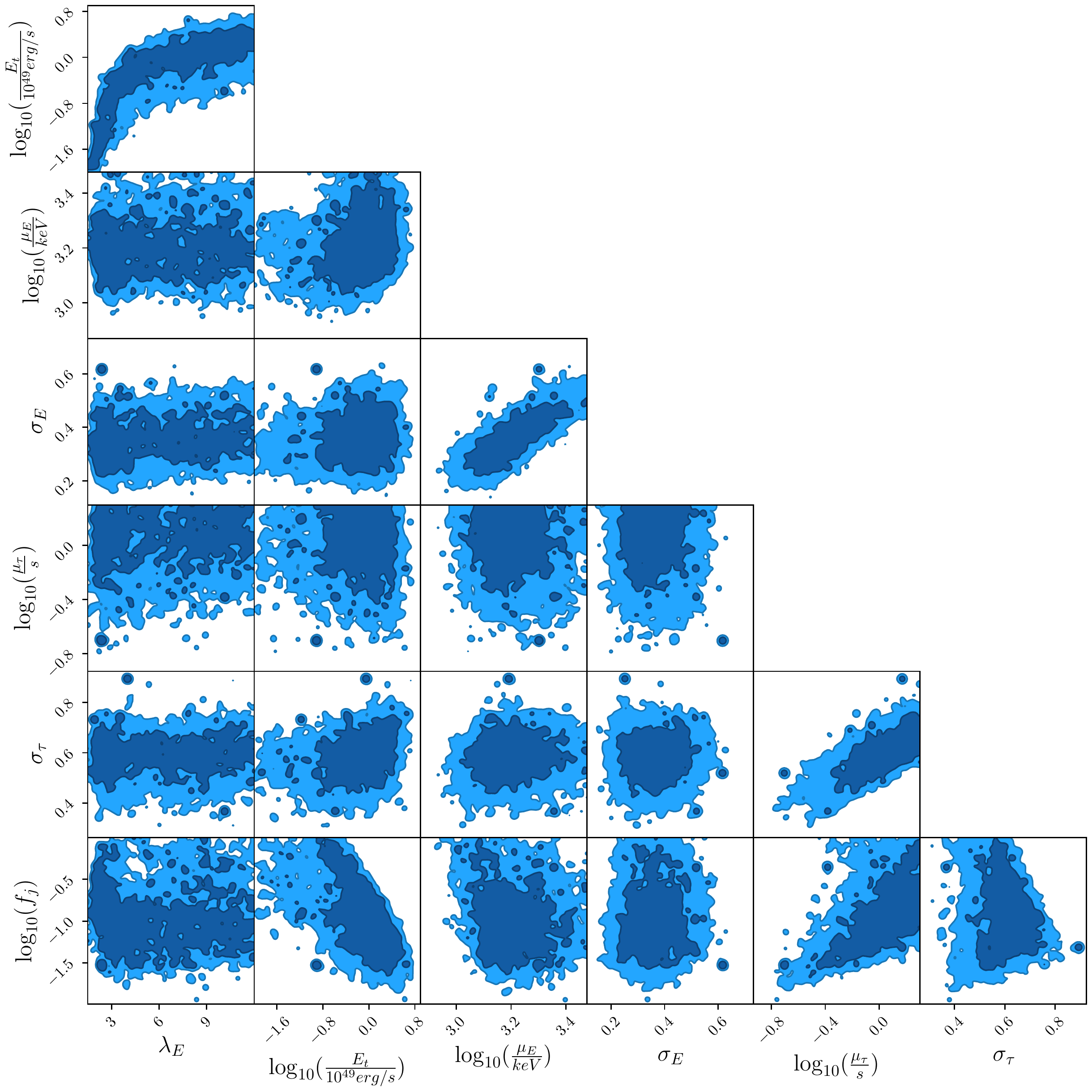}
    \caption{Same as Fig.~\ref{corner_s1}, adopting \emph{Stru2}}
    \label{corner_s2}
\end{figure*}
\end{appendix}

\end{document}